\documentclass[pra,letterpaper,aps,10pt,superscriptaddress,twocolumn,floatfix,showpacs]{revtex4}
\usepackage{graphicx}
\usepackage{amsmath}
\usepackage{amsfonts}
\usepackage{amssymb}
\usepackage{epsfig}
\usepackage{epstopdf}
\DeclareGraphicsExtensions{.pdf,.eps,.png,.jpg,.mps}
\usepackage[pdftex]{color}
\usepackage{amsmath,graphicx,amssymb,braket,xcolor,subfigure,upgreek}
\PassOptionsToPackage{subfigure}{tocloft}
\usepackage{fixmetodonotes}
\usepackage[colorlinks, linkcolor=blue, citecolor=blue, urlcolor=blue, breaklinks=true]{hyperref}
\usepackage{microtype}
\usepackage{bbm}
\usepackage{color}
\usepackage{braket}
\usepackage{dsfont}

\usepackage{minitoc}

\usepackage{amsmath,amssymb,amsfonts,latexsym,color,dcolumn,bm,amsbsy}
\usepackage{graphicx}
\usepackage[utf8]{inputenc}
\usepackage{textcomp}
\usepackage{braket}
\usepackage{mathbbol}

\DeclareSymbolFontAlphabet{\amsmathbb}{AMSb}%

\newcommand{\D}{{\cal{D}}}
\newcommand{\Dd}{{\cal{D}^{\dagger}}}
\newcommand{\bd}{b^{\dagger}}

\newcommand{\sd}{\sigma^{\dagger}}
\newcommand{\s}{\sigma}

\renewcommand{\thesubsection}{\Alph{subsection}}

\newcommand{\fref}[1]{Fig.~\ref{#1}}
\newcommand{\fsref}[1]{Figs.~\ref{#1}}
\newcommand{\eqsref}[1]{Eqs.~(\ref{#1})}
\newcommand{\ts}[1]{_\text{#1}}

\newcommand{\Hm}{\mathcal{H}}

\newcommand{\inpt}{\ensuremath{\mathrm{in}}}
\newcommand{\out}{\ensuremath{\mathrm{out}}}

\DeclareMathOperator{\sinc}{sinc}

\newcommand{\nocontentsline}[3]{}
\let\oldaddcontentsline\addcontentsline
\newcommand{\tocless}[2]{%
  \let\addcontentsline\nocontentsline
  #1{#2}
  \let\addcontentsline\oldaddcontentsline}

\begin{document}
\title{Cooperative quantum phenomena in light-matter platforms}
\author{Michael Reitz}
\affiliation{Max Planck Institute for the Science of Light, Staudtstra{\ss}e 2,
D-91058 Erlangen, Germany}
\affiliation{Department of Physics, University of Erlangen-Nuremberg, Staudtstra{\ss}e 7,
D-91058 Erlangen, Germany}
\author{Christian Sommer}
\thanks{Present Address: Alpine Quantum Technologies, 6020 Innsbruck, Austria}
\affiliation{Max Planck Institute for the Science of Light, Staudtstra{\ss}e 2,
D-91058 Erlangen, Germany}
\author{Claudiu Genes}
\affiliation{Max Planck Institute for the Science of Light, Staudtstra{\ss}e 2,
D-91058 Erlangen, Germany}
\affiliation{Department of Physics, University of Erlangen-Nuremberg, Staudtstra{\ss}e 7,
D-91058 Erlangen, Germany}
\date{\today}
\begin{abstract}
Quantum cooperativity is evident in light-matter platforms where quantum-emitter ensembles are interfaced with confined optical modes and are coupled via the ubiquitous electromagnetic quantum vacuum. Cooperative effects can find applications, among other areas, in topological quantum optics, in quantum metrology or in quantum information. This tutorial provides a set of theoretical tools to tackle the behavior responsible for the onset of cooperativity by extending open quantum system dynamics methods, such as the master equation and quantum Langevin equations, to electron-photon interactions in strongly coupled and correlated quantum-emitter ensembles.  The methods are illustrated on a wide range of current research topics such as the design of nanoscale coherent-light sources, highly reflective quantum metasurfaces or low intracavity power superradiant lasers. The analytical approaches are developed for ensembles of identical two-level quantum emitters and then extended to more complex systems where frequency disorder or vibronic couplings are taken into account. The relevance of the approach ranges from atoms in optical lattices to quantum dots or molecular systems in solid-state environments.
\end{abstract}


\maketitle

\tableofcontents

\section{Introduction}
\label{Sec1}

Some of the most intriguing phenomena in nature, both in the classical and quantum domains, are a product of \textit{cooperative} effects, i.e., they cannot be understood by sole consideration of the individual constituents as they arise from the interplay among them. While at the fundamental science level, the understanding of many of those problems in the quantum domain already poses a great intellectual challenge, there is an ever increasing interest to build, control and harness complex cooperative platforms for emerging quantum technologies~\cite{deutsch2020harnessing}.\\
\indent Light-matter platforms provide an optimal playground for the observation and exploitation of quantum cooperative effects. Quantum light, either multimode, as naturally arising in the quantum electromagnetic vacuum or single mode, as confined in the small volume of an optical resonator, can induce strong interactions among quantum emitters (QEs). Cooperativity then occurs e.g.~in free space under high-density conditions and manifests itself in a strongly modified material response owed to a continuous scattering and rescattering of photons between the matter constituents. Two main aspects brought on by the common coupling of an ensemble of $\mathcal{N}$ quantum emitters to an electromagnetic environment are dipole-dipole interactions, stemming from a virtual exchange of photons and collective radiative emission, stemming from the loss of excitation into the infinite number of the electromagnetic vacuum modes. While the former is included as a coherent effect, the latter is an incoherent one, observable either as an increase (\textit{superradiance}) or a decrease (\textit{subradiance}) in the collective spontaneous emission rate compared to that of an isolated emitter.\\
\indent Free-space near-field dipole-dipole interactions are utilized in structured \textit{subwavelength arrays} of quantum emitters where they allow for the hopping of surface excitations. Such arrays are ideal platforms for achieving strong light-matter interactions and high fidelities for photon storage capabilities~\cite{facchinett12016storing,asenjo2017exponential}. Furthermore, these platforms can allow for the propagation of photons or excitations to be protected against disorder and scattering caused by defects, in an approach dubbed \textit{topological quantum optics}~\cite{perczel2017topological,bettles2017topological,perczel2020topological}. An advantage over linear topological photonic systems is the intrinsic nonlinearity of the quantum emitters which could lead to a rich many-body physics dynamics. Individual addressing of a single qubit emitter has been suggested via quantum spin lenses~\cite{glaetzle2017quantum} and an approach to quantum networking with composite quantum systems comprised of many atomic arrays has been proposed building on theoretical results showing the production of a Bell entangled superposition quantum state for two distant arrays~\cite{guimond2019subradiant}.\\
\indent  Free-space subradiance properties are an important resource for applications ranging from quantum-information processing~\cite{chang2018quantum} to metrology~\cite{ostermann2013protected,facchinett12016storing,manzoni2018optimization}. Such cooperative dissipative effects have been extensively studied, especially in the direction of engineering robust many-particle quantum states characterized by extremely long lifetimes~\cite{asenjo2017exponential,zhang2020subradiant}. Subradiance has been experimentally employed to show near-unity ultrathin reflectors~\cite{rui2020asubradiant} with potential applications in nano-optomechanics~\cite{shahmoon2019collective,shahmoon2020quantum}, antiresonance spectroscopy~\cite{plankensteiner2017cavity,plankensteiner2019enhanced} and nonlinear quantum optics~\cite{bettles2020quantum}.\\
\indent Cavity quantum electrodynamics (cavity QED)~\cite{haroche1989cavity,berman1994cavity,walther2006cavity,haroche2013exploring} can additionally mediate and enhance emitter-emitter interactions by interfacing quantum emitters with confined optical resonances. The famous model of \textit{Dicke superradiance}~\cite{dicke1954coherence,gross1982superradiance}, showing the quick burst of spontaneous emission from $\mathcal{N}$ indistiguishable quantum emitters can be simulated in the context of cavity QED and finds application in the operation of superradiant lasers~\cite{bohnet2012asteadystate,norcia2016superradiance}. The combination of subwavelength arrays with plain mirrors allows for the design of hybrid cavities with one or two frequency-dependent end mirrors, superior in performance to ones made up by plain, frequency-insensitive mirrors~\cite{wild2018quantum,cernotik2019cavity}.\\
\indent While many light-matter platforms are designed at the level of quasi-pure quantum emitters, i.e., considering only electron-photon interactions, a variety of extremely promising directions in quantum engineering are utilizing more complex quantum emitters such as organic molecules, quantum dots or color centers. In particular at the level of organic molecules, the strong confinement of light modes in micro-cavities gives rise to a novel research direction into \textit{vacuum-dressed} or \textit{cavity-dressed} materials, i.e., vacuum-hybridized materials with enhanced properties. At the level of mesoscopic systems, changes in charge conductivity~\cite{orgiu2015conductivity,schachenmayer2015cavity,feist2015extraordinary,hagenmuller2017cavity,hagenmuller2018cavity}, energy transfer rates ~\cite{zhong2016non,zhong2017energy}, chemical reactivity~\cite{hutchinson2012modifying,galego2016suppressing,herrera2016cavity,galego2017many,martinezmartinez2018can} have been experimentally observed and theoretically studied. At the level of single molecules, the focus is in producing reliable single quantum emitters as single photon sources with applications in entanglement generation or in optical quantum computing \cite{lettow2010quantum}. Recent results have shown that a cavity-dressed molecule can act as an almost ideal quantum emitter, exhibiting a closed electronic transition~\cite{wang2019turning}. For two or more closely spaced molecules, cooperativity is naturally manifested in the process of F\"{o}rster resonance energy transfer based on energy exchange via near-field dipole-dipole interactions followed by quick vibrational relaxation~\cite{foerster1948zwischen,blankenship2014molecular}, a key process in photosynthetic light harvesting.\\
\indent Many of the aforementioned applications can be understood within the formalism of open quantum system dynamics extended to the more complex problem of correlated matter, such as occurring in a coupled quantum emitter ensemble. To this end, this tutorial utilizes two competing, interconnected approaches, one at the level of the density operator time evolution, i.e., the master equation (ME) and the other following time dynamics of system operators, i.e., the quantum Langevin equations (QLEs) approach. As the two formalisms are standard textbook methods ~\cite{breuer2002theory,weiss1999quantum,scully1997quantum,gardiner2004quantum,vogel2006quantum,walls2012quantum}, this tutorial proceeds with more complex aspects of light-matter interactions such as the emergence of a cooperative ME for coupled quantum-emitter systems in free space, introduced in Sec.~\ref{Sec2A} together with the consequential occurrence of subradiance and superradiance, which are tackled in Sec.~\ref{Sec2B}. Particular aspects such as the single excitation subspace and Dicke superradiance~\cite{dicke1954coherence} are discussed in Sec.~\ref{Sec2C} and Sec.~\ref{Sec2D}, respectively. The master equation allows for the derivation of equations of motion for $\mathcal{N}$ mutually coupled emitters, utilized in understanding on how energy dispersion relations and energy band gaps in one-dimensional (1D) arrays can be engineered (in Sec.~\ref{Sec3A}). Also in 1D chain and ring configurations, applications in quantum metrology, quantum information and lasing are presented in Sec.~\ref{Sec3B}. The optical response of 2D subwavelength arrays is derived in Sec.~\ref{Sec3C} to show the perfect reflection of incoming light around certain collective resonances.\\
\indent Fundamental aspects of cavity QED with correlated quantum emitters are introduced in Sec.~\ref{Sec4A} followed by the input-output theory for operators detailed in Sec.~\ref{Sec4B}. The question of frequency disorder, relevant in the case of more complex quantum-emitter ensembles affected, for example, by inhomogeneous broadening, is tackled in Sec.~\ref{Sec4C}. Applications of the formalism introduced in Sec.~\ref{Sec4} are described in Sec.~\ref{Sec5A} on 1D arrays towards antiresonance spectroscopy applications and in 2D highly reflective arrays employed in hybrid cavities exhibiting Fano-like narrow resonances.  Finally, for three-dimensional mesoscopic ensembles of invertable two-level systems, the theory of Dicke superradiance is applied to the characterization of superradiant lasers inside lossy cavities.\\
\indent The transition to increase the complexity of the considered quantum emitter systems is undertaken in Sec.~\ref{Sec6A} by providing a derivation for electron-vibron interactions in the form of a Holstein Hamiltonian~\cite{holstein1959study}, and emphasizing the application of QLEs at the level of polarons to characterize molecular emission and absorption~\cite{reitz2019langevin,reitz2020molecule}. The effect of cooperative behavior as mediated by dipole-dipole interactions is tackled in Sec.~\ref{Sec6B} to analytically characterize the FRET (F\"{o}rster resonance energy transfer) process as well as for the description of the physics of molecular dimers~\cite{diehl2014emergence}, a standard model used in quantum chemistry.\\

\section{Cooperativity of light and matter}
\label{Sec2}
We provide in this section a quick review of collective effects in an ensemble of free space standing $\mathcal{N}$ quantum emitters with emphasis on the dipole-dipole interactions and collective decay mediated by the electromagnetic vacuum. The emergent properties of subradiance, superradiance and dipole-dipole-induced collective energy shifts are then introduced under the weak excitation assumption (weak driving). Eventually, dynamics in the full extended Hilbert space of the $\mathcal{N}$ emitters is treated by introducing the collective Bloch sphere and the resulting Dicke superradiance regime. This formalism is the starting point for the understanding of applications exemplified in Sec.~\ref{Sec3} and Sec.~\ref{Sec5}.\\

\subsection{Collective radiative emission}
\label{Sec2A}
\indent The standard system we consider in the next sections is that of an ensemble of $\mathcal{N}$ identical quantum emitters modeled for simplicity as hydrogenlike atoms with nuclei fixed in random positions $\mathbf{R}_j$, where $j=1,\hdots,\mathcal{N}$. The (single) electronic degree of freedom for each emitter is described by quantized momentum $\hat{\mathbf{p}}_j$ and position $\hat{\mathbf{r}}_j$ (relative to its respective nucleus) operators appearing in the Hamiltonian $h_j=\hat{\mathbf{p}}_j^2/(2\mu)+V(\hat{\mathbf{r}}_j)$ consisting of the kinetic energy (where the reduced mass is denoted by $\mu$) and an electrostatic potential $V(\hat{\mathbf{r}}_j)$. Out of the infinite set of eigenvectors of $h_j$, we pick the lowest energy one (and set its energy to zero) $\ket{g_j}$ and assume that all the physics described in the following involves transitions to only one excited state $\ket{e_j}$ at energy $\omega_0$ (we set $\hbar=1$). This amounts to a two-level system approximation where the unity of the Hilbert space is a sum of only two projectors $\mathds{1}_j=\ket{g_j}\bra{g_j}+\ket{e_j}\bra{e_j}$. To quantify transitions between the two levels, we introduce  the standard ladder (Pauli) operators $\sigma_j = \ket{g_j}\bra{e_j}$, $\sigma^{\dagger}_j = \ket{e_j}\bra{g_j}$. Their commutator $[\sigma_j^\dagger,\sigma_j]=\sigma_z^{j}$ is the population difference operator. The free Hamiltonian can then be written as ${h}_j=\omega_0 \sigma_j^\dagger \sigma_j$ or alternatively as ${h}_j=\omega_0 \sigma_z^{j}/2$ (up to a constant energy shift).\\
\indent To describe the interaction of the emitters with the electromagnetic vacuum, one introduces a fictitious perfectly reflecting box of volume $\mathcal{V}=\ell^3$ and follows a standard quantization procedure for the electromagnetic field imposing periodic boundary conditions~\cite{breuer2002theory,weiss1999quantum,scully1997quantum,gardiner2004quantum,vogel2006quantum,walls2012quantum,loudon1973the}. This leads to a plane-wave expansion of the electric field operator
\begin{equation}
\label{efield}
\hat{\mathbf{E}}(\mathbf{R}) = \textstyle \sum_\mathbf{k} \mathcal{E}_k \left( a_\mathbf{k} e^{i\mathbf{k} \cdot \mathbf{R}} +  a^\dagger_\mathbf{k} e^{-i\mathbf{k} \cdot \mathbf{R}}\right){\boldsymbol{\epsilon}}_\mathbf{k},
\end{equation}
where the allowed $k$ vectors are multiples of $2\pi/\ell$ on each Cartesian direction and the index $\mathbf{k}$ runs over all possible $k$ vectors and also over the two orthogonal polarizations with unit vectors ${\boldsymbol{\epsilon}}_\mathbf{k}$. The bosonic operators follow the commutation $[a_\mathbf{k},a_\mathbf{k'}^\dagger]=\delta_{\mathbf{k}\mathbf{k}'}$ and their action is to create and destroy excitations in a given field mode. The zero-point electric field amplitude for a given mode with frequency $\omega_k=c k$ is defined as $\mathcal{E}_k=\sqrt{\omega_k/(2\epsilon_0 \mathcal{V})}$ (where $c$ is the speed of light and $\epsilon_0$ is the vacuum permittivity). The Hamiltonian for the field inside the box can then be written as a sum over an infinite number of bosonic modes $\mathcal{H}_\text{vac}=\sum_{\mathbf{k}}\omega_k( a_\mathbf{k}^\dagger a_\mathbf{k}+1/2)$.

\indent The coupling of light and matter occurs within the formalism of the minimal coupling Hamiltonian~\cite{breuer2002theory,weiss1999quantum,scully1997quantum,gardiner2004quantum,vogel2006quantum,walls2012quantum,loudon1973the}
\begin{equation}
\mathcal{H}= \sum_{j=1}^{\mathcal{N}} \frac{[\hat{\mathbf{p}}_j-e\hat{\mathbf{A}}(\mathbf{R}_j+\hat{\mathbf{r}}_j)]^2}{2\mu}+V(\hat{\mathbf{r}}_j)+\mathcal{H}_\text{vac},
\end{equation}
describing the total quantum system of charges and electromagnetic vacuum modes. Notice that the canonical momentum of the electron is shifted to $\hat{\mathbf{p}}_j=\mu\dot{\hat{\mathbf{r}}}_j+e\hat{\mathbf{A}}(\mathbf{R}_j+\hat{\mathbf{r}}_j)$ where $\hat{\mathbf{A}}$ is the vector potential operator of the electromagnetic field. The observation that the size of the electronic orbital is much smaller than the wavelength $\lambda_0=2\pi c/\omega_0$ associated with the optical transition between the two electronic orbitals allows for a simplified form of the Hamiltonian
\begin{equation}
\label{Hfreespace}
\mathcal{H}= \sum_{j=1}^{\mathcal{N}}h_j+\sum_{j=1}^{\mathcal{N}} \hat{\mathbf{d}}_j \cdot \hat{\mathbf{E}}(\mathbf{R}_j)+\mathcal{H}_\text{vac}.
\end{equation}
In this so-called \textit{dipole approximation}, the light-matter interaction Hamiltonian $\mathcal{H}_\text{int}=\textstyle{\sum_{j=1}^{\mathcal{N}} \hat{\mathbf{d}}_j \cdot \hat{\mathbf{E}}(\mathbf{R}_j)}$ involves only the dipole moment operator $\hat{\mathbf{d}}_j=-e \hat{\mathbf{r}}_j$ and the electric field operator evaluated at the position of the nuclei. Under the two-level assumption, the dipole operator is written as $\hat{\mathbf{d}}_j=\mathbf{d}_\text{eg}\sigma_j+\text{h.c.}$, where the transition dipole moment is computed between the two states $\mathbf{d}_\text{eg}=\bra{g_j}\hat{\mathbf{d}}_j\ket{e_j}$ (assuming identical emitters). Moreover, one performs an additional \textit{rotating-wave approximation} (RWA) where fast oscillating terms $\sigma_j a_\mathbf{k}$ and $\sigma_j^\dagger a_\mathbf{k}^\dagger$ (oscillating under free evolution with $\omega_0+\omega_k) $ are neglected. Under these approximations, the light-matter interaction part of the Hamiltonian can be written as
\begin{equation}
\mathcal{H}_\text{int}= \sum_{\mathbf{k}}\sum_{j=1}^{\mathcal{N}} \left(g_\mathbf{k} \sigma_j^\dagger a_\mathbf{k} e^{i\mathbf{k} \cdot \mathbf{R}_j} + g^*_\mathbf{k} a^\dagger_\mathbf{k} \sigma_j e^{-i\mathbf{k} \cdot \mathbf{R}_j}\right),
\end{equation}
describing photon-emitter energy exchanges at rates $g_\mathbf{k}=\mathcal{E}_k {\boldsymbol{\epsilon}}_\mathbf{k}\cdot\mathbf{d}_\text{eg}$. Notice that this interaction Hamiltonian conserves the number of excitations in the system.\\
\begin{figure}[t]
\includegraphics[width=0.99\columnwidth]{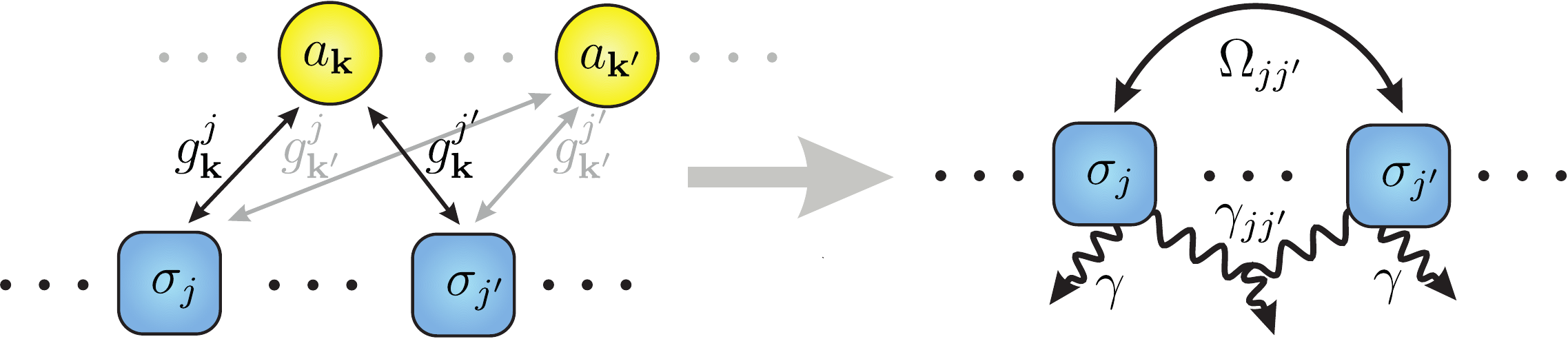}
\caption{Schematics of the procedure followed to obtain the open-system dynamics formulation for a system of $\mathcal{N}$ quantum emitters (denoted by $\sigma_j$ and $\sigma_{j'}$) subject to the quantum electromagnetic vacuum. The key point is that, for a pair of emitters, coupling to the \textit{common reservoir modes} (denoted by $a_\mathbf{k}$ and $a_{\mathbf{k}'}$) gives rise to separation-dependent dipole-dipole interactions at rate $\Omega_{jj'}$ as well as to collective decay rates $\gamma_{jj'}$.}
\label{fig21}
\end{figure}
\indent The Hamiltonian $\mathcal{H}$ governs unitary evolution in an infinite-dimensional Hilbert space (owing to the infinite number of electromagnetic modes). One could then, in principle, use a deterministic Schr\"{o}dinger equation or equivalently the von Neumann equation for the density operator $\rho_\text{tot}(t)$ to compute the state of the system at any time. However, this is an extremely complex task requiring an immense computational resource even for small size systems. Instead, in order to drastically reduce complexity, an open-system dynamics approach is usually followed, which consists in reducing the system of interest to the Hilbert space of dimension $2^\mathcal{N}$ of the matter part only (as schematically illustrated in Fig.~\ref{fig21}). A main observation allowing such reduction of complexity is that the time-dependent light-matter interaction Hamiltonian can be simply expressed in the interaction picture as $\mathcal{H}_\text{int}=\sum_{j=1}^\mathcal{N}[\sigma_j \mathcal{F}_j^\dagger(t)+\sigma_j^\dagger \mathcal{F}_j(t)]$ describing an excitation exchange between the emitters and the field, with the time-dependent operator
\begin{equation}
\label{Fspon}
\mathcal{F}_j(t)=\sum_{\mathbf{k}} g_\mathbf{k} a_\mathbf{k} e^{i\mathbf{k} \cdot \mathbf{R}_j} e^{-i(\omega_k-\omega_0)t},
\end{equation}
acting solely in the Hilbert space of the photon modes.\\
\indent The open-system dynamics procedure is then based on a weak-coupling assumption and implies that the electromagnetic degrees of freedom are traced over to obtain an equation of motion for $\rho(t)=\text{Tr}_\text{em}\left[\rho_\text{tot}(t)\right]$. The steps for this procedure are summarized in Appendix~\ref{A} closely following the standard textbook derivation in Ref.~\cite{breuer2002theory} but also widely covered in, among others, Refs.~\cite{weiss1999quantum,scully1997quantum,gardiner2004quantum,vogel2006quantum,walls2012quantum}. In the final master equation all system properties are derived from two-time correlations of the operators $\mathcal{F}_j(t)$. For $\mathcal{N}$ quantum emitters in the electromagnetic vacuum at zero temperature, the resulting collective quantum master equation reads
\begin{equation}
\label{rhored}
\dot{\rho}(t) =i\left[\rho(t), \sum_{j=1}^{\mathcal{N}}h_j+\mathcal{H}_{dd}\right] + \mathcal{L}_{e}[\rho].
\end{equation}
The last term in the equation above describes irreversible loss of excitation into the electromagnetic vacuum and is expressed in the form of a superoperator (an operator acting on density operators) defined as \cite{lehmberg1970radiation}
\begin{equation}
\label{Lcoll}
\mathcal{L}_e[\rho] = \sum_{j,j'=1}^{\mathcal{N}} \gamma_{jj'}\left[ 2\sigma_{j} \rho(t)\sigma^{\dagger}_{j'}- \sigma^{\dagger}_j\sigma_{j'} \rho(t) - \rho(t)\sigma^{\dagger}_j\sigma_{j'}\right].
\end{equation}
Notice that this is not in standard, diagonal Lindblad form~\cite{breuer2002theory,weiss1999quantum,scully1997quantum,gardiner2004quantum,vogel2006quantum,walls2012quantum} which, for a collapse operator $\mathcal{O}$ and collapse rate $\gamma_\mathcal{O}$, is defined as
\begin{equation}
\label{Lstandard}
\mathcal{L}_\gamma[\rho] = \gamma_{\mathcal{O}}\left[ 2\mathcal{O} \rho(t)\mathcal{O}^{\dagger}- \mathcal{O}^{\dagger}\mathcal{O} \rho(t) - \rho(t)\mathcal{O}^{\dagger}\mathcal{O}\right],
\end{equation}
and describes decay at rate $\gamma_\mathcal{O}$ through a single channel with operator $\mathcal{O}$.\\
The decay rates are $\gamma_{jj'}=(3\gamma/2) F(k_0 R_{jj'})$ where the relative distance vector is $R_{jj'}=|\mathbf{R}_j-\mathbf{R}_{j'}|$ and the function of position is defined as
\begin{eqnarray}
F(k R) = \left[1 + \frac{(\mathbf{e}_{\mathbf{d}} \cdot \nabla_{\mathbf{R}})^2}{k^2} \right]\frac{\sin(k R)}{k R},
\end{eqnarray}
with unit vector $\mathbf{e}_{\mathbf{d}}$ in the direction of $\mathbf{d}\ts{eg}$. The rate $\gamma = \omega^{3}_0 d_\text{eg}^2/(6\pi c^3 \epsilon_0)$ is the spontaneous emission rate of a single emitter (note that the decay rate of the excited-state population is given by $2\gamma $). The collective radiative decay properties are given by the oscillatory behavior of $F(k R)$ as illustrated in Fig.~\ref{fig22}b. Notice that, for emitters separated by much more than $\lambda_0$, the decay becomes purely diagonal as expected for noninteracting, independent emitters.\\
\indent The coherent term $\mathcal{H}_{dd}$ in Eq.~\eqref{rhored} instead describes a dipole-dipole interaction characterized by a virtual exchange of excitation via the vacuum modes without loss of photons
\begin{equation}
\label{Hdd}
\mathcal{H}_{dd} = \sum_{j,j':j\neq j'}^{\mathcal{N}} \Omega_{jj'}\sigma^{\dagger}_j\sigma_{j'}.
\end{equation}
The dipole-dipole exchange rate $\Omega_{jj'} = -(3\gamma/2) G(k_0 R_{jj'})$ can be obtained from the same function that characterizes the collective decay rates via the definition
\begin{eqnarray}
G(k_0 R) = \frac{c}{\pi\omega_0^3}\mathcal{P}\int dk\frac{(ck)^{3} }{c k- \omega_0}F(kR),
\end{eqnarray}
where $\mathcal{P}$ denotes the Cauchy principal value. The explicit functional dependence of $F(kR)$ and $G(kR)$ on distance is detailed in Appendix~\ref{B}. The behavior illustrated in \fref{fig22}b shows that the dipole-dipole interaction ceases at large distances, as expected, but diverges at close separations. This is, however, only an artefact of the initial assumptions that the dipole-electric field interaction is valid at any interparticle distance. This is of course not true, as for separations on the order of the size of the orbitals, one ends up with a fundamentally quantum many-body problem where the tunneling of electrons between neighboring emitters has to be taken into account (leading to molecule formation, hybridization of orbitals, etc). More involved models, based, for example, on a quantum electrodynamics density-functional formalism, can then be employed~\cite{flick2017atoms}.

\subsection{Superradiance and subradiance}
\label{Sec2B}

\begin{figure}[t]
\includegraphics[width=0.9\columnwidth]{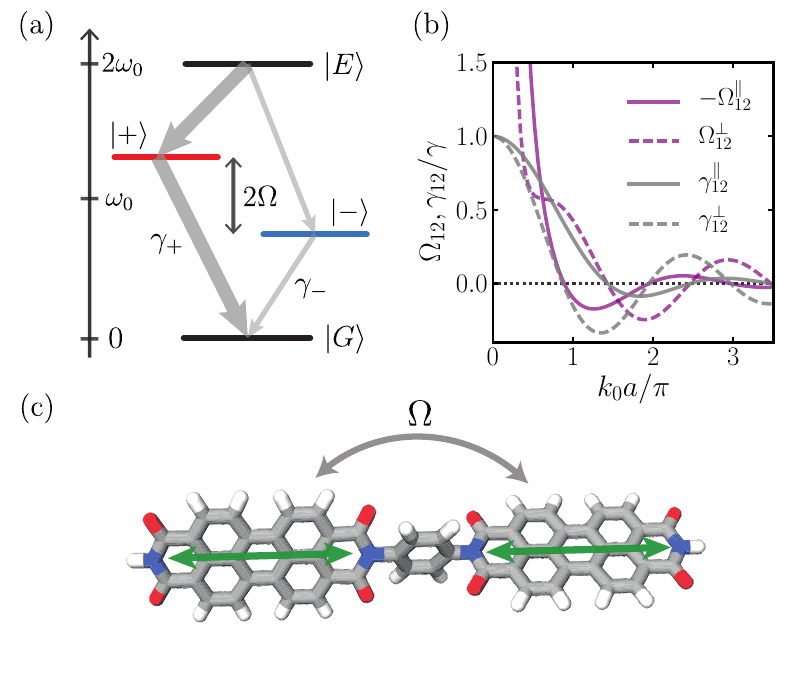}
\caption{(a) Decay and eigenstates in the collective two-emitter basis. Superradiant and subradiant decay channels emerge through the formation of symmetric and antisymmetric superposition states. (b) Dipole-dipole coupling $\Omega_{12}$ and  collective decay rate $\gamma_{12}$ scaling with normalized interatomic distance $k_0 a$ for parallel ($\parallel$) and perpendicular ($\perp$) orientation of dipoles (with respect to the connection vector $\mathbf{R}_{12}$). (c) Possible experimental realization of strong dipole-dipole couplings for two identical chromophores connected by an insulating bridge of length much smaller than $\lambda_0$. The green arrows represent the orientation of the transition dipoles. Adapted from Ref.~\cite{diehl2014emergence}.}
\label{fig22}
\end{figure}
The superoperator in Eq.~\eqref{Lcoll} describes nontrivial mutual decay characterized by the matrix $\mathbf{\Gamma}$ made up by the elements $\gamma_{jj'}$. However, this term is not in standard diagonal form (characterized by a single collapse operator) as it is not comprised of $\mathcal{N}$ independent decay channels. One can perform a basis transformation with a matrix $\textbf{T}$ (such that $\textbf{T}^{-1}=\textbf{T}^\top$) which diagonalizes $\boldsymbol{\Gamma}$ such that $\text{diag}\left(\tilde{\gamma}_1,...,\tilde{\gamma}_\mathcal{N}\right) =\textbf{T}^\top\boldsymbol{\Gamma}\textbf{T}$
where $\tilde{\gamma}_k$ is the $k$th eigenvalue of the decay matrix. In order to achieve the Lindblad form defined in Eq.~\eqref{Lstandard} one can define  a set of collapse operators $ \Pi_k = \sum_j T_{jk}\sigma_j$ such that
\begin{align}
\label{Ldiag}
\mathcal{L}_e[\rho] &= \sum_{k=1}^{\mathcal{N}}\tilde{\gamma}_k\left(2 \Pi_k\rho \Pi_k^\dagger - \Pi_k^\dagger \Pi_k \rho - \rho \Pi_k^\dagger  \Pi_k\right),
\end{align}
now describes $\mathcal{N}$ independent decay channels each with an associated collapse operator $\Pi_k$ and associated loss rate $\tilde{\gamma}_k$. Notice that the preparation of a collective quantum superposition that decays at one of the rates $\tilde{\gamma}_k$ would require the application of the Hermitian conjugate of the corresponding collapse operator $\Pi_k^\dagger$ to the collective ground state $\ket{G}=\ket{g}_1\otimes \ket{g}_2\otimes ...\otimes \ket{g}_\mathcal{N}$ of the system. In practice this is, however, not a straightforward task.\\

\noindent \textbf{Two coupled emitters} - Let us first make the connection between collective radiative effects and the symmetry of quantum superpositions by considering the simplest example of two quantum emitters separated by distance $a$. Diagonalization of the decay matrix leads to superradiant/subradiant decay channels characterized by $\gamma_\pm=\gamma \pm \bar{\gamma}$ and at the same time renders the Hamiltonian in diagonal form with eigenenergies $\omega_\pm=\omega_0\pm\Omega$. The eigenstates are symmetric/antisymmetric superpositions $\ket{\pm}=(\ket{e}_1\otimes\ket{g}_2\pm\ket{g}_1\otimes\ket{e}_2)/\sqrt{2}$. This is illustrated in Fig.~\ref{fig22}a in the collective basis where the other two states are the fully excited one $\ket{E}=\ket{e}_1\otimes\ket{e}_2$ and the ground state $\ket{G}=\ket{g}_1\otimes\ket{g}_2$. The splitting between the two levels in the collective basis $2\Omega=2\Omega_{12}(a)$ as well as the magnitude and the sign of the mutual decay rate $\bar{\gamma}=\gamma_{12}(a)$ depend strongly on the particular choice of dipole orientations as well as on separation (as illustrated in Fig.~\ref{fig22}b). However, for distances below half a wavelength $a<\lambda_0/2$, the antisymmetric (symmetric) states always have a subradiant (superradiant) character. One can then understand the connection between state symmetry and radiative properties in terms of an destructive (constructive) interference of radiative paths.\\
\indent The strong scaling of the near-field dipole-dipole coupling with the interparticle distance renders such a simple system valuable for experimental applications in superresolution imaging. For example, experimentally it has been shown that optical resolution of fluorescent molecules can be achieved at distances as small as $12\,\mathrm{nm}$~\cite{hettich2002nanometer}. In addition, for strongly coupled emitters, chemical or mechanical means can be employed to correct energy shifts and render the closely spaced emitters indistinguishable such that a source of indistinguishable photons can be achieved. Finally, we remark that subwavelength separations (even at the level of less than $10\,\mathrm{nm}$) could be achieved in assembled molecular dimers~\cite{diehl2014emergence} where two chromophores are coupled via an insulating bridge (illustrated in Fig.~\ref{fig22}c). The connection between the dipole-dipole exchange for two-level systems and a vibronic dimer model for chromophores is detailed in Sec.~\ref{Sec6B}.\\

\subsection{The single excitation subspace}
\label{Sec2C}

 While the simultaneous diagonalization of the Lindblad term and the Hamiltonian is generally not possible for $\mathcal{N}>2$, one can still get some intuition in the nature of cooperative decay in the particular case of $\mathcal{N}$ equally spaced quantum emitters in a 1D chain configuration. We start by analyzing the eigenstates and eigenvalues of the dipole-dipole Hamiltonian under the nearest-neighbor approximation and only up to a single excitation shared in the whole system (reducing the Hilbert space from $2^\mathcal{N}$ to only contain $\mathcal{N}+1$ states). We then follow with an analytical derivation of the dissipation rates of such collective states. As a consequence of the fact that the Lindblad term (i.e., the decay matrix $\boldsymbol{\Gamma}$) and the Hamiltonian do not commute, dissipation is not diagonal in the collective basis.\\

\begin{figure}[t]
\includegraphics[width=1\columnwidth]{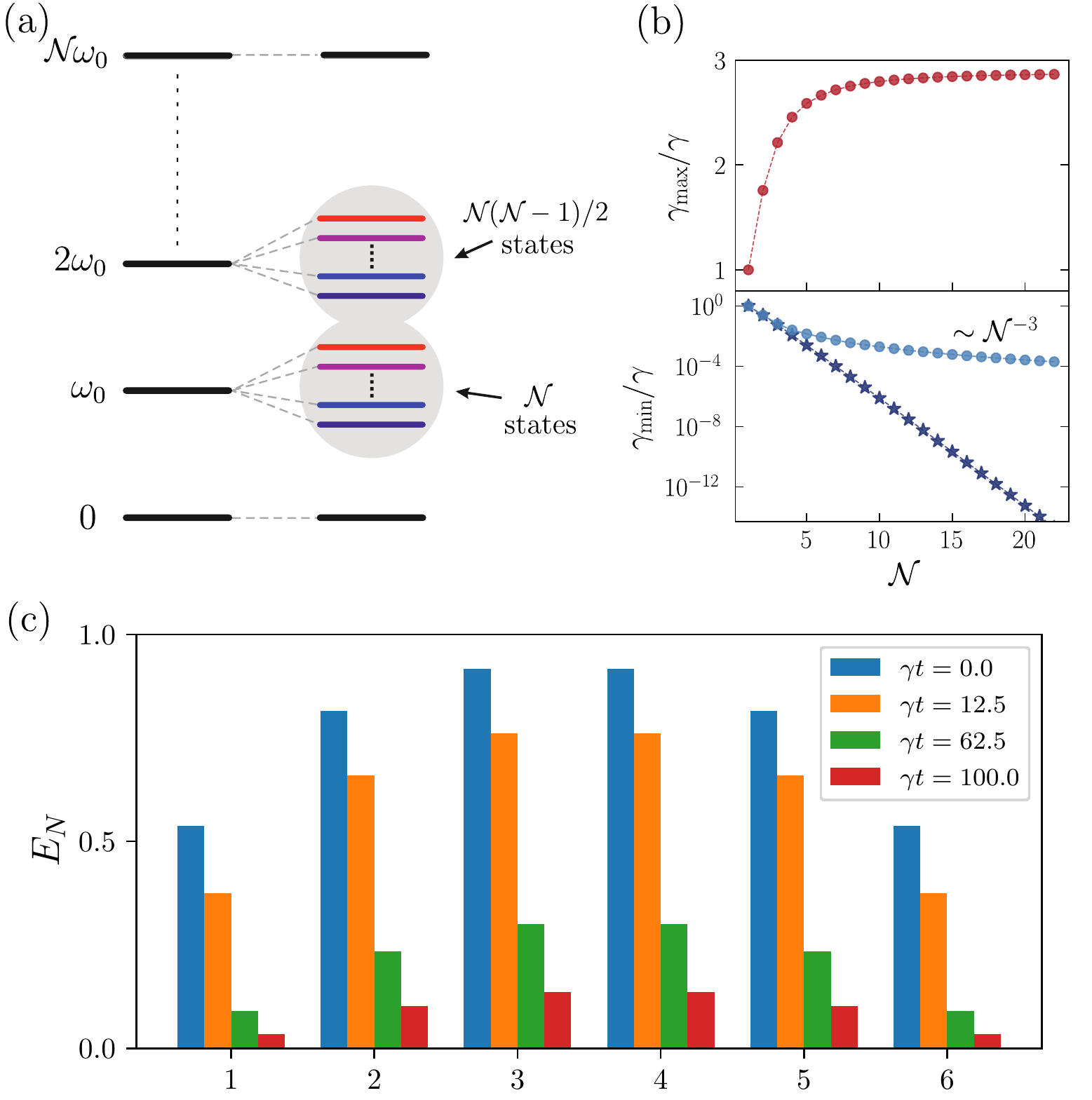}
\caption{(a) Energy scheme of the collective $\mathcal{N}$-emitter system. The $C_n^\mathcal{N}$-fold degeneracy of a given $n$-excitation manifold is lifted by the dipole-dipole interactions and contains states with superradiant (red) and subradiant (blue) character. (b) Decay rates of most superradiant $\gamma_{\text{max}}$ and most subradiant state $\gamma_{\text{min}}$ for an equidistant chain with lattice constant $a=\lambda_0/4$ as a function of $\mathcal{N}$ in the single-excitation manifold. The decay rate of the most subradiant state shows a scaling with $\mathcal{N}^{-3}$ (light blue circles). From the diagonalization of the decay matrix only, one would obtain an exponential scaling  of the subradiant state with $\mathcal{N}$ (dark blue stars).  (c) Entanglement of six-emitter equidistant chain with separation $a=0.1\lambda_0$ which is initialized in the most subradiant state $\tilde{\ket{\mathcal{N}}}$. The logarithmic negativity of each emitter with respect to the $\mathcal{N}-1$ other emitters is plotted for different points in time.    }
\label{fig23}
\end{figure}

\noindent \textbf{Eigenstates of the dipole-dipole Hamiltonian} -  Let us first inspect the Hamiltonian in the analytically solvable case where a nearest-neighbor (NN) approximation is performed. This is justified as the dipole-dipole interactions scales as $R_{jj'}^{-3}$ for distances smaller than a wavelength and thus the nearest neighbor coupling is almost an order of magnitude larger than that of the next to nearest neighbor one. The resulting Hamiltonian is then in the form of a tridiagonal Toeplitz matrix
\begin{align}
\mathcal{H} = \omega_0 \sum_j\sigma_j^\dagger\sigma_j + \sum_{j}\Omega (\sigma_j^\dagger \sigma_{j+1}+\sigma_{j+1}^\dagger \sigma_j),
\end{align}
where $\Omega$ is the coupling between two neighbors. The free Hamiltonian has eigenenergies of degeneracy $C_n^\mathcal{N}=\mathcal{N}!/[(\mathcal{N}-n)!n!]$ for a given level of energy $n\omega_0$ where $n$ ranges from
$0$ for the ground state to $\mathcal{N}$ for the highest excited
state. We illustrate this in Fig.~\ref{fig23}a where we plot the eigenvalues of the Hamiltonian up to the second excitation manifold. The reduced single-excitation manifold on which we focus is characterized by the particle basis $\ket{j}=\sigma_j^\dagger\ket{G}$ with $j=1,\hdots,\mathcal{N}$ and the collective ground state $\ket{G}$. The resulting eigenenergies of the Hamiltonian are
\begin{equation}
\label{eigdipole}
\epsilon_k=\omega_0+2\Omega\cos{\left[\frac{\pi k}{\mathcal{N}+1}\right]},
\end{equation}
for an index $k$ running from $1$ to $\mathcal{N}$ (and the trivial energy of the ground state is $0$). The corresponding eigenstates of the Hamiltonian are then described in the collective exciton basis
\begin{align}
\tilde{\ket{k}} = \sum_{j} \sqrt{\frac{2}{\mathcal{N}+1}} \sin{\left(\frac{\pi k j}{\mathcal{N}+1}\right)} \ket{j}=\sum_{j} f_{jk} \ket{j}.
\end{align}
Notice that the reverse transformation is straightforward $\ket{j}=\sum_k f_{jk} \tilde{\ket{k}}$. The transformation also holds at the level of operators $\tilde{\sigma}_k=\sum_j f_{jk} \sigma_j$ (in the single excitation manifold) which sees the diagonal form of the Hamiltonian as $\mathcal{H}=\sum_k \epsilon_k \tilde{\sigma}_k^\dagger \tilde{\sigma}_k$.\\

\noindent \textbf{Collective dissipation} - To compute the decay rates of the Hamiltonian eigenstates, we use the master equation to derive the equation of motion for the population component $ \rho_{kk}=\bra{\tilde{k}} \rho\ket{\tilde{k}}$. We arrive at
\begin{equation}
\dot{\rho}_{kk}=-\sum_{k'} \left\{\sum_{jj'} \gamma_{jj'} f_{jk} f_{j'k'}\right\} (\rho_{kk'}+\rho_{k'k}),
\end{equation}
which shows that the eigenstates of the Hamiltonian couple within the first excitation manifold in addition to the decay to the ground state. The diagonal elements can then be estimated by setting $k=k'$ to derive $\tilde{\gamma}_{k}=\sum_{jj'}\gamma_{jj'} f_{jk} f_{j'k}$ and more explicitly
\begin{align}
\tilde{\gamma}_{k}=\frac{2}{\mathcal{N}+1}\sum_{jj'}\gamma_{jj'}\sin{\left(\frac{\pi k j}{\mathcal{N}+1}\right)}\sin{\left(\frac{\pi k j'}{\mathcal{N}+1}\right)}.
\end{align}
From this expression, one can derive a scaling of subradiant states with roughly $\mathcal{N}^{-3}$ (see lower panel of  Fig.~\ref{fig23}b for the scaling of the most subradiant state) which will be utilized for cavity antiresonance spectroscopy applications in Sec.~\ref{Sec5A}. More generally, scalings with $\mathcal{N}^{-s}$ up to $s=5$ can be reached~\cite{zhang2020subradiant}. Also, superradiance emerges, which for small distances $a$ is characterized by a rate proportional to $\mathcal{N}$ but eventually saturates with increasing $\mathcal{N}$ (around $\mathcal{N}=\lambda_0/a$, as illustrated in Fig.~\ref{fig23}b). Notice that generally, for $k\neq k'$, the dissipative couplings $ \textstyle \sum_{jj'} \gamma_{jj'} f_{jk} f_{j'k'}$ are not vanishing, which is the direct consequence of the fact that the Lindblad term and Hamiltonian do not commute.  \\
\indent Up to here we have only considered the dissipative properties of the eigenstates of the Hamiltonian. Instead, one can directly inspect the Lindblad term by a diagonalization of the decay matrix in Eq.~\eqref{Ldiag}. Numerical results in Fig.~\ref{fig23}b indicate that in this case subradiant states are present which are characterized by decay rates scaling exponentially down with $\mathcal{N}$. As these states are not eigenstates of the Hamiltonian, they are not easily addressable. The targeting scheme would then involve some excitation scheme that applies the collective operator $\Pi^\dagger_k$ to the ground state $\ket{G}$. In Sec.~\ref{Sec3B} we will show how a combination of a magnetic field gradient together with a pulsed-laser excitation could instead be used to perform such an action.\\
\indent In Sec.~\ref{Sec3A} we provide an alternative treatment in terms of excitations propagating on the 1D support by hopping between neighboring sites via the dipole-dipole exchange. The discrete index $k$ then describes the quasimomentum of an individual collective mode and its location with respect to the light cone distinguishes between superradiant versus subradiant modes.\\

\noindent \textbf{Entanglement properties} - The collective eigenstates of the dipole-dipole Hamiltonian (both super- and subradiant ones) commonly feature non-classical correlations~\cite{plankensteiner2015selective,hebenstreit2017subradiance}, rendering them as an interesting resource for quantum information processing; highly subradiant states are of course even more useful due to the increased lifetime of correlations. To this end we analyze the logarithmic negativity~\cite{plenio2005logarithmic}, which is an entanglement monotone. For a bipartite system consisting of the subsystems $A$ and $B$, it is defined as
\begin{align}
E_{N}(\rho) &= \log_2\left(|\rho^{T_A}|\right),
\end{align}
where $\rho^{T_A}$ denotes the partial transpose with respect to the subsystem $A$ and $|\cdot|$ is the tracenorm. In Fig.~\ref{fig23}c, we show the time evolution of the state with $k=\mathcal{N}$. At distinct time points, we compute the logarithmic negativity for each emitter (i.e., we choose our bipartite system to consist of the $i$th emitter and the rest of the chain). One can see, that the amount of bipartite entanglement is  significantly larger in the center of the chain, even in the initial state. Over time, this behavior is retained and correlation is only slowly lost due to excitation loss of the chain. Even at $t=100\gamma^{-1}$ there still is considerable entanglement in the system.\\

\subsection{The collective Bloch sphere: Dicke superradiance}
\label{Sec2D}

\begin{figure}[t]
\includegraphics[width=1\columnwidth]{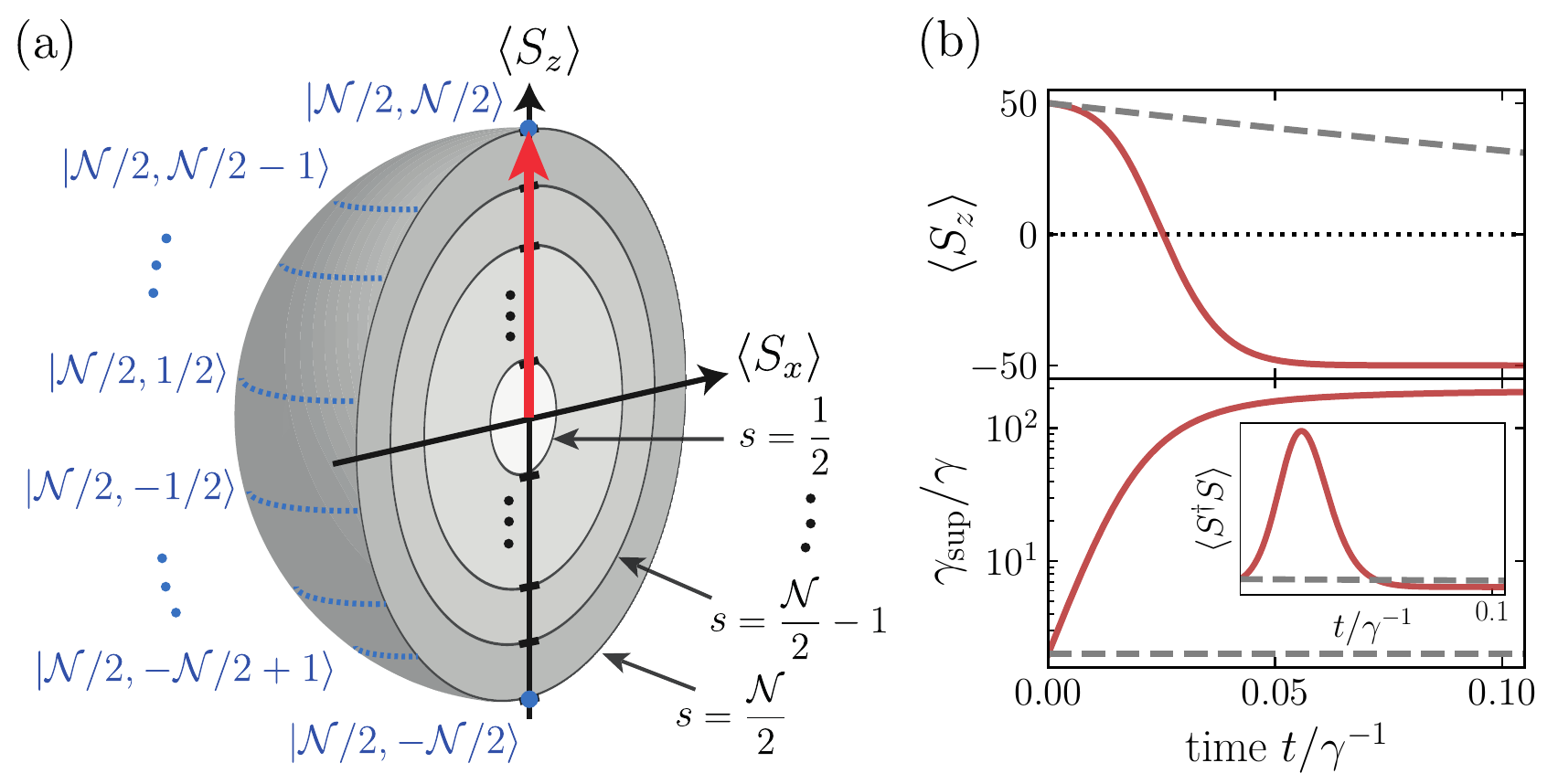}
\caption{ (a) Cross section of collective $\mathcal{N}$-emitter Bloch sphere (for an odd number of spins) showing an onionlike structure with Bloch vector (red) pointing towards the north pole representing the fully excited state $\ket{E}$. The lines of latitude (blue dotted lines) correspond to the $\ket{\mathcal{N}/2,m_{\mathcal{N}/2}}$ states. (b) Dicke superradiance for $\mathcal{N}=100$ emitters initialized in the fully excited state $\ket{E}$: The top shows time dynamics of $\braket{S_z}$ while bottom shows the normalized decay rate $\gamma_{\text{sup}}=-\partial_t\braket{S_z}/(\braket{{S}_z}+\mathcal{N}/2)$. The inset shows the time evolution of $\braket{S^\dagger S}$ with superradiant `burst' reaching a maximum when $\braket{S_z}=m_s$ reaches zero. The dashed gray lines show the result for pure exponential decay of independent, uncoupled emitters. }
\label{fig24}
\end{figure}

\indent Let us now go beyond the single-excitation manifold and consider a famous example introduced by Dicke~\cite{dicke1954coherence} which shows the generation of a superradiant pulse from an ensemble of indistinguishable quantum emitters. The model assumes an idealized case of $\mathcal{N}$ quantum emitters within a very small volume and neglects their dipole-dipole interactions. While this is per se an unrealistic assumption (a densely packed ensemble of QEs exhibits large dipole-dipole frequency shifts), we indicate later in Sec.~\ref{Sec5C} how this model can be realized in the context of cavity QED and is relevant for the physics of lasing in what is known as bad cavity superradiant lasers~\cite{bohnet2012asteadystate}.\\
\indent In the following we will make use of the Bloch-sphere illustration in Fig.~\ref{fig24}a, where $\mathcal{N}$ two-level systems can be described by a collective spin of length $\mathcal{N}/2$. This follows from $1/2\otimes1/2\otimes...\otimes1/2=0\oplus1\oplus...\oplus\mathcal{N}/2$ for $\mathcal{N}$ even and $1/2\otimes1/2\otimes...\otimes1/2=1/2\oplus3/2\oplus...\oplus\mathcal{N}/2$ for $\mathcal{N}$ odd (the tensor product space of dimensions $2^\mathcal{N}$ of $\mathcal{N}$ spins $1/2$ can be written as a direct sum of spaces of different sizes). We can then introduce collective spin operators $S_z=\textstyle \sum_{j}\sigma_z^j/2$ and $S=\textstyle \sum_{j}\sigma_j$ and the total spin
$\mathbf{S}^2=S_z^2+(S^\dagger S+S S^\dagger)/2$. As in the standard angular momentum algebra, it is then possible to find a collective basis, known as the \textit{Dicke basis} of $\mathbf{S}^2$ and $S_z$, indexed by two quantum numbers $\ket{s,m_s}$
\begin{subequations}
\begin{align}
\mathbf{S}^2\ket{s,m_s}&=s(s+1)\ket{s,m_s},\\
S_z\ket{s,m_s}&=m_s\ket{s,m_s}.
\end{align}
\end{subequations}
The spin quantum number $s$ can take either integer or half-integer values and is bounded by $0,1/2\leq s\leq \mathcal{N}/2$ where the zero holds for even and the $1/2$ holds for odd values of $\mathcal{N}$.  The so-called inversion quantum number $m_s$ which measures the projection of the collective spin onto the $z$ axis can take values from $-s\leq m_s\leq s$ and equals $-\mathcal{N}/2$ in the ground state $\ket{G}=\ket{g}^{\otimes \mathcal{N}}$ and $\mathcal{N}/2$ in the fully excited state $\ket{E}=\ket{e}^{\otimes \mathcal{N}}$. The ladder operators $S$ and $S^\dagger$ act on the Dicke states as
\begin{subequations}
\begin{align}
S^\dagger \ket{s,m_s}&=\sqrt{(s-m_s)(s+m_s+1)}\ket{s,m_s+1},\\
S\ket{s,m_s}&=\sqrt{(s+m_s)(s-m_s+1)}\ket{s,m_s-1}.
\end{align}
\end{subequations}
In a closed system interacting with the electromagnetic field, the interactions are mediated by $S$ and $S^\dagger$. Thereby the selection rules for optical transitions are given by $\Delta m_s= \pm 1$, $\Delta s=0$ and the Hilbert space splits into non-interacting subspaces defined by the quantum number $s$  with dimension $2s+1$ (illustrated in Fig.~\ref{fig24}a).\\
\indent Let us now describe the phenomenon of Dicke superradiance in the ideal case where all mutual decay rates are equal and maximal $\gamma_{jj'}=\gamma$. A trivial observation is now that the Lindblad decay term in Eq.~\eqref{Lcoll} assumes a very simple form with a single superradiant decay channel and collective collapse operator
\begin{align}
\label{Dicke}
\mathcal{L}_e[ \rho] &= \gamma\left[2S \rho S^\dagger- S^\dagger S  \rho -  \rho S^\dagger S\right].
\end{align}

We will follow the time evolution under such a Lindblad term for an initially fully excited ensemble characterized by the state vector $\ket{\mathcal{N}/2,\mathcal{N}/2}$. Notice that the action of the collapse operator cannot take the system out of the $s=\mathcal{N}/2$ symmetric manifold containing $\mathcal{N}+1$ Dicke states. We can derive an equation of motion for the population difference operator average $\braket{\dot{S}_z}=-2\gamma \braket{S^\dagger S}$
which simply states that the loss rate is proportional to the emitted intensity. For a single quantum emitter one has  $\braket{\sigma^\dagger \sigma}=\braket{\sigma_z}/2+1/2$, implying that the spontaneous emission always follows an exponential law. The population difference of a collective Dicke state $\ket{s,m_s}$ decays instead at a state-dependent rate $2\gamma \braket{S^\dagger S}=2\gamma(s+m_s)(s-m_s+1)$. For the initially fully excited state, the decay rate is $2\gamma \mathcal{N}$ which is the same as expected for $\mathcal{N}$ independently decaying emitters. However, particle-particle correlations start building up during the evolution and by the time the $\ket{\mathcal{N}/2,0}$ state (for even $\mathcal{N}$) is reached, the emission is superradiant and scales approximately as $\gamma \mathcal{N}^2/2$ (see \fref{fig24}b). Notice that the crucial effect of correlations can be distinguished by rewriting $\braket{S^\dagger S}=\braket{\sum_{i} \sigma_i^\dagger\sigma_i}+\braket{\sum_{i\neq j}\sigma_i^\dagger\sigma_j}$.
Using that the sum over populations is given by $\braket{\sum_{i}\sigma_i^\dagger\sigma_i}=s+m_s$, the dipole-dipole correlation between emitters $i$ and $j$ can be estimated as $\braket{\sigma_i^\dagger\sigma_j}=(s^2-m_s^2)/(\mathcal{N}(\mathcal{N}-1))$,
reaching a maximum of approximately $1/4$ for $m_s=0$ and becoming zero for $m_s=\pm s$.\\

\section{Subwavelength quantum emitter arrays}
\label{Sec3}

The interplay between dipole-dipole interactions and collective radiance in quantum-emitter ensembles leads to a multitude of applications of 1D and 2D subwavelength arrays such as in nonlinear quantum optics~\cite{bettles2020quantum, wild2018quantum}, nano-optomechanics~\cite{shahmoon2020quantum}, the design of quantum metamaterials with magnetic response at optical frequencies \cite{alaee2020quantum, ballantine2020optical}, as platforms for quantum information processing~\cite{glaetzle2017quantum, guimond2019subradiant, bekenstein2020quantum} or as chiral light-matter interfaces \cite{grankin2018freespace}. These applications are based on the fact that such structures can support collective surface resonances that can interact in a controllable fashion with impinging fields.

We will start by studying the dispersion relations of the surface modes on 1D platforms by means of a Bloch ansatz, showing the occurrence of band gaps and Dirac points. This direction has recently emerged showing the promise of subwavelength arrays for topological quantum optics implementations~\cite{perczel2017topological,bettles2017topological}. We then proceed by providing illustrations on 1D emitter chains and rings aimed at showing the usefulness of subradiance as a resource for: i) improved frequency sensitivity (as shown in Refs.~\cite{ostermann2013protected,ostermann2014protected}), ii) robust quantum memories~\cite{plankensteiner2015selective} and iii) the design of nanoscale coherent light sources as recently proposed in Ref.~\cite{holzinger2020nanoscale}. Finally, we derive the optical response of two dimensional subwavelength arrays around certain confined surface-mode resonances~\cite{bettles2016enhanced,shahmoon2017cooperative,bettles2020quantum} and describe a regime recently experimentally tackled showing close to unity reflectivity for arrays of optically trapped atoms~\cite{rui2020asubradiant}.\\

\subsection{Band structure and topology of 1D chains}
\label{Sec3A}

\begin{figure*}[t]
\includegraphics[width=2.00\columnwidth]{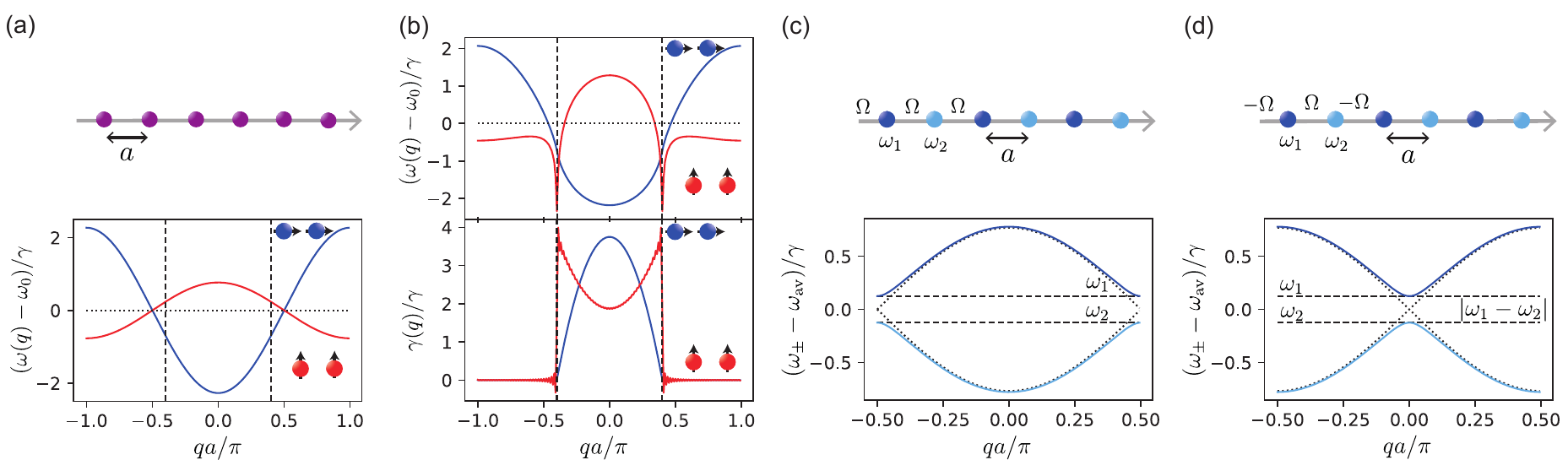}
\caption{(a) Energy band structure for a simplified model of a chain of nearest neighbor coupled identical emitters, for two different alignments of dipoles (perpendicular and parallel to the chain). (b) Top: band structure considering the exact expression of near-field coupling $\Omega_{jj'}$ and in the presence of collective radiative decay $\gamma_{jj'}$. Bottom: corresponding decay rate as a function of the quasi momentum $q$. The light cone characterized by $\omega = c|q|$, here plotted at the two level resonance $\omega_0 = ck_0$ is indicated by the dashed black lines. Parameters are $a/\lambda_0 = 0.2$ with $\mathcal{N} = 200$ as in \cite{asenjo2017exponential}. (c) Band structure for a chain of equidistant emitters with two different alternating species. A band gap of size $|\omega_1-\omega_2|$ around $\omega_\text{av}=(\omega_1+\omega_2)/2$ is opened. (d) Additionally, when the interaction between the emitters has an alternating sign, we obtain a Dirac point at $q=0$ (for the case of degenerate frequencies $\omega_1=\omega_2$).}.
\label{fig31}
\end{figure*}

Up to this point, we have solved for collective resonances and their associated collective radiation rates by a direct diagonalization of the Hamiltonian with near-field coupling terms. One can however take a solid-state approach instead where the array provides a crystalline structure for the quasiexcitations propagating on its surface. As the simplest example, let us first revisit the one-dimensional equidistant chain of $\mathcal{N}$ emitters with identical frequencies as already discussed in Sec.~\ref{Sec2C} described by the Hamiltonian
\begin{equation}
\mathcal{H} = \omega_0 \sum_j \sigma^{\dagger}_j\sigma_j + \sum_{j' \neq j}\Omega_{jj'}\sigma^{\dagger}_j\sigma_{j'}.
\end{equation}
We first assume nearest neighbor (NN) coupling $\Omega_{jj'} = \Omega \delta_{j,j'\pm 1}$ which insures translational symmetry with periodicity $a$ (equivalently stated, the unit cell contains one site only). We impose \textit{periodic boundary conditions} (PBCs) to simulate the mesoscopic case by asking that at the edge $\Omega_{1\mathcal{N}} = \Omega$. We then ask the question: what kind of excitations can propagate in this chain, i.e., what kind of dispersion relations such excitations will exhibit. To this end, we start with the equations of motion for the expectation values $\beta_j=\braket{\sigma_j}$ which can be obtained from $\braket{\dot{\sigma}_j}=\mathrm{Tr}[\sigma_j\dot{\rho}]$ where $\dot{\rho}=-i[\mathcal{H},\rho]$ (neglecting radiative emission in a first step)
\begin{equation}
\label{Eq.Band1}
\dot{\beta}_j = -i\omega_0 \beta_j - i\Omega \beta_{j-1} - i\Omega \beta_{j+1}.
\end{equation}
We furthermore assume a weak excitation limit $\braket{{\sigma}_z^j}\approx -1$.
In a compact matrix formulation, we rewrite $\dot{\mathbf{v}} = M\mathbf{v}$ where $\mathbf{v} = ( \beta_1, \cdots, \beta_\mathcal{N})^{\top}$ and the drift matrix is expressed as a circulant Toeplitz matrix
\begin{equation}
M = -i\left(\begin{array}{cccccc} \omega_0 & \Omega & 0 & \cdots & 0 & \Omega \\ \Omega & \omega_0 & \Omega & \cdots & 0 & 0 \\ 0 & \Omega & \omega_0 & \cdots & 0 & 0 \\ \vdots & \vdots & \vdots & \ddots & \vdots & \vdots \\ 0 & 0 & 0 & \cdots & \omega_0 & \Omega \\
\Omega & 0 & 0 & \cdots & \Omega & \omega_0 \end{array} \right).
\end{equation}
The direct diagonalization of the matrix above gives the expected $\mathcal{N}$ collective modes listed in Sec.~\ref{Sec2C} with a state index from $1$ to $\mathcal{N}$. However, we now instead look for a dispersion relationship with a quasimomentum $q$, to which end we plug the ansatz $\beta_j = A_q e^{i(qaj - \omega t)}$ into Eq.~\eqref{Eq.Band1} which straightforwardly leads to the following dispersion relation
\begin{equation}
\omega(q) = \omega_0 + 2\Omega\cos(qa).
\end{equation}
From the application of the PBCs we have $\beta_1=\beta_{\mathcal{N}+1}$ which leads to $qa\mathcal{N}=2\pi m$ where $m$ is an integer. This indicates that the allowed quasimomenta are $q = 2\pi m/ (\mathcal{N}a)$ and we can fix the first Brillouin zone to $m \in \{-\mathcal{N}/2, \mathcal{N}/2 \}$ corresponding to $q$ from $-\pi/a$ (left propagating wave) to $\pi/a$ (right propagating wave). The resulting dispersion relations for two arrangements of dipoles (parallel and perpendicular to the chain axis) are indicated in Fig.~\ref{fig31}a.\\

\noindent \textbf{Beyond NN with dissipation} - The results presented above describe a simplified model where only nearest-neighbor interactions are employed and spontaneous emission is disregarded. We now proceed by analyzing the full equations of motion including dissipation which are obtained from Eq.~\eqref{rhored} and read
\begin{equation}
\dot{\beta}_j  = -i(\omega_0 - i\gamma)\beta_j -\sum_{j' \neq j}(i\Omega_{jj'} + \gamma_{jj'})\beta_{j'}.
\end{equation}
These equations take into account the exact behavior of $\Omega_{jj'}$, $\gamma_{jj'}$ (as defined in Sec.~\ref{Sec2A} and App.~\ref{B}) and lead to the following dispersion relation
\begin{eqnarray}
\omega(q)\! -\!i\gamma (q)\! &=&\! \omega_0 \!- \!i\gamma +\! \sum_{k \neq j} (\Omega_{jk}\! -\! i\gamma_{jk})e^{iqa(k-j)}.
\end{eqnarray}
The result is numerically illustrated in Fig.~\ref{fig31}b and compared with the approximation analytically obtained for NN coupling in Fig.~\ref{fig31}a. The corresponding collective decay is plotted in the lower panel of Fig.~\ref{fig31}b. The vertical lines indicate the location of the light cone, where the quasimomentum of the excitation propagating on the surface equals the wave vector of the photon $k_0=2\pi/\lambda_0$, which for the chosen distance $a=\lambda_0/5$ leads to $q_0a/\pi=0.4$. Notice that the interpretation in terms of the location with respect to the light cone is straightforward as waves with quasimomentum larger (in absolute value) than that of the photon cannot escape the chain and therefore are subradiant (in agreement with results presented in Ref.~\cite{asenjo2017exponential}).\\

\noindent \textbf{Emerging band gaps} - Let us now consider a slightly more complex structure with alternating emitter frequencies $\omega_1$ and $\omega_2$ and identical NN couplings at rate $\Omega$ (as illustrated in Fig.~\ref{fig31}c). One then immediately notices that this implies a lattice with a double unit cell. We denote the two types of emitters by an upper index such that their amplitudes are $(\beta^{(1)}_j, \beta^{(2)}_j)$ and the site index $j$ runs from $1$ to $\mathcal{N}/2$ (such that we keep the chain length at $\mathcal{N} a$). The dipole-dipole interaction now couples the two kinds of emitters with each other within the unit cell and also between neighboring unit cells leading to the following set of coupled equations
\begin{subequations}
\begin{align}
\dot{\beta}^{(1)}_j&= -i\omega_1 \beta^{(1)}_j - i\Omega \beta^{(2)}_{j-1} -i\Omega \beta^{(2)}_{j}, \\
\dot{\beta}^{(2)}_j &= -i\omega_2 \beta^{(2)}_j - i\Omega \beta^{(1)}_{j+1} -i\Omega \beta^{(1)}_{j}.
\end{align}
\end{subequations}
We now ask for propagating waves with the following profile $\beta^{(1)}_{j} = A_q e^{i(2qaj - \omega t)}$ and $\beta^{(2)}_{j} = B_q e^{i(2qaj - \omega t)}$ and plug in this ansatz into the above equations to result in the following eigenvalue problem
\begin{equation}
\left(\begin{array}{cc} \omega_1 - \omega(q) & \Omega\left(1+e^{-i2qa}\right) \\ \Omega\left(1+e^{i2qa}\right) & \omega_2 - \omega(q)  \end{array} \right)\left(\begin{array}{c} A_q \\ B_q \end{array}\right) = 0.
\end{equation}
Two types of eigenvalues $\omega_{\pm}(q)$ result from diagonalization of the above matrix corresponding to two distinct energy bands
\begin{equation}
\omega_{\pm}(q) = \frac{\omega_1 + \omega_2}{2} \pm \sqrt{\left(\frac{\omega_1 - \omega_2}{2} \right)^2 + 4\Omega^2\cos^2(qa)}.
\end{equation}
The particularity of the system is that the two bands exhibit a band gap $|\omega_1-\omega_2|$ at the edge of the first Brillouin zone (as depicted in Fig.~\ref{fig31}c). The PBC now impose that $\beta^{(1)}_{1}=\beta^{(1)}_{\mathcal{N}/2+1}$ which leads to the condition $q = \pi m /(\mathcal{N}a)$. The first Brillouin zone is now defined by $m \in \{-\mathcal{N}/2,\mathcal{N}/2\}$ corresponding to $q$ varying between $-\pi/(2a)$ to $\pi/(2a)$.\\

\noindent \textbf{Dirac points} - Further complexity in the band structure can be engineering by assuming alternating coupling strengths (for example from $\Omega$ to $-\Omega$ for consecutive pairs  as in Fig.~\ref{fig31}d). Noticing that this situation is again characterized by a double unit cell, we follow the same steps as above to ask for two kind of propagating waves and reach the following eigenvalue problem
\begin{equation}
\left(\begin{array}{cc} \omega_1 - \omega(q) & \Omega\left(1-e^{-i2qa}\right) \\ \Omega\left(1-e^{i2qa}\right) & \omega_2 - \omega(q)  \end{array} \right)\left(\begin{array}{c} A_q \\ B_q \end{array}\right) = 0,
\end{equation}
which leads to the following dispersion relation for the two branches
\begin{equation}
\omega_{\pm}(q) = \frac{\omega_1 + \omega_2}{2} \pm \sqrt{\left(\frac{\omega_1 - \omega_2}{2} \right)^2 + 4\Omega^2\sin^2(qa)}.
\end{equation}
The quasimomentum varies as above between $-\pi/(2a)$ to $\pi/(2a)$. The resulting dispersion curve plotted in Fig.~\ref{fig31}d shows the emergence of an avoided crossing in the center of the Brillouin zone which exhibits a Dirac point with linear dispersion relation at frequency degeneracy $\omega_1=\omega_2$.\\

\noindent \textbf{Berry phase} - A particularly interesting case occurs for identical frequency emitters (transition frequency $\omega_0$) with alternating interaction strengths $\Omega_1$ and $\Omega_2$. This can be mapped onto the Su-Schrieffer-Heeger model (SSH)~\cite{su1979solitons}, which for electrons describes the emergence of bulk insulating phases distinguished by topological invariants and which in nature occurs in polyacetylene molecules.
\begin{figure}[b]
\includegraphics[width=0.58\columnwidth]{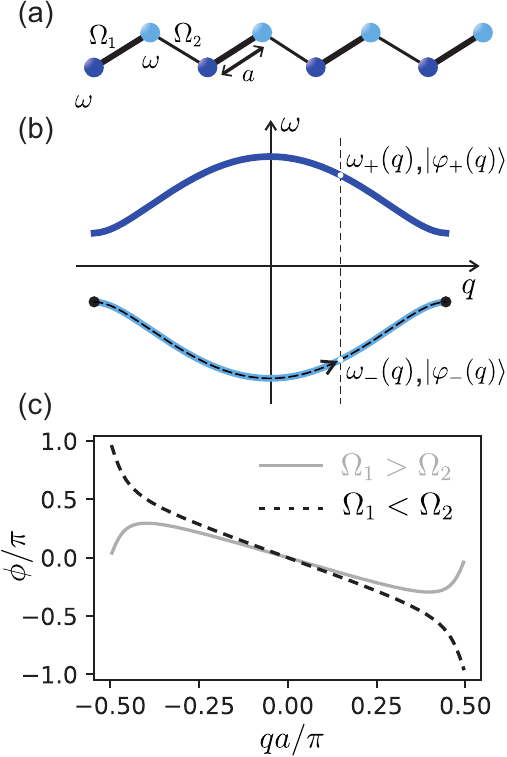}
\caption{(a) Chain with alternating dipole-dipole couplings forming the geometry of the Su-Schrieffer-Heeger (SSH) model. (b) Two distinct bands emerge, separated by a band gap $2|\Omega_1-\Omega_2|$ and with shape independent of the relation $\Omega_1 \lessgtr \Omega_2$. The vertical dashed line shows the corresponding energies and eigenvectors corresponding to a given parameter $q$. The arrow in the lower band shows the path chosen for the integration to derive the accumulated phase from one end to the other one of first Brillouin zone. (c) The accumulated phase on the lower energy branch shows considerable differences between the case of $\Omega_1 > \Omega_2$ and that of $\Omega_1< \Omega_2$.}.
\label{fig32}
\end{figure}
As before we write the equations of motion
\begin{subequations}
\begin{align}
\dot{\beta}^{(1)}_j&= -i\omega_0 \beta^{(1)}_j - i\Omega_2 \beta^{(2)}_{j-1} -i\Omega_1 \beta^{(2)}_{j}, \\
\dot{\beta}^{(2)}_j &= -i\omega_0 \beta^{(2)}_j - i\Omega_2 \beta^{(1)}_{j+1} -i\Omega_1 \beta^{(1)}_{j}.
\end{align}
\end{subequations}
and testing for the two kinds of propagating waves results in the following eigenvalue problem
\begin{equation}
\left(\begin{array}{cc} \omega_0- \omega & \Omega_1+\Omega_2e^{-i2qa} \\ \Omega_1+\Omega_2e^{i2qa} & \omega_0- \omega  \end{array} \right)\left(\begin{array}{c} A_q \\ B_q \end{array}\right) = 0.
\end{equation}
Here we will focus both on the energy dispersion curves but also on the characteristics of their corresponding eigenstates. The eigenvalues are given by
\begin{equation}
\omega_{\pm}(q) = \omega_0\pm \sqrt{(\Omega_1 - \Omega_2)^2 + 4\Omega_1\Omega_2 \cos^2(qa)}.
\end{equation}
While the band structure is insensitive to the exchange of $\Omega_1$ and $\Omega_2$, the eigenvectors are not. The normalized eigenstates are analytically expressed in vector form as
\begin{equation}
\label{Top4}
\ket{\varphi_\pm (q)} = \frac{1}{\sqrt{2}}\left( \begin{array}{c} \pm e^{-i\phi(q)} \\ 1 \end{array}\right),
\end{equation}
where the $q$-dependent phase is given by the following relation
\begin{equation}
e^{-i\phi(q)}=\frac{\Omega_1+\Omega_2e^{-i2qa}}{\sqrt{\Omega_1^2+\Omega_2^2+2\Omega_1\Omega_2\cos{2qa}}}.
\end{equation}
The eigenvalue problem can now be written formally as $\mathcal{H}(q)\ket{\varphi_n (q)} =\omega_n(q)\ket{\varphi_n (q)}$ (where $n$ stands for $\pm$). We now proceed by assuming that a path $q(t)$ is taken, with $q(t=0)=-\pi/(2a)$ and $q(\tau)=\pi/(2a)$; this is illustrated in Fig.~\ref{fig32}b, particularized to the lower energy band. By moving adiabatically slow, tunneling to the orthogonal eigenstate is not allowed. Generally, we can then write the state of the system at any time $t$ as $\ket{\psi(t)}=e^{-i\theta(t)}\ket{\varphi_n (q(t))}$ where the time-dependent phase could be computed directly by the application of the time-ordered evolution operator with a time-dependent Hamiltonian to the initial state. However, a more elegant solution comes from simply writing the Schr\"odinger equation $\mathcal{H}(q(t))\ket{\psi(t)} = i\partial_t \ket{\psi(t)}$ explicitly, to arrive at an equation for the phase
\begin{equation}
\left(\partial_t\theta(t)\right) \ket{\varphi_n (q(t))} = \omega_n(q(t))\ket{\varphi_n (q(t))} -i\partial_t \ket{\varphi_n (q(t))}.
\end{equation}
We sandwich the equation above with $\bra{\varphi_n (q(t))}$ and integrate to obtain two distinct contributions. The integral of the energy over the whole band vanishes as the final and initial point are equal in energy. The second contribution is path dependent and reads
\begin{equation}
\theta(\tau) = -i\int^{\tau}_0 \braket{\varphi_n (q(t))|\partial_{t}|\varphi_n (q(t))}dt.
\end{equation}
The integration over time can be turned into a path integration such that
\begin{align}
\theta =-i\int^{q(\tau)}_{q(0)} \braket{\varphi_n (q)|\partial_{q}|\varphi_n (q)}dq.
\end{align}
This describes the path dependent Berry phase where $A_n(q) = \braket{\varphi_n (q)|\partial_{q}|\varphi_n (q)}$ can be identified as the Berry potential. The Berry phase is gauge invariant for closed integration loops. In general, the temporal aspect of the adiabatic deformation is a more or less fictitious process but it is helpful to unravel the topological structure.
Integrating the Berry phase for our Hamiltonian $\mathcal{H}(q)$ over the Brillouin zone $(-\pi/(2a), \pi/(2a))$ for the lower band results in
\begin{equation}
\nu = \frac{1}{\pi}\oint A_{-}(q)dq = \left\{ \begin{array}{cr} 0 & \Omega_1 > \Omega_2 \\ 1 & \Omega_1 < \Omega_2 \\ \mathrm{undef.} & \Omega_1 = \Omega_2 \end{array} \right.
\end{equation}
where the matrix element is easily computed from the vector expression of the eigenstates in Eq.~\eqref{Top4} to result in $A_{-}(q) =  (1/2)d\phi(q)/dq$. For differentiation we use the following expression for $\phi(q) = \mathrm{arctan2}(\Omega_1+\Omega_2\cos(2qa),\Omega_2\sin(2qa))$ and we note that $\nu$ is similar to a winding number. The result of the integral is a topological invariant showing that the two cases $\Omega_1 > \Omega_2$ and $\Omega_1 < \Omega_2$ with the same band structure outcome are topologically different which is expressed by the Berry phase for a closed parameter path (see Fig.~\ref{fig32}c). In order to go from one topological phase $\nu$ in the bulk to another here given by smoothly varying $\Omega_1$ and $\Omega_2$, one needs to cross a point $(\Omega_1 = \Omega_2)$ where the band gap is zero which would violate adiabaticity. This shows that the Hamiltonians for the two insulating phases occurring for $\Omega_1 > \Omega_2$ and $\Omega_1 < \Omega_2$ are not adiabatically equivalent which is manifested in the difference of the topological invariant. In the case of open boundary conditions one finds that the number of edge states forms a topological invariant as well~\cite{asboth2016a}.\\

\subsection{Applications of quantum-emitter rings and chains}
\label{Sec3B}

Subradiance can be exploited towards applications in quantum metrology for sensitive frequency detection as well as in quantum information for the engineering of robust quantum memories. Moreover, subradiance in symmetric arrangements, such as rings, can be utilized to design nanoscale light sources acting as thresholdless nanolasers. To describe these effects, we make use here of the collective master equation in the single excitation regime introduced in Sec.~\ref{Sec2C} and of the Bloch sphere representation introduced in Sec.~\ref{Sec2D}.\\
\begin{figure*}[t]
\includegraphics[width=1.98\columnwidth]{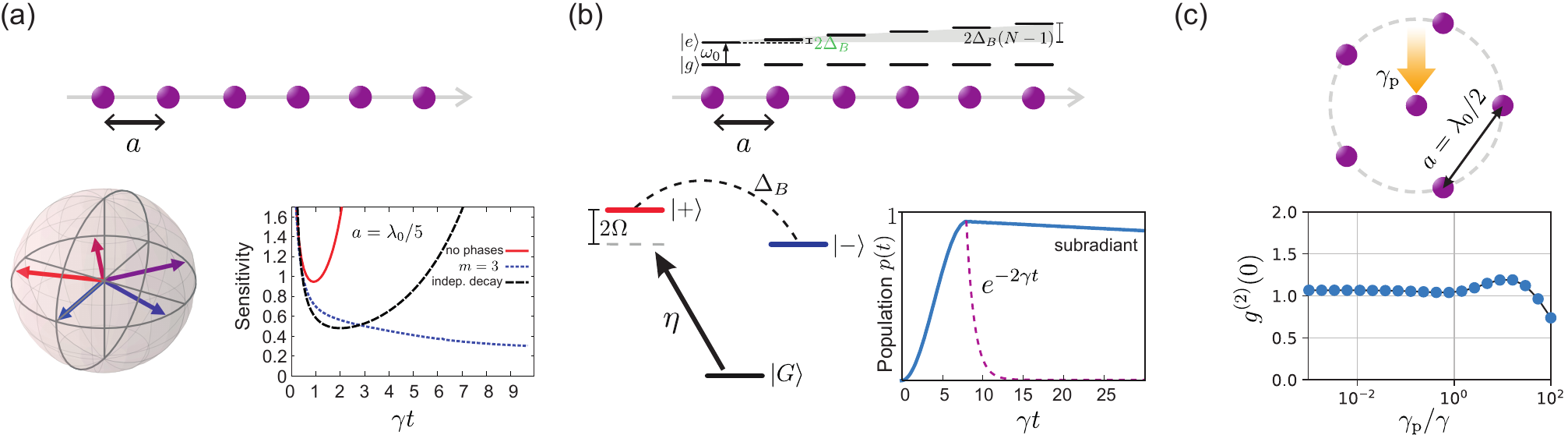}
\caption{(a) Ramsey metrology with phased excitation (shown on the individual spin Bloch sphere) for a chain of six emitters. Minimum sensitivity is plotted against interrogation time $\tau$ showing improved sensitivity for $m=3$ (blue, dotted curve) as opposed to the independent case (black, dashed) or symmetric illumination (red, full line). (b) Phased excitation achieved via magnetic field gradient. For two coupled emitters, adiabatic population transfer to the robust state is achieved via detuned illumination of the symmetric state and subsequent transfer to the robust antisymmetric state via the magnetic field. The population reaches values close to unity and decays much slower than $e^{-2\gamma t}$. (c) Subwavelength structure ($a = \lambda_0/2$) of a ring resonator made of five quantum emitters surrounding an incoherently pumped central emitter. The second-order correlation function at zero delay $g^{(2)}(0)$ is presented as a function of the pumping rate $\gamma_\text{p}$.}.
\label{fig33}
\end{figure*}

\noindent \textbf{Quantum metrology} - Ramsey interferometry is routinely used in quantum metrology for the most sensitive measurements of optical clock frequencies. The conventional method of separated oscillatory fields~\cite{Ramsey1990molecular} assumes an ensemble of atoms initially in the ground state illuminated by a laser of frequency $\omega_\ell$ in two consecutive steps separated by time $\tau$. The two pulses are assumed to be instantaneous and tuned as $\pi/2$ pulses that can be visualized as $\pi/2$ rotations around the $y$ axis on the collective Bloch sphere of radius $\mathcal{N}/2$ (introduced in Sec.~\ref{Sec2D}). After the interrogation time $\tau$ the population difference $\braket{S_z}(\omega_\ell)$ is monitored as a function of the laser frequency. Its behavior is sinusoidal where the argument is the total accumulated phase $(\omega_0-\omega_\ell)\tau$ stemming from the mismatch of the (drifting, variable) laser frequency $\omega_\ell$ and the constant frequency separation of the emitter ground and excited states $\omega_0$. Analysis of the monitored population difference curve indicates a minimal sensitivity:
\begin{equation}\label{sensdef}
\delta \omega =\min \left[\frac{\Delta S^z (\omega,\tau)}{\left|
\partial_{\omega}\langle S^z\rangle (\omega,\tau) \right| }\right],
\end{equation}
where the minimization is performed with respect to $\omega=\omega_0-\omega_\ell$. For weakly decaying emitters ($\gamma\ll\tau^{-1}$), the sensitivity is simply $1/(\tau\sqrt{\mathcal{N}})$ and can obviously be optimized by using longer interrogation times and more emitters. However, longer interrogation times bring decay into play, while operation at higher densities achieved by increasing $\mathcal{N}$ indefinitely are marked by the onset of cooperative effects such as dipole-dipole shifts and super/subradiance.\\
\indent For independently decaying emitters, the application of the master equation introduced in Sec.~\ref{Sec2} leads to a simple analytical solution for both $\braket{S_z}$ and $\Delta S^z$ allowing one to estimate $\left[ \delta\omega \right]_{\text{indep}} = e^{\gamma \tau}/(\tau \sqrt{\mathcal{N}})$.
Further optimization with respect to the interrogation time gives an
optimal $\tau_{\text{opt}}=1/\gamma$ and optimal sensitivity $e\gamma
/\sqrt{\mathcal{N}}$, which shows that the main impediment of Ramsey
interferometry subject to radiative loss is the limited interrogation times available. This expression hints towards a deterioration of the sensitivity in the case of equally illuminated dense ensembles where symmetric collective states are addressed which are typically characterized by superradiant behavior.\\
\indent To protect against such detrimental effects, Refs.~\cite{ostermann2013protected,ostermann2014protected} introduced an alternative procedure which uses an additional step in the Ramsey sequence aimed at hiding the collective states into decoherence-free subspaces. To this end, one complements the $\pi/2$ pulse with a phase distribution pulse, which for a particular atom $j$ is represented
by a rotation around the $z$-direction with the angle
$\varphi_{j}^{(m)}=2\pi m  (j-1)/\mathcal{N}$, where $m=1,\hdots,[\mathcal{N}/2]$ and
$[\mathcal{N}/2]$ is the integer before $\mathcal{N}/2$. Rotations in the single emitter subspace are defined as $\mathcal{R}^{(j)}_\mu [\varphi] = \exp \left( i \varphi \,
\sigma^j_\mu / 2 \right)$ where $\mu \in \{ x, y, z \}$ and $\sigma^j_x=\sigma_j+\sigma_j^\dagger$ and $\sigma^j_y=i(\sigma_j-\sigma_j^\dagger)$. The first generalized Ramsey pulse operator of such an asymmetric Ramsey technique is then
\begin{equation}
  \mathcal{R}_1 = \, \bigotimes_j \mathcal{R}^{(j)}_z \left[ \varphi_{j}^{(m)} \right] \cdot \mathcal{R}^{(j)}_y \left[ \frac{\pi}{2} \right].
\end{equation}
The action of this phase distribution pulse is illustrated in Fig.~\ref{fig33}a on the single particle Bloch sphere. In the second step, after the free evolution where robustness is now expected owing to the folding of the collective state into the subradiant part of the Bloch sphere, at time $\tau$ the phase spread is (instantaneously) reversed and a $\pi/2$ pulse follows leading to the second generalized Ramsey pulse
\begin{equation}
   \mathcal{R}_2 = \bigotimes_j \, \mathcal{R}^{(j)}_y \left[ \frac{\pi}{2} \right] \cdot \mathcal{R}^{(j)}_z \left[- \varphi_{j}^{(m)} \right].
\end{equation}
Finally, detection takes place as before and the sensitivity is minimized with respect to frequency to obtain results as shown in Fig.~\ref{fig33}a. Here, the sensitivity for symmetric illumination is worse than that for independent emitters while the interrogation times and consequently the minimum frequency sensitivity can be considerably increased for phased excitations with $m\neq0$. Further results and analytical considerations can be found in Refs.~\cite{ostermann2013protected,ostermann2014protected}. \\

\noindent \textbf{Quantum memories} -  We now restrict the discussion to the single-excitation manifold of a 1D emitter chain. Here, we aim at targeting collective subradiant states as they show both robustness against decoherence while exhibiting multipartite entanglement. As pointed out in Sec.~\ref{Sec2C}, eigenstates of the dipole-dipole Hamiltonian with lower energy typically exhibit subradiance. To access such states, one can proceed by first selecting them with individual addressing by tailoring the laser light amplitude to fit the shape of the collective state one wishes to address. For example, for a collective state $\tilde{\ket{k}}$ introduced in Sec.~\ref{Sec2C} one can provide geometrical matching with the following driving Hamiltonian
\begin{equation}
\mathcal{H}_k=\eta\sum_{j=1}^{\mathcal{N}}\sin{\left(\frac{\pi k j}{\mathcal{N}+1}\right)}\left(\sigma_j e^{i \omega_\ell t}+\sigma_j^\dagger e^{-i \omega_\ell t}\right).
\end{equation}
Moreover, imposing the condition for resonance by setting $\omega_\ell=\epsilon_k$ ensures that states far enough from the desired targeted one are only weakly populated. In Ref.~\cite{plankensteiner2015selective} it is shown that enhanced lifetimes much larger than $(2\gamma)^{-1}$ can be reached by such tailored excitation.\\
\indent While tailored phase excitation with subwavelength resolution might pose great challenges, alternative methods could be envisioned: for example, symmetric addressing could be combined with the application of a magnetic field gradient. The effect of magnetic field gradient applied along the direction of the emitter chain is to progressively shift the excited state by a quantity $\Delta_j=2\Delta_\text{B} (j-1)$ from the first emitter with $j=1$ to the end of the chain $j=\mathcal{N}$. During the duration $\tau$ of an applied laser pulse, this amounts to a rotation around the $z$ axis of the Bloch vector of each emitter by the angle $\Delta_j \tau$. For a conveniently chosen duration $\tau$, the effective phase difference between neighboring emitters can be controlled: for example with the choice $\Delta_\text{B} \tau=\pi$ a completely antisymmetric superposition can be constructed. For example, for two coupled emitters, Fig.~\ref{fig33}b shows results for pulsed adiabatic transfer of population from the ground state to the antisymmetric state by a simultaneous off-resonant drive of the symmetric state and the action of the magnetic field gradient. The population then ends up in the protected antisymmetric state which then decays much slower than the timescale defined by $(2\gamma)^{-1}$. More results and analytical calculations for larger systems sizes can be found in Ref.~\cite{plankensteiner2015selective}.\\

\noindent \textbf{Nanoscale coherent-light sources} - Subwavelength spaced ensembles of quantum emitters in ring-like configurations resemble the structure of certain biological light-harvesting complexes (LHCs), which have been shown to act as an extremely efficient system of antennae in the photosynthetic process~\cite{cogdell2006architecture}. While the complexity of such biological systems is incredibly hard to tackle even at the computational level, a number of theoretical works have proposed simplified phenomenological models where, for example, the combined effects of disorder, vibronic coupling, electron-phonon couplings, are included as time-dependent frequency shifts (in the Hamiltonian) or as dephasing (as an additional Lindblad term)~\cite{bourneworster2019structure, caycedosoler2017quantum}.\\
\indent Inspired by such naturally occurring systems, a recent theoretical proposal has shown the possibility of designing a thresholdless laser, i.e., a coherent light source~\cite{holzinger2020nanoscale} in the configuration depicted in Fig.~\ref{fig33}d. Here, a central emitter acts as a gain medium and is pumped by incoherent light at some rate $\gamma_\text{p}$. The role of the optical resonator (as present in standard lasing systems) is then taken by the optical modes defined by the geometry of the ring emitters (indexed by $1,\hdots,\mathcal{N}$). In order to test the properties of the system (emission rate, coherence of light, etc.) one can proceed to solve the master equation under the assumption that the incoherent pumping can be modeled by a Lindblad term with a $\sigma\ts{p}^\dagger$ collapse operator (that, opposed to the spontaneous emission case, takes population from the ground state to the excited state).\\
\indent A first observation is that the physics of the system can be reduced to solely the interaction of the symmetric mode $\ket{\psi_{\mathrm{sym}}} = (1/\sqrt{\mathcal{N}})\sum_j \ket{j}=\sigma^{\dagger}_{\mathrm{sym}}\ket{G}$ of the ring resonator with the central atom. The reduced Hamiltonian is then
\begin{equation}
\mathcal{H}_\text{ring} = \Omega_{\mathrm{sym}}\sigma^{\dagger}_{\mathrm{sym}}\sigma_{\mathrm{sym}} + \sqrt{\mathcal{N}}\Omega (\sigma^{\dagger}_{\mathrm{sym}}\sigma_\text{p} + \mathrm{h.c.}),
\end{equation}
where $\Omega\ts{sym}=\sum_{j=2}^\mathcal{N}\Omega_{1j}$ is the dipole energy shift of the symmetric state and the decay of the system is governed by $\mathcal{L}[\rho] = \mathcal{L}_{\gamma}[\rho] + \mathcal{L}_{\gamma_{\text{p}}}[\rho] + \mathcal{L}_{\mathrm{sym}}[\rho]$. For favorable geometries, it is possible to obtain a subradiant decay rate for the symmetric mode of the ring resonator. Additionally, such configurations can reduce the dissipative coupling of the central atom, strengthening the lossless transport of population from the central atom to the ring. We can now analyze the emitted light properties by using the fact that the far field is proportional to the sum of the dipole operators such that we can define the normalized second-order correlation function at zero time delay
\begin{equation}
g^{(2)}(0) = \frac{\sum_{ijkl}\braket{\sigma^{\dagger}_i\sigma^{\dagger}_j\sigma_k\sigma_l}}{\left|\sum_{mn}\braket{\sigma^{\dagger}_m\sigma_n} \right|^2}.
\end{equation}
Numerical simulations, plotted in Fig.~\ref{fig33}c show a close to unity $g^{(2)}(0)$, indicating a coherent state as an output of such a thresholdless laser. For higher pumping rates, antibunching of light is expected. Further results and considerations can be found in Refs.~\cite{holzinger2020nanoscale}. \\

\subsection{Optical response of 2D subwavelength mirrors}
\label{Sec3C}

\begin{figure*}[t]
\includegraphics[width=1.6\columnwidth]{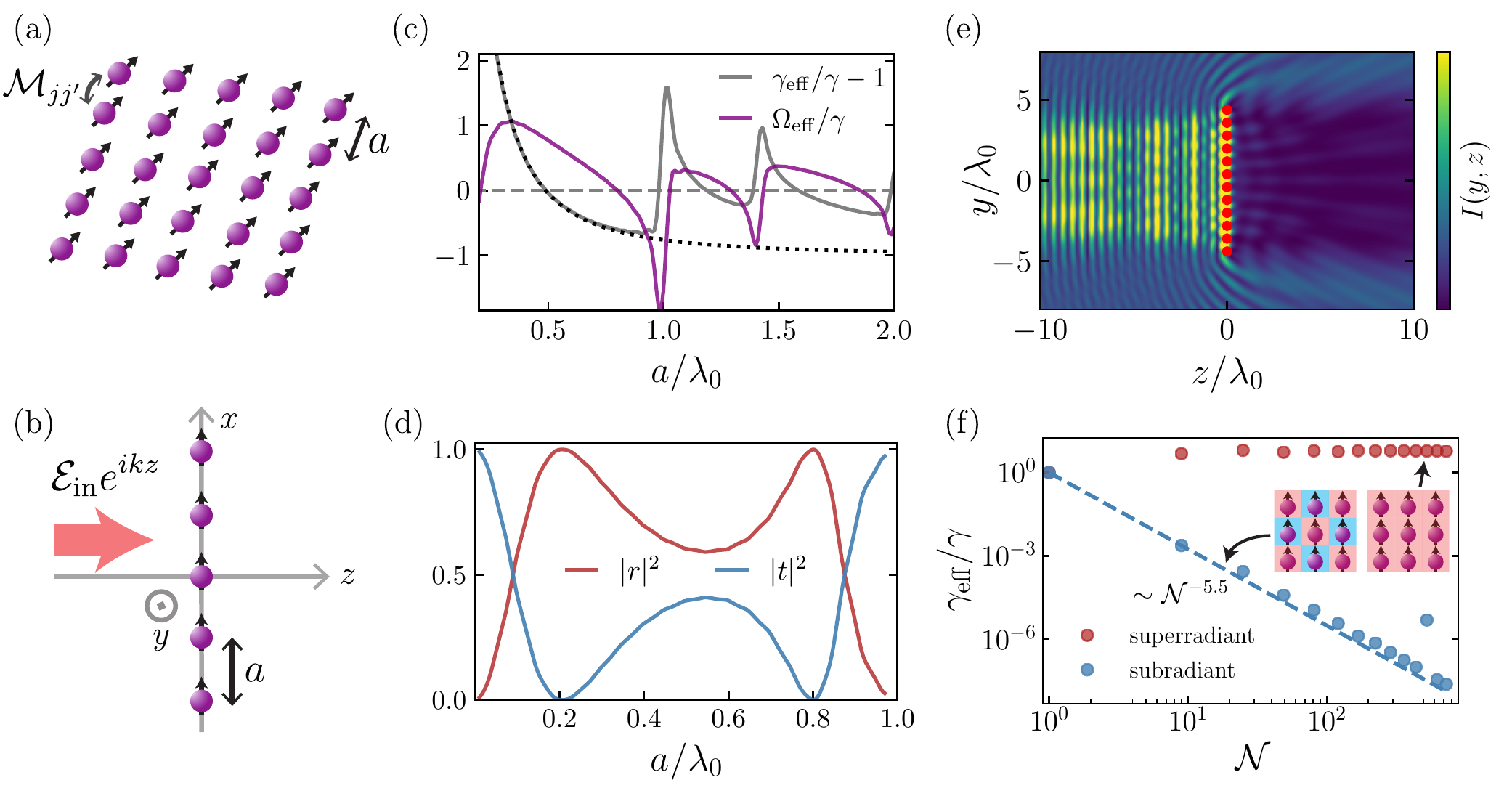}
\caption{(a) Two dimensional quantum-emitter array illustrated as a regular pattern of near-field coupled dipoles with lattice constant $a$. (b) The array is placed in the $x$-$y$ plane and is illuminated by a plane wave with wave number $k$. (c) Effective decay rate $\gamma_\text{eff}$ and frequency shift $\Omega_\text{eff}$ of an array consisting of $20\times20$ atoms as a function of $a/\lambda_0$. The dotted black curve shows the approximation $\gamma_\text{eff}=\gamma\frac{3}{4\pi}(\lambda_0/a)^2$ valid for lattice constants $a<\lambda_0$. (d) Reflection $|r|^2$ and transmission coefficient $|t|^2=|1+r|^2$ of  an atomic array as a function of the lattice constant $a$ (for resonant illumination of the emitters). (e) Calculated intensity profile $I(y,z)$ (in arbitrary units) in the $y$-$z$ plane for an array of $12\times 12$ atoms (indicated by red dots) with lattice constant $a=0.8\lambda_0$ which is resonantly illuminated by a plane wave. (f) Targeted addressing of super- and subradiant states of a 2D array with $a=0.2\lambda_0$ as a function of the number of emitters in the array $\mathcal{N}$. As indicated in the inset, subradiant states are targeted by illumination with antisymmetric phases $f_{nm}=(-1)^{n+m} $ in a checkerboard pattern while superradiant states are obtained from uniform, symmetric illumination. The effective decay rate of the subradiant states scales proportional to $\mathcal{N}^{-5.5}$ as shown by the dashed line. }
\label{fig34}
\end{figure*}

Let us now consider the situation depicted in~\fsref{fig34}a,b where a 2D periodic array of quantum emitters is positioned in the $x$-$y$ plane. To derive the optical response of such a structure, we consider excitation in the form of a plane wave with wavenumber $k=\omega_\ell/c=2\pi/\lambda$ impinging at normal incidence, along the $z$ axis.\\
\indent The electric field (source field) at some position $\mathbf{R}$ emitted by a collection of emitters, each located at $\mathbf{R}_j$, takes the following expression in the far-field limit~\cite{plankensteiner2019enhanced, shahmoon2017cooperative} (only positive frequency component and in the direction of the emitter dipole moment)
\begin{align}
\label{dipolefield}
\mathcal{E}_\text{dip}(\mathbf{R},t)\!=\!-\frac{3\gamma}{2 d_\text{eg}}\sum_j \beta_j (t)\left(\frac{e^{ik|\mathbf{R}-\mathbf{R}_j|}}{k|\mathbf{R}-\mathbf{R}_j|}\right)e^{-i \omega_\ell t}.
\end{align}
The total amplitude is given by the sum of dipole-radiated fields, which, in the linear regime, are proportional to the expectation value of the individual particle dipole operators $\beta_j$ (in a frame rotating at the laser frequency). We aim at describing regimes where plane waves propagating through the 2D structure either get reflected or transmitted only in the $z$ direction while scattering in other directions is inhibited. To this end, we assume a constant phase illumination (denoting this regime as a \textit{symmetric driving} case) where the electric field amplitude does not depend on $x$ and $y$. The total field can then be written as the sum of the incident field and the field radiated by the emitters
\begin{align}
\label{Efieldz}
\mathcal{E}(z)=\mathcal{E}_\text{in} e^{ikz}+\mathcal{E}_\text{dip}(z).
\end{align}
Let us first look at the case of just a single emitter where the radiated field is a dipole pattern with the $z$-direction amplitude
\begin{align}
\mathcal{E}(z)=\mathcal{E}_\text{in}\left[e^{ikz}+\frac{\alpha\ts{p}\pi}{\epsilon_0\lambda^2}\frac{e^{ik|z|}}{|z|}\right],
\end{align}
falling off with increasing distance $z$. We have introduced the single-atom polarizability $\alpha\ts{p}=-d_{\text{eg}}\beta/\mathcal{E}_\text{in}$ which can be expressed in terms of the resonant wavelength, linewidth and $\omega_\ell$ as $\alpha\ts{p}=-(3\epsilon_0\lambda_0^3)/(4\pi^2)\gamma/[(\omega_\ell-\omega_0)+i\gamma]$.
The extinction cross section of a quantum emitter $\sigma_\text{ext}$ (i.e., the effective area seen by an impinging photon) can be related to the polarizability via $\sigma_\text{ext}(\omega_\ell)=k\Im(\alpha\ts{p})/\epsilon_0$ which in the case of resonant illumination simply becomes $\sigma_\text{ext}(\omega_\ell=\omega_0)=3\lambda_0^2/(2\pi)$. Notice the interesting aspect that the cross section is much larger (proportional to $\lambda_0^2$) than the square of the actual size of the electronic orbital which can be a milion times smaller.\\
\indent For many emitters, the driving for the symmetric illumination case can be included as $\mathcal{H}_\ell=\sum_{j=1}^\mathcal{N}\eta (\sigma_j^\dagger e^{-i \omega_\ell t}+\sigma_j e^{i \omega_\ell t})$ with the Rabi frequency $\eta=d_\text{eg}\mathcal{E}_\text{in}$. Transforming into a frame rotating at the laser frequency $\omega_\ell$ yields the equations of motion for the dipole amplitudes
\begin{align}
\label{betaj}
\dot{\beta}_j= i(\omega_\ell-\omega_0) \beta_j-\sum_{j'=1}^\mathcal{N} \mathcal{M}_{jj'}\beta_{j'}-i\eta.
\end{align}
Here, the collective effects coming from the coherent and incoherent emitter-emitter interactions are contained in the off-diagonal elements of the matrix $\mathcal{M}_{jj'}=(i\Omega_{jj'}+\gamma_{jj'})$ (while the diagonal elements $\mathcal{M}_{jj}=\gamma_{jj}$ give the independent decay dynamics).
The assumption of constant drive $\eta$ over the whole array combined with the assumption that the array is quasi-infinite leads to a trivial solution where $\beta_j=\beta$ for any $j=1,...\mathcal{N}$ can be easily estimated from the equation above as $\beta=\eta/[(\omega_\ell-\omega_0)+i\sum_{j'=1}^\mathcal{N} \mathcal{M}_{1j'}]$. This in practice means that only a symmetric superposition of all emitters is excited (only the symmetric surface mode is activated). Notice that the sum $\sum_{j'=1}^\mathcal{N} \mathcal{M}_{jj'}$ does not depend on the index $j$ which is why we set it to $1$ above and the result can be cast into the form $i\sum_{j'=1}^\mathcal{N} \mathcal{M}_{1j'}=-\Omega_\text{eff}+i\gamma_\text{eff}$. The term $\Omega_\text{eff}$ leads to a shift of the collective resonance of the array (as also observed in 1D configurations \cite{glicenstein2020collective}), while $\gamma_\text{eff}$ describes the effective decay rate of the array. Also notice that, as we will also detail more in the end of the subsection, the solution $\beta_j=\beta$ holds solely under constant illumination conditions. Generally, for a laser drive of the form $\mathcal{E}(z)f(x,y)$, i.e., showing phase or spatial imprinting in the transverse direction, different surface modes of the array would be driven (depending on the overlap integral between the function $f(x,y)$ and the transverse profile of the surface mode.\\
\indent The sum over the spatial distribution along the $z$-direction can be estimated by a plane-wave expansion as (for derivation see Appendix~\ref{C})
\begin{align}
\label{Emfield}
\sum_j \frac{e^{ik|z\mathbf{e}_z-\mathbf{R}_j|}}{|z\mathbf{e}_z-\mathbf{R}_j|}=\frac{2\pi i}{ a^2}\sum_{m,n=-\infty}^\infty \frac{e^{i k_{mn}|z|}}{k_{mn}},
\end{align}
where $k_{mn}=\sqrt{k^2-q_m^2-q_n^2}$ with $q_m=(2\pi m/a)$ ($q_n$ analogously). Here, $q_m$ and $q_n$ represent the quasimomenta of the surface modes on the array propagating along the $x$- and $y$-direction (Appendix~\ref{C}). The wave number $k_{mn}$ is real if $(a/\lambda)^2\geq m^2+n^2$. For a subwavelength lattice, this can then obviously only be fulfilled with the choice $m,n=0$: this means that only a mode with a $k$ vector equal in amplitude to the impinging laser can propagate in the $z$-direction. This is exactly the symmetric mode which is also the only one that constant illumination can activate. Notice that, even if the illumination phase would slightly vary over the array, any surface mode which is accidentally excited would radiate in directions other than $z$. One can finally express the source field radiated by the dipoles in the far-field limit as
\begin{align}
\mathcal{E}_\text{dip}(z)=i\pi \left(\frac{\lambda}{a}\right)^2\frac{\alpha\ts{p}^\text{eff}}{\epsilon_0\lambda_0^3}\mathcal{E}_\text{in} e^{ik|z|},
\end{align}
with a renormalized effective polarizability summed over the whole ensemble response as $\alpha\ts{p}^\text{eff}= -d_{\text{eg}}\sum_j\beta_j/(\mathcal{N}\mathcal{E}_\text{in})$ and expressed as
\begin{align}
\alpha\ts{p}^\text{eff}=-\frac{3\epsilon_0\lambda_0^3}{4\pi^2}\frac{\gamma}{(\omega_\ell-\omega_0-\Omega_\text{eff})+i\gamma_\text{eff}}.
\end{align}
We illustrate in Fig.~\ref{fig31}c the behavior of the collective rates $\Omega\ts{eff},\gamma\ts{eff}$ as a function of the lattice constant $a/\lambda_0$. We note that similarly to the two-particle interactions discussed in Sec.~\ref{Sec2A}, the real and imaginary parts are not independent but can be connected by a Kramers-Kronig relation. Regions with $\gamma_\text{eff}<\gamma$ (negative values in \fref{fig34}c) correspond to collective subradiant behavior and are of particular interest. One can see that a special operation point occurs e.g.~at $a/\lambda_0\approx 0.8$ where the collective frequency shift vanishes while a pronounced subradiant behavior remains, facilitating the experimental realization of subradiant optical mirrors with cold atoms in an optical lattice~\cite{rui2020asubradiant}.\\
\indent Notice that the effective cross section per emitter can be considerably increased as, on resonance, one has $\sigma_\text{ext}^\text{eff}(\omega_\ell)=\sigma_\text{ext}(\omega_\ell)\gamma/\gamma_\text{eff}$. For $a<\lambda_0$ one can furthermore approximate the effective decay rate as $\gamma_\text{eff}=3\gamma(\lambda_0/a)^2/(4\pi)$ [see \fref{fig34}c]~\cite{shahmoon2017cooperative}, showing a decrease by a factor of roughly $2.68$ around the optimal operation point at $a/\lambda_0\approx 0.8$. This can also be connected to an increase in the overall reflectivity of the array. To derive this, we write $\mathcal{E}_\text{dip}(z)=r (\omega_\ell) \mathcal{E}_\text{in}e^{ik|z|}$ where the complex reflectivity amplitude reads (considering small detunings around the resonance)
\begin{align}
\label{reflectivity}
r(\omega_\ell)=-i\frac{\gamma_\text{eff}}{(\omega_\ell-\omega_0-\Omega_\text{eff})+i\gamma_\text{eff}},
\end{align}
while the transmission amplitude is obtained as $ t=1+r$. In \fref{fig34}d we plot the absolute square of these quantities as a function of the lattice constant for resonant illumination $\omega_\ell=\omega_0$. One can see that, while generally the reflectivity of the array is high over a broad range of separations, for certain values of $a/\lambda_0 \approx ( 0.2, 0.8 )$ the atomic dipoles can even act as a perfect mirror and reflect the entire input field with unit efficiency \cite{shahmoon2017cooperative}. Moreover, for the whole region where the approximation $\gamma_\text{eff}=3\gamma(\lambda_0/a)^2/(4\pi)$ holds, the mirror shows no losses (quantified by the scattering outside the $z$ axis mode), i.e., $|r|^2+|t|^2=1$. This is of course only valid in the absence of any other channels of nonradiative decay (at rate $\gamma_\text{nr}$), in which case the denominator of $r$ acquires an extra term $i(\gamma_\text{eff}+\gamma_\text{nr})$. The resulting intensity distribution of the total electric field is shown in \fref{fig34}(e), revealing that the emitters can indeed shut off the transmission of an incident plane wave for $z>0$.\\
\indent Let us stress that the expressions above are only valid under \textit{symmetric illumination} conditions, which is the relevant experimental situation as tackled in Ref.~\cite{rui2020asubradiant}. As optical lattices have interatomic distances at the level of $\lambda_0/2$ or larger, the simplest operation is around a point where $\Omega_\text{eff}$ vanishes (at $a=0.8\lambda_0$) and unit reflectivity is reached as soon as the laser is resonant to the emitter transition $\omega_\ell=\omega_0$. However, the subradiance of such a symmetrically excited collective state is only a factor of around $4\pi\times0.8^2/3\approx2.68$ (at $a=0.8\lambda_0$) smaller than $\gamma$. To fully exploit the scalability of subradiance with $\mathcal{N}$, one could instead assume \textit{antisymmetric, or phased} illumination conditions. For example, Fig.~\ref{fig34}f shows that extremely narrow resonances scaling as $\mathcal{N}^{-5.5}$ can be reached for very dense arrays at $a=0.1\lambda_0$ with antisymmetric phases in a checkerboard pattern where $f_{nm}=(-1)^{n+m}$ (for $n=1,...,\sqrt{\mathcal{N}}$ and $m=1,...,\sqrt{\mathcal{N}}$) (deviations from the scaling are due to imperfect addressing of subradiant states). Let us only sketch how the derivation above will change in such a case. First, Eq.~\eqref{Efieldz} is changed to include the $x$ and $y$ spatial dependence of the incoming field in some function $f(x,y)$ with the periodicity equal to the lattice constant $a$. Then, the combination of surface modes which are excited by such an illumination is computed from the steady state of Eqs.~\eqref{betaj} where each $\beta_{nm}$ is driven by an amplitude $\eta f_{nm}$. Finally, replacing the new solution for $\beta_{nm}$ into Eq.~\eqref{dipolefield}, the sum in Eq.~\eqref{Emfield} needs to be recomputed. Finally, the specific modes which can propagate into the far field in the $z$ direction will depend on the specific phased illumination pattern chosen. \\

\subsection{Further remarks}
\label{Sec3D}
One- and two-dimensional ensembles of coupled quantum emitters arranged in regular patterns provide an ideal platform for achieving strong light-matter interactions and high fidelities for photon storage capabilities~\cite{facchinett12016storing,asenjo2017exponential}. Their subradiance properties are an important resource for applications ranging from quantum-information processing~\cite{chang2018quantum} and metrology~\cite{ostermann2013protected,facchinett12016storing,manzoni2018optimization} to excitation transfer~\cite{moreno2019subradiance,needham2019subradiance,ballantine2020subradiance} and it is envisioned that one can build quantum matter in a bottom-up approach from nanoscopic lattices of atoms and photons~\cite{chang2018quantum}. A single subradiant array has been shown to act as an ideal quantum memory with efficient storage and retrieval~\cite{manzoni2018optimization} and it has been suggested~\cite{rui2020asubradiant} that this geometry could lead to vast improvements in the error bound of quantum memories. Regular or honeycomb lattices of three-level systems in a $V$-configuration, where time-reversal symmetry is broken by the application of a magnetic field, have been proposed as platforms for studying topological phenomena with strongly interacting photons~\cite{perczel2017topological,bettles2017topological} and further improvements have been proposed in the form of interfacing the arrays with two-dimensional photonic crystals~\cite{perczel2020topological}. A crucial aspect of these proposals that distinguish them from linear topological photonic systems is the intrinsic nonlinearity of the quantum emitters which could lead to a rich many-body physics dynamics on such subradiant lattices. Furthermore, composite quantum systems comprised of many atomic arrays could find applications in quantum networking: at the level of two distant layers, nonlocal entangled Bell superposition states have been shown to exist~\cite{guimond2019subradiant}. While quantum information processing at the level of atomic layers require qubit encoding on delocalized spin states over the whole array, quantum spin lenses have been recently introduced, where incoming flying qubit photons can be mapped and stored in single atoms~\cite{glaetzle2017quantum}.\\

\section{Cooperativity in cavity QED}
\label{Sec4}
\indent Light-matter interactions can be greatly enhanced by using optical elements which confine electromagnetic fields in very small volumes. This is obvious from the scaling of the zero-point electric field amplitude $\mathcal{E}_k=\sqrt{\omega_k/(2\epsilon_0 \mathcal{V})}$ which indicates that stronger field amplitudes per photon mode are achieved for smaller mode volumes. This led to the development of cavity QED as a subfield of quantum optics specializing in the description of coherent light-matter interactions in cavities~\cite{haroche1989cavity,berman1994cavity,walther2006cavity,haroche2013exploring}. This section provides fundamental concepts and tools of cavity QED with coupled quantum emitter systems, which are then utilized to describe applications in Sec.~\ref{Sec5}.\\
\indent We start by introducing a master equation approach to intracavity light-matter interactions encompassing both collective emitter loss and cavity photon loss. We then move on to a quantum Langevin equations approach supplemented with an input-ouput formalism allowing access to the correlations of cavity output field operators. Finally, to address inhomogeneous emitter ensembles as often present in experimental setups, we tackle the question of strong frequency disorder effects on light-matter interactions.
\subsection{Cavity QED with coupled quantum emitters}
\label{Sec4A}
\indent The simplest example of an optical cavity is the co-planar design known as a Fabry-P\'{e}rot cavity comprised of two highly reflective parallel mirrors. For a distance $\ell$ between the mirrors and assuming at first perfect reflectivity, such a setup defines resonances conditioned by $\ell=n\lambda/2$. The fundamental mode has, therefore, a wavelength of $2\ell$ with frequency $\pi c/\ell$ and higher harmonics are multiples of this mode. The quantity $\omega_{\text{FSR}}=\pi c/\ell$ is known as the free spectral range and gives the frequency difference between consecutive optical resonances of the structure. Assuming two mirrors positioned at $z=0$ and $z=\ell$ and a transverse area $\mathcal{S}$ of the supported optical resonances, one can proceed in quantizing the field inside the optical resonator by associating bosonic operators $a_n$ to all resonances and writing the total Hamiltonian as $\mathcal{H}_c=\sum_n \omega_n a_n^\dagger a_n$. The electric field operator can then be decomposed as
\begin{equation}
\hat{\mathbf{E}}(z) = \textstyle \sum_n \mathcal{E}_n \left( a_n +  a_n^\dagger\right)\sin{(k_n z)}{\boldsymbol{\epsilon}}_n,
\end{equation}
where the sum runs over all allowed wave vectors $k_n=n\pi/\ell$  and the zero-point electric field amplitude $\mathcal{E}_n=\sqrt{\omega_n/(\epsilon_0 \ell \mathcal{S})}$ and all polarizations ${\boldsymbol{\epsilon}}_n$ are assumed orthogonal to the $z$-direction. For all situations we will consider in the following, a single cavity mode suffices. We denote the mode by the operator $a$ at frequency $\omega_c$ such that the free Hamiltonian becomes $\mathcal{H}_c=\omega_c a^\dagger a$.\\

\noindent \textbf{ME for a driven, lossy cavity mode} - Perfect mirrors define infinitely sharp resonances; in order to allow the in-coupling of light into the optical resonator side mirrors with slightly less than unit reflectivity are used. For a double-sided optical resonator, the localized resonances can then couple to the infinite number of modes to the left and right of the cavity, leading to a loss of the intracavity photons. This loss can be described within the master equation formalism by following a phenomenological model where photons from mode $a$ can tunnel to the left $b(\omega)$ and right $c(\omega)$ continuum of free radiation modes via an excitation exchange Hamiltonian. The standard open-system dynamics approach detailed in Appendix~\ref{A} leads to a Lindblad form of the cavity photon decay
\begin{equation}
\mathcal{L}_\kappa[ \rho] = \kappa \left[2a  \rho a^\dagger -a^\dagger a \rho - \rho a^\dagger a\right],
\end{equation}
where $\kappa=\kappa_R+\kappa_L$ encompasses total losses via both the right-side and left-side mirrors.\\
\indent To allow for the driving of the cavity mode let us now assume a continuous-wave laser with power $\mathcal{P}$ entering the cavity from the left side. This can be included in the following Hamiltonian
\begin{equation}
\mathcal{H}_\ell=i\eta(a^\dagger e^{-i \omega_\ell t}-ae^{i \omega_\ell t}),
\end{equation}
where the drive amplitude is $\eta=\sqrt{2\kappa_L \mathcal{P}/\omega_\ell }$. The equation of motion for the expectation value of the cavity operator can be derived from the master equation (with Hamiltonian $\mathcal{H}_c+\mathcal{H}_\ell$ and Lindblad term $\mathcal{L}_\kappa[ \rho]$) leading to
\begin{equation}
\label{ava}
\braket{\dot{a}}=-\kappa \braket{a} -i(\omega_c-\omega_\ell) \braket{a}+ \eta.
\end{equation}
The steady state of the equation above (reached by requiring that $\braket{\dot{a}}=0$) describes the expected Lorentzian response $\braket{a}_\text{ss}=\eta/(\kappa+i(\omega_c-\omega_\ell))$ of the cavity field exhibiting a linewidth $\kappa$ and resonance frequency $\omega_c$.\\

\noindent \textbf{Tavis-Cummings Hamiltonian} - Let us consider now $\mathcal{N}$ nonidentical quantum emitters with transition frequencies $\omega_j$ placed inside the optical cavity around a single optical mode of interest. Assuming the field to be varying only along the cavity axis, for emitter $j$ positioned at $z_j$ within the cavity volume, the dipolar interaction is characterized by a position-dependent coupling strength $g_j=\mathcal{E} d\ts{eg} \sin(kz_j)$. The light-matter Hamiltonian describing excitation exchange between the emitters and the cavity mode can then be cast in the standard Tavis-Cummings~\cite{tavis1968exact} form
\begin{align}
\label{HTC}
\mathcal{H}_\text{TC}= \sum_{j=1}^{\mathcal{N}} \omega_j\sigma^\dagger_j \sigma_j+\omega_c a^\dagger a+  \sum_{j=1}^{\mathcal{N}} \left(g_j a\sigma^\dagger_j+ g_j^* \sigma_j  a^\dagger\right).
\end{align}
This interaction is a particular form of the free-space case in Eq.~\eqref{Hfreespace} where the coupling to the cavity mode $a$ is enhanced while the coupling to all other modes is inhibited. At the single emitter level, the interaction is known as the Jaynes-Cummings Hamiltonian~\cite{jaynes1963comparison}. Notice that such Hamiltonian conserves the excitation number and can be exactly solved~\cite{tavis1968exact}. The excitation nonconserving case (known also as counter-rotating terms) for $\mathcal{N}=1$ is the exactly integrable~\cite{Braak2011integrability}, Rabi Hamiltonian~\cite{Rabi1936on}. The extension of the Rabi model to $\mathcal{N}>1$ for identical emitters $\omega_j=\omega_0$ and identical couplings $g_j=g$ is a simplified case describing exchange between a single collective operator $\sum_j \sigma_j$ and the cavity mode. This is known as the Dicke model, generally nonintegrable~\cite{Emary2003quantum,Braak2013solution} except for when $\omega_0=0$, and in the thermodynamic limit with $\mathcal{N}\rightarrow\infty$ ~\cite{hepp1973superradiant,brankov1975asymptotically,bogolubov1976a} where phase transitions have been identified~\cite{Wang1973phase}.\\

\noindent \textbf{Linear optical response} - For dense, interacting ensembles, the dipole-dipole interactions $\mathcal{H}_{dd}$ defined in Eq.~\eqref{Hdd} and the collective decay terms as defined in Eq.~\eqref{Lcoll} play an important role. To follow the system evolution, we then write the master equation with the total Hamiltonian $\mathcal{H}_{\text{TC}}+\mathcal{H}_{dd}+\mathcal{H}_\ell$ and Lindblad term $\mathcal{L}_\kappa[ \rho] +\mathcal{L}_e[ \rho]$. From here, one can derive equations of motion for single operator amplitudes. The emerging set of coupled equations is not closed, i.e., expectation values of single operators are coupled to two-operator correlations which again couple to three or more operator correlations and so on. However, in a first approximation, assuming weak driving conditions ($\eta\ll\gamma$) some factorization of operators under the approximation $\braket{\sigma_z^j}\approx-1$ can occur. This approximation is sufficient to describe the linear optical response as used throughout Sec.~\ref{Sec5A}. In Sec.~\ref{Sec5C} instead, we will go beyond this approximation to describe population-inverted systems exhibiting lasing behavior.\\
\indent Let us denote the expectation values by $\alpha=\braket{a}$ and $\beta_j=\braket{\sigma_j}$ and derive
\begin{subequations}
\label{alphabeta}
\begin{align}
\dot{\alpha}&=-(\kappa+i\omega_c)\alpha-i\sum_{j=1}^\mathcal{N}g_j \beta_j+\eta e^{-i \omega_\ell t}\\
\dot{\beta}_j&= -i\omega_j \beta_j-i g_j \alpha-\sum_{j'=1}^\mathcal{N} \mathcal{M}_{jj'}\beta_{j'}.
\end{align}
\end{subequations}
In Sec.~\ref{Sec4B} these equations will be extended to the quantum regime within the quantum Langevin equations formalism.\\

\noindent \textbf{Regimes of cavity QED} -  Let us quickly review a few standard concepts of cavity QED with non-interacting ensembles by setting $\Omega_{jj'}=0$ and $\gamma_{jj'}=\gamma\delta_{jj'}$. In steady state, the normalized transmission of the cavity $t=\kappa \alpha/\eta$ can be easily computed from Eqs.~\eqref{alphabeta} to lead to
\begin{align}
t=\kappa\left[\kappa+i (\omega_c-\omega_\ell)+\sum_{j=1}^{\mathcal{N}}\frac{|g_j|^2}{\gamma+i (\omega_j-\omega_\ell)}\right]^{-1}.
\end{align}
The cavity transmission can then show different regimes either characterized by a single Lorentzian peak (empty cavity or weak light-emitter coupling), a peak with a narrow dip (the Purcell, or antiresonant regime for intermediate couplings) or a polaritonic regime, where two cavity resonances are resolved in transmission. These regimes are dependent on the magnitude of the loss rates relative to the coupling strength. Mathematically, they can be easily deduced from the eigenproblem defined by the evolution matrix of Eqs.~\eqref{alphabeta}. For $\mathcal{N}=1$, diagonalization of Eqs.~\eqref{alphabeta} leads to normal mode splitting when $g>|\kappa-\gamma|/2$ with new frequencies $\omega_\pm=\omega_0\pm\sqrt{g^2+(\kappa-\gamma)^2/4}$ (on resonance $\omega_0=\omega_c$). The difference $\omega_+-\omega_-$ is known as the vacuum Rabi splitting (VRS). The newly formed quantum states in such a strong coupling regime are hybrid light-matter ones, i.e., polaritons $(\ket{g1}\pm\ket{e0})/\sqrt{2}$ and can be read in the cavity transmission $t$ as two well-separated peaks. For $\mathcal{N}>1$, the collective strong coupling regime can be defined when $g_\mathcal{N}>|\kappa-\gamma|/2$ with a collective coupling rate $g_\mathcal{N}=\sqrt{\sum_j{|g_j|^2}}$. The loss rates of the polaritons are equal in this regime $(\kappa+\gamma)/2$. One can also identify a weak coupling regime for which $\gamma<g_\mathcal{N}<\kappa$ but with a strong cooperativity $\mathcal{C}_\mathcal{N}=g_\mathcal{N}^2/(\kappa \gamma)\gg 1$. For many emitters equally coupled to the cavity mode, the cooperativity shows a collective enhancement proportional to $\mathcal{N}$. For a single emitter, this regime (known as the Purcell regime) leads to a renormalization of the intrinsic spontaneous emission rate to $\gamma(1+\mathcal{C})$ understood as the addition of a new decay channel via loss through the cavity. A scan of $t$ as a function of laser frequency shows in such a case a dip in the cavity transmission around $\omega_\ell=\omega_0$ known as a cavity antiresonance. The case of $\mathcal{N}$ coupled emitters will be presented in Sec.~\ref{Sec5A} showing that the addressing of collective subradiant resonances can lead to a superlinear $\propto \mathcal{N}^4$ scaling of $\mathcal{C}_\mathcal{N}$.

\subsection{Input-output formalism for operators}
\label{Sec4B}

The master equation formalism previously utilized is based on discarding the state of the environment and focusing on the dynamics of the reduced density operator of the much smaller system of interest. An alternative approach are quantum Langevin equations (also called Heisenberg-Langevin equations)~\cite{gardiner2004quantum,Jacobs2010Stochastic} which follow the evolution at the level of system operators instead of system states, i.e.,~its density operator. In essence, the formalism consists in supplementing the Heisenberg equations of motion with the proper dissipation and fluctuation terms.\\

\noindent \textbf{Mapping the ME onto QLEs} - For open-system dynamics described by loss in Lindblad form, a direct transformation between the master equation and the QLE formulation exists that reads~\cite{gardiner2004quantum}
\begin{align}
\label{diagonal_HLE}
\dot{\mathcal{O}} =\frac{i}{\hbar}[\mathcal{H},\mathcal{O}] &-[\mathcal{O},c^\dagger]\left\{\gamma\ts{c} c+\sqrt{2\gamma\ts{c}}c\ts{in}\right\}\\
&+\left\{\gamma\ts{c} c^\dagger +\sqrt{2\gamma\ts{c}}c\ts{in}^\dagger\right\}[\mathcal{O},c].
\nonumber
\end{align}
The equation above for a generic system operator $\mathcal{O}$ is applied to each individual Lindblad collapse operator $c$ acting at rate $\gamma\ts{c}$ and with associated input noise $c\ts{in}$. The input noise is a stochastic operator which is zero averaged $\braket{c\ts{in}}=0$, delta-correlated in time $\braket{c\ts{in}(t)c\ts{in}^\dagger(t')}=\delta(t-t')$ (all other correlations vanish) and satisfies the following commutation relation $[c\ts{in}(t),c\ts{in}^\dagger(t')]=\delta(t-t')$. For linear evolution, such an equation can be formally integrated and expectation values of correlations containing any number of operators can be obtained.\\
\indent We will proceed by first exemplifying the derivation and solution to a QLE applied to a lossy bosonic field corresponding to a decaying and externally driven optical cavity mode. Then we introduce the input-output relations allowing the readout of intracavity light-matter interactions via the cavity output field. Finally, we introduce a set of coupled linear QLEs for an ensemble of coupled, collectively decaying quantum emitters within the same optical cavity mode and show how quantum correlations can be computed. We will then utilize QLEs in Sec.~\ref{Sec5A} for the description of the optical response of subradiant arrays in optical cavities and in Sec.~\ref{Sec5B} for the description of hybrid cavities. An extension of the QLEs to include electron-vibron coupling will be introduced in Sec.~\ref{Sec6} to analytically tackle molecule-photon dynamics.\\

\noindent \textbf{QLEs and input-output relations.} - For a driven, lossy, single-sided optical cavity, Eq.~\eqref{diagonal_HLE} yields the following QLE
\begin{equation}
\dot{a}=-(\kappa+i\omega_c)a+\eta e^{-i \omega_\ell t}+\sqrt{2\kappa} a\ts{in}.
\end{equation}
Notice that an expectation value of the equation above simply reproduces Eq.~\eqref{ava} which has been alternatively derived directly from the master equation. The more general Langevin equation also allows for an exact solution obtainable by formal integration. We separate this solution into a transient and steady state part $a(t)=a_\text{tr} (t)+a_\text{ss}(t)$. While the transient part $a_\text{tr}(t)$ decays away on a timescale defined by the cavity loss $\kappa^{-1}$, the steady state part $a_\text{ss}(t)$ dominates in the long time (steady state) limit
\begin{subequations}
\begin{align}
a_\text{tr} (t)&=a(0) e^{-\kappa t}e^{-i\omega_c t},\\
a_\text{ss} (t)&= \int_0^t dt'(\eta e^{-i\omega_\ell t'}+\sqrt{2\kappa}a\ts{in}(t'))e^{-(\kappa+i\omega_c) (t-t')}.
\end{align}
\end{subequations}
It is important to notice that the inclusion of the noise terms in such equations is crucial: while the commutator $[a_\text{tr}(t),a_\text{tr}^\dagger (t)]=e^{-2\kappa t}$ decays away, the commutator $[a_\text{ss}(t),a_\text{ss}^\dagger (t)]=1-e^{-2\kappa t}$ maintains the standard bosonic commutation relation $[a(t),a^\dagger (t)]=1$ at every time during evolution.\\
In addition to the equation of motion for the cavity field amplitude operators, the full description of the optical cavity includes the input-output relation
\begin{align}
a\ts{out}+a\ts{in}=\sqrt{\kappa}a(t),
\end{align}
which allows for the derivation of the properties of the cavity output mode. For example, the average shows that the cavity transmission can be simply written as: $t=\braket{a\ts{out}}/\eta=\sqrt{\kappa}\braket{a}/\eta$ (as already used and consistent with the result of Eq.~\eqref{ava} in Sec.~\ref{Sec4A}).\\

\noindent \textbf{QLEs for cooperative cavity QED} - We now assume a cavity mode coupled to $\mathcal{N}$ interacting emitters, as a system described by the Hamiltonian $\mathcal{H}_{\text{TC}}+\mathcal{H}_{dd}+\mathcal{H}_\ell$ and loss rates incorporated in the terms $\mathcal{L}_\kappa[\rho] +\mathcal{L}_e[\rho]$ (as introduced in Sec.~\ref{Sec4A}). In a first step, we notice that the decay terms are not in Lindblad form, thus we perform a diagonalizing transformation to bring it into the form of $\mathcal{N}$ independent decay channels as described by Eq.~\eqref{Ldiag} (such that Eq.~\eqref{diagonal_HLE} can then be directly applied). We also only consider the low excitation limit $\braket{\sigma_z^j}\approx-1$. Notice that this approximation discards the nonlinear response of the dipoles, which could, for example, lead to a collective Kerr nonlinear effect~\cite{plankensteiner2019enhanced}. One then obtains a set of coupled linear differential equations
\begin{subequations}
\label{asigma}
\begin{align}
\dot{a} &= -\left(\kappa - i\Delta_c\right) a - i \sum_j g_j \sigma_j +\sqrt{2\kappa} a\ts{in}+\eta,\\
\dot{\sigma}_j &= i\Delta \sigma_j -i g_j a - \sum_{j'} \mathcal{M}_{jj'} \sigma_{j'} + \sqrt{2\gamma}\sigma\ts{in}^j,
\end{align}
\end{subequations}
with the emitter and cavity detunings defined as $\Delta=\omega_\ell-\omega_0$ and $\Delta_c=\omega_\ell-\omega_c$ respectively. The input noise terms $\sigma\ts{in}^j$ (stemming from the coupling of the emitters to the electromagnetic modes outside the solid-state angle covered by the cavity field) are considered here as zero-averaged and delta-correlated in time $\braket{\sigma\ts{in}^j(t)\sigma\ts{in}^{j',\dagger}(t')}=(\gamma_{jj'}/\gamma)\delta(t-t')$ as a consequence of the low excitation assumption (for the more complex case of high excitation we refer to Refs.~\cite{gardiner2004quantum,plankensteiner2019enhanced}). We can write the equations above in a compact form in terms of the matrix $\textbf{M}(\Delta)= -i\Delta\mathbbm{1}+i\boldsymbol{\Omega}+\boldsymbol{\Gamma}$ and the coupling vector $\textbf{G}=(g_1,g_2,...,g_\mathcal{N})^\top$.
Additionally defining the vectors $\boldsymbol{\sigma}=(\sigma_1,\hdots,\sigma_\mathcal{N})^\top$ and $\boldsymbol{\sigma}_{\text{in}}=(\sigma_{\text{in}}^1,\hdots,\sigma_{\text{in}}^\mathcal{N})^\top$, \eqsref{asigma} express as
\begin{subequations}
\label{asigmavec}
\begin{align}
\dot{a}&=-(\kappa-i\Delta_c)a-i\mathbf{G}^\top\boldsymbol{\sigma}+\sqrt{2\kappa}a_\text{in}+\eta,\\
\dot{\boldsymbol{\sigma}}&=-\mathbf{M}(\Delta)\boldsymbol{\sigma}-i\mathbf{G}a+\sqrt{2\gamma}\boldsymbol{\sigma}_\text{in}.
\end{align}
\end{subequations}
This is the starting point for the derivations in Sec.~\ref{Sec5A} where we compute the classical and quantum response of a weakly excited transverse 1D or 2D array of coupled quantum emitters to a driven cavity mode.\\

\noindent \textbf{Quantum correlations} - From the equations of motion for operators Eqs.~\eqref{asigmavec}, one can go a step further and fully analyze the quantum properties (second-order correlation function for light $g^{(2)}$, bipartite entanglement, squeezing properties, etc.) of both the photon and matter counterparts. To this end, let us first consider only fluctuation operators only by expansion around expectation values $a= \alpha+\delta a$, $\sigma_j =\beta_j+\delta\sigma_j$. We can then cast~\eqsref{asigmavec} in convenient vector form as  $\dot{\mathbf{v}}=\mathbf{A}\mathbf{v}+\mathbf{N}\mathbf{v}_\text{in}$ for the vectors $\mathbf{v}=(\delta a,\delta a^\dagger,\delta\boldsymbol{\sigma},\delta\boldsymbol{\sigma}^\dagger)^\top$, $\mathbf{v}_\text{in}=(a_\text{in}, a_{\text{in}}^\dagger,\boldsymbol{\sigma}_{\text{in}}, \boldsymbol{\sigma}_{\text{in}}^\dagger)^\top$ with the drift matrix expressed in compact form as (for details also see Ref.~\cite{plankensteiner2019enhanced})
\begin{align}
\mathbf{A} &= \begin{pmatrix}
-\left(\kappa - i\Delta_c\right) & 0 & -i\textbf{G}^\top & \textbf{0}^\top  \\
0 & -(\kappa + i\Delta_{c}) & \textbf{0}^\top & i\textbf{G}^\top \\
-i\textbf{G} & \textbf{0} & -\textbf{M}(\Delta) & \underline{\textbf{0}}  \\
\textbf{0} & i\textbf{G} & \underline{\textbf{0}} & -\textbf{M}^*(\Delta)
\end{pmatrix}.
\end{align}
We have defined the vector $\textbf{0}$ containing $\mathcal{N}$ zeros and $ \underline{\textbf{0}}$ is a $\mathcal{N}\times\mathcal{N}$ matrix with only zeros. The matrix multiplying the input noise operators is
\begin{align}
\mathbf{N}= \begin{pmatrix}
\sqrt{2\kappa} & 0 & \textbf{0}^\top & \textbf{0}^\top  \\
0 &\sqrt{2\kappa} & \textbf{0}^\top & \textbf{0}^\top\\
\textbf{0} & \textbf{0} & \sqrt{2\gamma}\mathbbm{1} & \underline{\textbf{0}}  \\
\textbf{0} &\textbf{0} & \underline{\textbf{0}} & \sqrt{2\gamma}\mathbbm{1}
\end{pmatrix}.
\end{align}
Formally integrating this system of linearly coupled equations gives the solution
\begin{align}
\mathbf{v}(t)=e^{\mathbf{A}t}\mathbf{v}(0)+\int_0^t dt' e^{\mathbf{A}(t-t')}\mathbf{N}\mathbf{v}_\text{in}(t').
\end{align}
If all eigenvalues of the drift matrix are negative, the system is stable and will go towards a steady state (i.e., the transient term $e^{\mathbf{A}t}$ vanishes). From the steady state solution only, one can then define a correlation matrix $\mathbf{V}=\braket{\mathbf{v}(t)\mathbf{v}^\top(t)}$ which can be expressed as
\begin{align}
\mathbf{V}=\int_0^t dt' e^{\mathbf{A}(t-t')}\mathbf{D}e^{\mathbf{A}^\top (t-t')},
\end{align}
where we used that $\braket{\mathbf{v}_\text{in}(t')\mathbf{v}_{\text{in}}^\top(t'')}=\mathbf{C}\delta(t'-t'')$ and defined the diffusion matrix as $\mathbf{D}=\mathbf{N}\mathbf{C}\mathbf{N}^\top$. The matrix containing the correlations can be readily obtained as
\begin{align}
\mathbf{C}= \begin{pmatrix}
0 & 1 & \textbf{0}^\top & \textbf{0}^\top  \\
0 &0  & \textbf{0}^\top & \textbf{0}^\top\\
\textbf{0} & \textbf{0} &  \underline{\textbf{0}} & \mathbf{\Gamma}/\gamma \\
\textbf{0} &\textbf{0} & \underline{\textbf{0}} & \underline{\textbf{0}}
\end{pmatrix}.
\end{align}
\begin{figure}[t]
\center
\includegraphics[width=0.6\columnwidth]{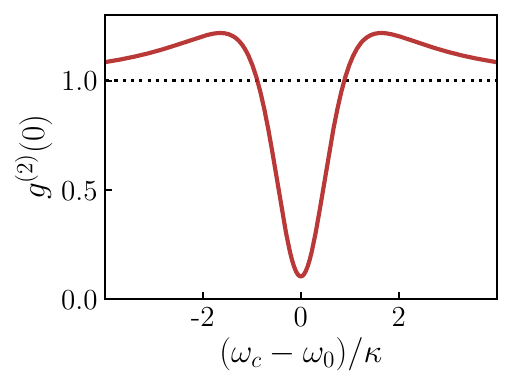}
\caption{Second-order optical correlation function $g^{(2)}(0)$ of a driven emitter-cavity system as a function of the cavity-emitter detuning $\omega_c-\omega_0$. Parameters: $\gamma=0.1\kappa$, $\eta=5\cdot10^{-3}\kappa$, $g=0.3\kappa$.}
\label{fig41}
\end{figure}
For the covariance matrix, one can then derive the \textit{Lyapunov equation} using integration by parts
\begin{align}
\nonumber
\mathbf{A}\mathbf{V}&+\mathbf{V}\mathbf{A}^\top=\int_0^t dt' \mathbf{A}e^{\mathbf{A}(t-t')}\mathbf{D}e^{\mathbf{A}^\top(t-t')}+\mathbf{V}\mathbf{A}^\top=\\
&\nonumber=-e^{\mathbf{A}(t-t')}\mathbf{D}e^{\mathbf{A}^\top(t-t')}\Big|_0^t-\mathbf{V}\mathbf{A}^\top+\mathbf{V}\mathbf{A}^\top=\\
&=-\mathbf{D}.
\end{align}
Alternatively, one can also perform a Fourier analysis to turn the system of differential equations to an algebraic set of coupled equations $i\omega \mathbf{v}(\omega)=\mathbf{A}\mathbf{v}(\omega)+\mathbf{N}\mathbf{v}_\text{in}(\omega)$, which allows to relate the intracavity operators to the input noise via
\begin{align}
\mathbf{v}(\omega)=\left(i\omega\mathbbm{1}-\mathbf{A}\right)^{-1}\mathbf{N}\mathbf{v}_\text{in}(\omega).
\end{align}
Furthermore, one can relate the input fields directly to the output fields by using the fact that in the time domain $\mathbf{v}_\text{out}(t)=\mathbf{N}^\top\mathbf{v}(t)-\mathbf{v}\ts{in}(t)$ and therefore one can express $\mathbf{v}\ts{out}(\omega)=\mathbf{F}(\omega)\mathbf{v}_\text{in}(\omega),$
where $\mathbf{F}(\omega)=\left[\mathbf{N}^\top(i\omega\mathbbm{1}-\mathbf{A})^{-1}\mathbf{N}-\mathbbm{1}\right]$. In the frequency domain, the $\delta$-correlations are preserved for the output fields $\braket{\mathbf{v}\ts{out}(\omega)\mathbf{v}\ts{out}^\top(\omega')}=\boldsymbol{\mathcal{S}}\ts{out}(\omega)\delta(\omega+\omega')$,
where the two-operator correlations are completely encoded in the spectrum matrix given by
\begin{align}
\boldsymbol{\mathcal{S}}\ts{out}(\omega)=\mathbf{F}(\omega)\mathbf{C}\mathbf{F}^\top(-\omega).
\end{align}
While this suffices to characterize the two-operator correlations of the system, also higher-order correlations (e.g.~four-operator correlations) can be of interest. We note that all higher order correlations can be expressed as a sum over products of two-point correlations via the Isserlis' theorem. An example of such a correlation is the the $g^{(2)}$-function already considered in Sec.~\ref{Sec3C}, characterizing the photon statistics of the system. It is defined in steady state as
\begin{align}
g^{(2)}(\tau)=\frac{\braket{a^\dagger (t)a^\dagger (t+\tau)a(t+\tau)a(t)}\ts{ss}}{\braket{a^\dagger (t)a(t)}\ts{ss}^2}.
\end{align}
Nonclassical light sources show sub-Poissonian statistics $g^{(2)}(0)<1$ (antibunching), implying that it becomes unlikely for photons to be detected in pairs. In \fref{fig41} this is illustrated for the coupled cavity-emitter system where the mechanism originates from the anharmonicity of the Jaynes-Cummings ladder, allowing only single photons inside the cavity.

\subsection{Cavity QED with disordered ensembles}
\label{Sec4C}

In the most general case of Eqs.~\eqref{alphabeta} the frequencies of the emitters are not identical. This is often encountered in experiments, as the coupling of electronic systems to embedding matrices can lead to strong inhomogeneous broadening, such as it is the case for quantum dots or molecular systems. To quantify frequency disorder, we assume that the frequencies $\omega_j$ are distributed around $\omega_0$ according to the distribution function $p(\delta)$ normalized to unity $\int_{-\infty}^{\infty}p(\delta)d\delta=1$. In particular we choose a Gaussian distribution of frequencies $p(\delta) = (1/\sqrt{2\pi w^2})e^{-\delta^2/(2w^2)}$. We write the transition frequencies as $\omega_j=\omega_0+\delta_j$ with vanishing classical average $\braket{\delta_j}_\text{cl}=0$ and variance $\braket{\delta_j^2}_\text{cl}=w^2$. For simplicity, we restrict the discussion to the case of identical couplings $g_{j} = g$ (for all $j$) while more generally we refer the reader to Ref.~\cite{sommer2021molecular}.\\
\indent For $w=0$ we note that the cavity couples only to a symmetric superposition $\hat{B}=\textstyle \sum_j \sigma_j/\sqrt{\mathcal{N}}$, i.e., a \textit{bright state}, with a collective coupling strength $g_\mathcal{N} = \sqrt{\mathcal{N} }g$. The other $\mathcal{N} -1$ combinations define \textit{dark states} which are obtainable by a Gram-Schmidt algorithm that leads to all vectors orthogonal to the bright state and to each other. However, a simple choice of coefficients is indicated by a discrete Fourier transform $\hat{D}_{k} = 1/\sqrt{\mathcal{N} }\textstyle \sum^{\mathcal{\mathcal{N} } }_{j = 1}e^{-i2\pi jk/\mathcal{N}}\sigma_j$. We index the dark-state manifold for $k=1,\dots,\mathcal{N} -1$ and note that for $k=\mathcal{N} $ we have $\hat{D}_{\mathcal{N} } = \hat{B}$. The equations of motion for all variables $\mathcal{D}_k=\braket{\hat{D}_k}$ and $\alpha=\braket{a}$ become (in a frame rotating at $\omega_0$)
\begin{subequations}
\begin{align}
\dot{\mathcal{D}}_k &= -\gamma \mathcal{D}_k-i\sum^{\mathcal{N} }_{k' = 1}\Delta_{kk'} \mathcal{D}_{k'} -i g_\mathcal{N} \alpha \delta_{k\mathcal{N}},\\
\dot{\alpha} &= -(\kappa-i\Delta_c)\alpha-ig^{*}_\mathcal{N}  \mathcal{D}_\mathcal{N} +\eta,
\end{align}
\end{subequations}
where the couplings between collective states are defined as $\Delta_{kk'} = 1/\mathcal{N} \textstyle\sum^{\mathcal{N} }_{j = 1} \delta_j e^{-i 2\pi j(k-k')/\mathcal{N} }$. For $k=\mathcal{N} $, the equations above indicate that the bright state couples to the cavity mode with the standard $g_\mathcal{N}$ rate, while also being coupled to all dark states.\\

\begin{figure}[t]
\center
\includegraphics[width=0.7\columnwidth]{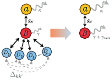}
\caption{In the case of disorder of the electronic transition frequencies, the dark modes are coupled to the bright mode and to each other while the bright mode couples to the cavity mode with rate $g_{\mathcal{N}}$. Elimination of the dark, cavity uncoupled modes then  leads to an effective two hybridized mode problem (cavity mode and bright collective state) with renormalized loss rates.}
\label{fig42}
\end{figure}

\noindent \textbf{Non-Markovian regime} - We proceed (see Fig.~\ref{fig42}) to eliminate the dark-state manifold in an exact way without making a Markovian approximation (which would imply that the dark state reservoir has no memory and therefore it would allow to set all derivatives of $\mathcal{D}_k$ to zero). Instead, we formally integrate the equations for $\mathcal{D}_k$ to obtain
\begin{equation}
\dot{\mathcal{B}}(t) = -(\gamma +i \bar{\delta})\mathcal{B}(t)-\int_{-\infty}^{\infty}dt' f(t-t')\mathcal{B}(t') -i g_{\mathcal{N}}\alpha,
\end{equation}
where the memory kernel generally describes a non-Markovian loss process. In the mesoscopic limit $\mathcal{N}\to\infty$, one finds \begin{equation}
f(t-t')\approx\Theta(t-t')w^2e^{-i(\bar{\delta}-i\gamma)(t-t')} \sinc(2w(t-t')).
\end{equation}
This now allows to identify the Markovian regime discussed in the next paragraph, where the condition that the disorder $w\gg\gamma$ implies that the kernel $f(t-t')$ tends towards a delta function.
\\

\noindent \textbf{Markovian limit} -  In the Markovian limit, the elimination of the dark-state manifold is straightforward as all the derivatives $\dot{\mathcal{D}}_k$ for $k\neq \mathcal{N}$ can be set to zero. This means $\mathcal{D}_k=-\sum_{k'}({\mathcal{A}^{-1}})_{kk'}\Delta_{k'\mathcal{N} } \mathcal{B}$ with the matrix
\begin{equation}
  \mathcal{A} =\left(
    \begin{array}{cccc}
      (\bar{\delta}-i\gamma) & \Delta_{12} & \ldots & \Delta_{1(\mathcal{N}-1)} \\
      \Delta_{21} & (\bar{\delta}-i\gamma) & \ldots & \Delta_{2(\mathcal{N}-1)}  \\
      \vdots & \vdots & \ddots & \vdots \\
      \Delta_{(\mathcal{N} -1)1} & \Delta_{(\mathcal{N} -1)2} & \ldots & (\bar{\delta}-i\gamma) \\
    \end{array}
  \right),
\end{equation}
and $\bar{\delta}=\Delta_{kk}=\textstyle\sum^{\mathcal{N} }_{j = 1} \delta_j/\mathcal{N} $. The Markovian kernel is explicitly $f(t-t')=(i\delta_\text{dark}+\gamma_\text{dark})\delta(t-t')$,
containing an effective frequency shift $\delta_\text{dark}$ and a loss rate $\gamma_\text{dark}$ derived from $\delta_\text{dark}+i \gamma_\text{dark}=\sum_{k,k'=1}^{\mathcal{N} -1}\Delta_{\mathcal{N} k}(\mathcal{A}^{-1})_{kk'}\Delta_{k'\mathcal{N} }$.
In the mesoscopic limit, one can further simplify the expression of the loss rate and derive scaling laws where $\gamma_\text{dark} =w^2/\gamma$ for $\gamma \gg w$ and $\gamma_\text{dark}=\pi w /4$ for $w\gg \gamma$. Moreover, both $\bar{\delta}$ and $\delta_\text{dark}$ vanish in this regime. The dependence of the VRS on disorder can then be obtained by diagonalizing the dynamics in the reduced cavity-bright state subspace
\begin{eqnarray}
\label{theoryvrs}
\text{VRS} \approx \Im \left\{2\sqrt{\left(\gamma+\gamma_\text{dark}-\kappa\right)^2/4-g_\mathcal{N}^2}\right\}.
\end{eqnarray}
The expression above shows that a large degree of disorder (on the order of $\kappa$) can lead to a strong reduction of the VRS and consequently pull the system out of the collective strong coupling regime. In Ref.~\cite{sommer2021molecular} a more detailed derivation is provided that shows that particles which have a large detuning with respect to the cavity frequency, or a very lossy behavior are effectively pulled out of the macroscopic superposition participating in the strong coupling condition. This in turn leads to the degradation of the VRS described by the equation above. \\

\section{Applications in cavity QED}
\label{Sec5}

We now discuss a few applications of cavity QED with coupled quantum-emitter ensembles. For 1D and 2D arrays of emitters, we now have systems exhibiting cooperative surface resonances, as already analyzed for free space applications in Sec.~\ref{Sec3}, but with an extra degree of freedom introduced by a cavity-confined field mode. In the limit of low reflectivity, such systems have been shown to act as quick frequency switchers~\cite{plankensteiner2017cavity} and to give rise to enhanced optical nonlinearities~\cite{plankensteiner2019enhanced}. In the high reflectivity limit, such as theoretically predicted in Refs.~\cite{bettles2016enhanced,shahmoon2017cooperative} and experimentally proven in Ref.~\cite{rui2020asubradiant}, we introduce the necessary steps, as laid out in Ref.~\cite{cernotik2019cavity}, to derive the correct input-output formalism for optical resonators made up of subwavelength arrays as end-mirrors.\\
\indent In the weak reflectivity limit, the array reacts as a strongly dispersive optical element within a short frequency window but it does not, at the same time, considerably change the spatial profile of the cavity mode (owing to its weak reflectivity). The treatment is then perturbative, within the standard Tavis-Cummings formalism and nontrivial effects occur such as enhanced optical nonlinearities and the reach of an enhanced collective Purcell effect with a cooperativity scaling up to $\propto \mathcal{N}^4$ (for a 1D arrangement). \\
\indent In the opposite case, where the emitter array acts as a near-unity reflectivity end-mirror, a \textit{hybrid cavity} design emerges where asymmetric transmission profiles can be achieved, potentially much narrower than those obtained with frequency-independent mirrors of comparable reflectivity. For such designs, the standard input-output theory and the master equation for the photon mode losses (from Sec.~\ref{Sec4A}) loses validity, as the tunneling rate of photons strongly depends on their frequency. We present a general roadmap to derive the correct input-output relations for optical cavities comprised of either one or two of such end-mirrors. We remark that this is not limited to subradiant emitter arrays but also extendable to patterned subwavelength gratings or photonic crystal structures~\cite{Miroshnichenko2010,Chang-Hasnain2012,Zhou2014,Limonov2017} and semiconducting monolayers~\cite{Zeytinoglu2017,Back2018,Scuri2018}.\\
\indent Finally, for ensembles of cavity-embedded quantum emitters, externally and incoherently pumped, we provide a minimal theory of lasing to illustrate how the laser threshold condition can be easily derived using the formalism introduced in Sec.~\ref{Sec4A} and by going beyond the linear regime. In particular, we analyze the physics of superradiant lasers, where coherence is being stored in the gain medium instead of in the cavity photon field. Finally, we connect the dynamics of such a laser to the Dicke superradiance model introduced in Sec.~\ref{Sec2D}.\\

\subsection{Antiresonance spectroscopy with 1D or 2D arrays}
\label{Sec5A}

Let us first consider a weakly reflecting array placed transversely to the axis of a standard optical cavity. In the presence of near-field couplings, the system undergoes dynamics describable at the level of averages by Eqs.~\eqref{alphabeta} and at the fully quantum level by Eqs.~\eqref{asigma}. We will assume a vector $\textbf{G}$ containing all the cavity-emitter couplings and later particularize to symmetric coupling (where all couplings are equal to $g$) and antisymmetrically-phased coupling (with alternating $(-1)^j g$ couplings). Assuming a single cavity drive, one can then easily deduce the cavity transmission as
\begin{align}
t= \displaystyle\frac{\kappa}{i\Delta_{c}+\kappa + \textbf{G}^\top\textbf{G}/\left[i\Omega_\text{eff}(\Delta) + \gamma_\text{eff}(\Delta)\right]},
\end{align}
where the effective $\Delta$-dependent collective energy shifts and linewidths are derived from the matrix $\textbf{M}(\Delta)= -i\Delta\mathbbm{1}+i\boldsymbol{\Omega}+\boldsymbol{\Gamma}$ as real and imaginary parts
\begin{align}
\gamma_\text{eff}(\Delta)+i\Omega_\text{eff}(\Delta)=\frac{\textbf{G}^\top\textbf{G}}{\textbf{G}^\top\textbf{M}^{-1}(\Delta)\textbf{G}}.
\end{align}
These expressions are the equivalents of the decay rates and shifts for free space arrays given in Sec.~\ref{Sec3C} where the addressing of collective resonances is now controlled via the choice of cavity couplings instead of the external drive.\\
\begin{figure}[t]
\includegraphics[width=0.95\columnwidth]{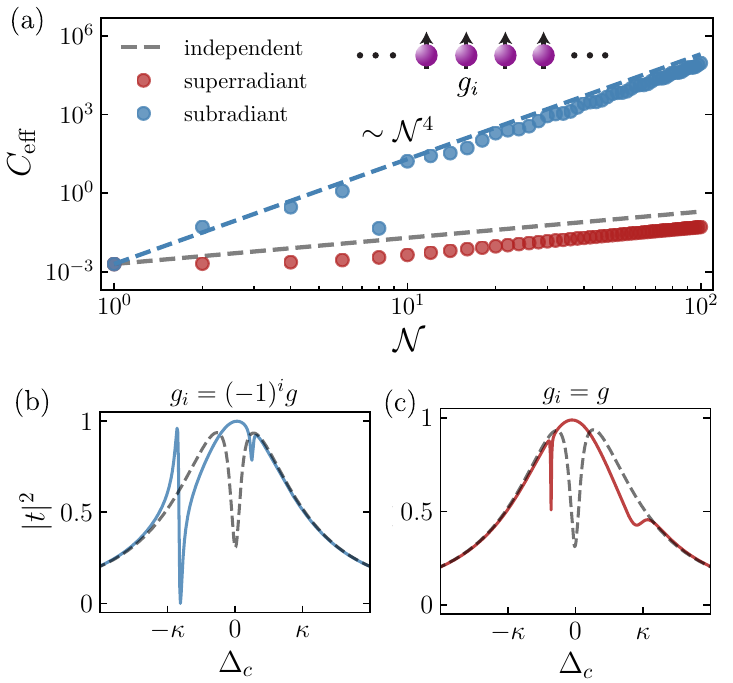}
\caption{(a) Effective cooperativity $C_\text{eff}$ of a cavity-embedded chain of emitters as a function of $\mathcal{N}$ for the superradiant (red, symmetric addressing $g_i=g$) and subradiant (blue, antisymmetric addressing $g_i=(-1)^i g$) case. Parameters: $a=0.1\lambda$, $\gamma=\kappa/20$, $g=\kappa/100$. The bottom half shows the cavity transmission functions for (b) antisymmetric and (c) symmetric excitation of four equally spaced emitters for  $a=0.08\lambda$, $g=\kappa/10$. The gray dashed curves shows the result for independent emitters. In all plots, the dipole moments are chosen perpendicular to the chain.}
\label{fig51}
\end{figure}

\noindent \textbf{Enhanced cooperativity} -  Notice that one can now proceed by introducing a modified $\mathcal{N}$ emitter effective cooperativity
\begin{align}
\label{CN_def}
C\ts{eff}(\Delta) = \frac{\textbf{G}^\top\textbf{G}}{\kappa\gamma_\text{eff}(\Delta)}.
\end{align}
As mentioned in Sec.~\ref{Sec4A}, for a single emitter the cooperativity is independent on the emitter properties (dipole moment, decay rate) and simply depends on the cavity finesse. For $\mathcal{N}$ uncoupled emitters the same holds true and only a linear increase with $\mathcal{N}$ will be obtained. For coupled emitters, an interesting decoupling of the dipole moment in the direction of the cavity (quantified by the term $\textbf{G}^\top\textbf{G}$) from the collective radiative properties $\gamma_\text{eff}(\Delta)$ can be achieved. As $\gamma_\text{eff}$ is not a natural constant of the ensemble, but strongly dependent on the relative positioning and phase of individual emitters, one can reach subradiant states with $\gamma_\text{eff}\ll\gamma$. By proper design of the cavity transverse field amplitude profile, the numerator can at the same time be maximized, resulting in a scaling up of $C\ts{eff}$ well above the independent emitter case $\mathcal{N} g^2/(\kappa \gamma)$. This is illustrated in Fig.~\ref{fig51}a where a scaling with $\mathcal{N}^4$ for a 1D chain is shown possible. In Figs.~\ref{fig51}b and c, we consider scans of the cavity transmission for a system of four coupled emitters. The system will have three subradiant and one superradiant state with resonances given by Eq.~(\ref{eigdipole}) (the superradiant state is located at $\omega_0+2\Omega\cos(\pi/5)$). We adress the system either antisymmetrically with $\mathbf{G}=(g,-g,g,-g)^\top$ [Fig.~\ref{fig51}b] or symmetrically with $\mathbf{G}=(g,g,g,g)^\top$ [Fig.~\ref{fig51}c]. While antisymmetric excitation overlaps with two subradiant states, symmetric excitation overlaps with one subradiant and one superradiant state. In Ref.~\cite{plankensteiner2017cavity} it is shown that phased excitation could be realized by using the higher-order TEM-modes of the optical cavity. Alternatively, one could use symmetric addressing around the optimal points utilized experimentally in Ref.~\cite{rui2020asubradiant} or symmetric addressing combined with a magnetic field gradient as in Ref.~\cite{plankensteiner2015selective}.\\

\subsection{Hybrid cavities with 2D subwavelength mirrors}
\label{Sec5B}

\begin{figure*}[t]
\center
\includegraphics[width=1.85\columnwidth]{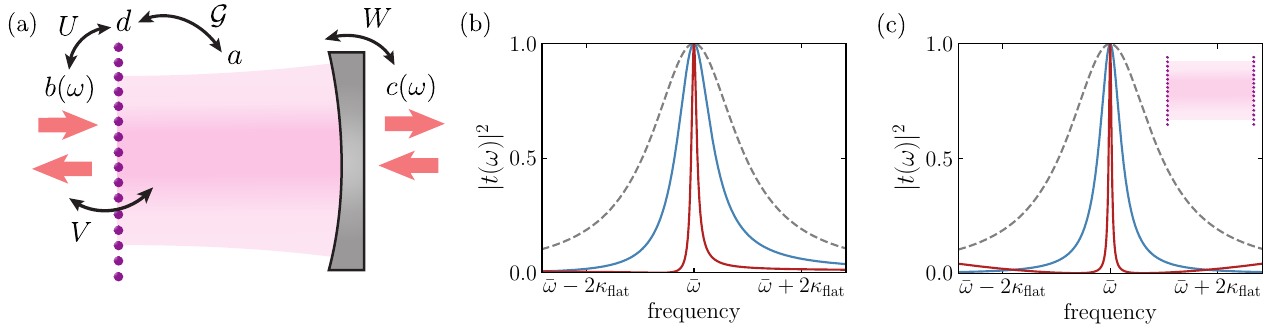}
\caption{(a) A hybrid cavity comprised of an atomic array (left mirror) and a standard, flat frequency mirror (right). The cavity mode $a$ interacts with two external continua $b(\omega)$, $c(\omega)$ and the surface-confined mode on the array $d$. (b) Cavity response for $\zeta_0=10$. The gray dashed line is the expected Lorentzian profile with linewidth $\kappa_\text{flat}=\omega_\text{FSR}/(2\pi \zeta_0^2)$ in the Markovian regime where $\gamma_d=\omega_\text{FSR}$. For narrower mirror linewidths $\gamma_d=\omega_\text{FSR}/10$ (blue line) and $\gamma_d=\omega_\text{FSR}/100$ (red line) the cavity shows an asymmetric transmission profile with a strongly reduced linewidth at the level of $\gamma_d/\zeta_0$. (c) Transmission profile for a double-sided hybrid cavity, where also the right mirror is replaced by an atomic array (same values for $\gamma_d$ as in (b)) shows a symmetric response with an even narrower linewidth.}
\label{fig52}
\end{figure*}

We now aim to provide a theoretical framework for hybrid cavities where subwavelength arrays are to be used as end mirrors as suggested in Refs.~\cite{shahmoon2017cooperative,cernotik2019cavity,rui2020asubradiant}. The standard approach for cavity loss and input-output relations taken in Sec.~\ref{Sec4} is based on the assumption that the tunneling rate of photons between a cavity-confined mode $a$ and the outside continuum of modes ($b(\omega)$ to the left and $c(\omega)$ to the right) is flat around the cavity resonance. This assumption is strongly modified in the case of subwavelength reflective arrays and the presence of a surface-confined resonance (mode $d$) has to explicitly be taken into account (see illustration in Fig.~\ref{fig52}a). The roadmap to construct an input-output quantum theory for such hybrid cavities follows the steps described in Ref.~\cite{cernotik2019cavity} where it has been applied to photonic crystal mirrors. The procedure consists of three steps: i) the derivation of equations of motion for cavity field operators based on a coupled modes model where the photon-exchange processes between modes $a$ and $d$ are included, ii) the derivation of the cavity amplitude transmission from a classical transfer-matrix approach based on the expression of the reflectivity of the array $r(\omega)$ derived in Sec.~\ref{Sec3C} and iii) the extraction of the phenomenologically introduced parameters by matching the predictions of the two theories. We apply this procedure here to a hybrid cavity made of one flat mirror and one subwavelength array and point out expected improvements in the cavity finesse where the hybrid design makes use of two emitter array mirrors.\\

\noindent \textbf{Coupled modes theory} - We consider the situation depicted in Fig.~\ref{fig52}a where the cavity mode $a$ is coupled to the surface-confined mode $d$ at some complex rate $\mathcal{G}$ and the photon tunneling rates $U,V,W\in\mathbb{R}$ are frequency independent. Tunneling will then give rise to loss rates $\kappa_L = \pi V^2$, $\kappa_R = \pi W^2$, $\gamma_d = \pi U^2$ (see Appendix~\ref{A}).
The dynamics is then described by the Langevin equations
\begin{subequations}\label{eq:EOM}
\begin{align}\label{eq:EOMCavity}
    \dot{a} &= -(i\omega_a+\kappa)a - \mathcal{G}d + \sqrt{2\kappa_L}b_\inpt + \sqrt{2\kappa_R}c_\inpt, \\
    \dot{d} &= -(i\omega_d+\gamma_d)d - \mathcal{G}a + \sqrt{2\gamma_d}b_\inpt\label{eq:EOMFano},
\end{align}
\end{subequations}
showing that the mode $a$ is subject to two types of noises: $b_\inpt$, entering through the left and $c_\inpt$, entering through the right mirror. In a first approximation, we will consider that the surface-confined mode reacts to the same noise term $b_\inpt$ consisting of modes on the left of the cavity but extra independent noise terms stemming for example from nonradiative losses could be added. The total decay rate is $\kappa = \kappa_L+\kappa_R$ and the input fields have the usual nonvanishing correlation functions $\braket{b_\inpt(t)b_\inpt^\dagger(t')} = \delta(t-t')$ [and similarly for $c_\inpt(t)$].
The associated output fields follow the input-output relations $b_\out = b_\inpt-\sqrt{2\kappa_L}a-\sqrt{2\gamma_d}d,$ and $c_\out = c_\inpt -\sqrt{2\kappa_R}a$.\\
\indent We are mainly interested here in deriving the free parameters in the coupled modes model, i.e., $\mathcal{G}$, $\omega_a$ and $\kappa_L$. For more in-depth discussions highlighting the non-Markovianity of such hybrid cavities we refer to Ref.~\cite{cernotik2019cavity} where a full analysis of photonic crystal mirror cavities is provided.

The transmission coefficient $t(\omega)=\braket{c_\out(\omega)}/\braket{b_\inpt(\omega)}$ is easily expressed as  $t(\omega)=-\sqrt{2\kappa_R} \braket{a(\omega)}/\braket{b_\inpt(\omega)}$ in the case of sole driving through the left cavity port and can be analytically derived from Eqs.~\eqref{eq:EOM} in the frequency domain
\begin{equation}
\label{tcm}
    t(\omega) = \frac{\sqrt{2\kappa_R}[\epsilon_d(\omega)\mathcal{G}\sqrt{2\gamma_d}-\sqrt{2\kappa_L}]}{\epsilon_a^{-1}(\omega)-\mathcal{G}^2\epsilon_d(\omega)},
\end{equation}
where we introduce the susceptibilities $\epsilon_a^{-1}(\omega) = \kappa+i(\omega_a-\omega)$ and $\epsilon_d^{-1}(\omega) = \gamma_d+i(\omega_d-\omega)$. This is the main result of the coupled modes theory, which at this point still retains complete generality; in the following we will apply it to the particular scenario of a subwavelength reflective array.\\

\noindent \textbf{Transfer matrix results} - The transfer matrix approach, on the other hand, consists in solving the classical one-dimensional wave propagation in a one-dimensional setup with two mirrors parametrized by the polarizabilities $\zeta_0$ (right mirror) and $\zeta_L(\omega)$ (left, subwavelength array mirror). In linear response theory, one can find the transmission function of the setup for any incoming plane wave at a given frequency $\omega$. We use the parametrization $\zeta_L(\omega)=-ir(\omega)/(1+r(\omega))=\gamma_d/(\omega_d-\omega)$ with $r(\omega)=-i\gamma_d/((\omega-\omega_d)+i\gamma_d)$ for the subwavelength array [cf.~Eq.~\eqref{reflectivity}] and choose a fixed $\zeta_0\gg1$ (corresponding to a close to unity reflectivity) for the right mirror. The classical transmission coefficient (see Appendix~\ref{D}) then reads
\begin{equation}\label{eq:TransTM}
    \tilde{t}(\omega)
    = \frac{1}{(1-i\zeta_0)[1-i\zeta_L(\omega)]e^{-i\theta}+\zeta_0\zeta_L(\omega)e^{i\theta}},
\end{equation}
where $\theta = \omega \ell /c$. For $\zeta_L(\omega)$ and $\zeta_0$ infinite, the expression above reaches unit absolute value at resonances $\omega_m=m\times\omega_\text{FSR}$ for any positive integer $m$. The resonances are separated by the free spectral range $\omega_\text{FSR}=c\pi/\ell$. Let us first fix a given resonance number $m$ such that $\omega_m$ lies in the neighborhood of the mirror resonance and see that for finite $\zeta_0$ and assuming $\zeta_L(\omega)$ is flat in frequency and equal to $\zeta_0$, the transmission can reach unity at a shifted $\omega'_m=\omega_m+\omega_\text{FSR}/(\pi \zeta_0)$. The linewidth of such a resonance is then $\kappa_\text{flat}=\omega_\text{FSR}/(2 \pi \zeta_0^2)$.\\
\indent When the two mirrors have unequal reflectivities, the transmission of the optical resonator is always less than unity. Therefore, for variable susceptibility $\zeta_L$ we ask now that a resonance $\bar{\omega}$ should be reached for which $|\zeta_L(\bar{\omega})|=\zeta_0$ such that the transmission is unity. There are two solutions for this equation and we pick the one where the mirror resonance sits to the left of the cavity resonance. The reason is that the reflectivity at $\omega_d$ is exactly unity which means the cavity transmission reaches zero. With the choice $\bar{\omega}=\omega_d+\gamma_d/\zeta_0$ the zero of the cavity sits on the left of the cavity resonance. We also fix $\bar{\omega}=\omega_m$, which in practice means that one first determines $\gamma_d, \zeta_0$ and then adjusts the cavity length.\\
\indent For varying $\gamma_d$, Fig.~\ref{fig52}b shows a scan of the cavity resonance. For $\gamma_d\gg\kappa_\text{flat}$, the expected Lorentzian response is obtained (gray dashed line) while for decreasing $\gamma_d$ at the level of $\kappa_\text{flat}$, a very narrow (around $\gamma_d/\zeta_0$) asymmetric Fano profile is obtained. The zero of the hybrid cavity transmission is at $\omega_d$ while unity is reached at $\omega_m$.\\
\indent Notice that the double hybrid cavity has an even narrower linewidth as illustrated in Fig.~\ref{fig52}c for the same regimes as provided for  Fig.~\ref{fig52}b. In addition, the transmission profile is symmetric and has the advantage of presenting two Fano resonances, situated symmetrically with respect to the cavity resonance.\\

\noindent \textbf{Extraction of parameters} - Comparison of the two expressions in Eq.~\eqref{eq:TransTM} and Eq.~\eqref{tcm} can allow for the identification of $\mathcal{G}$, $\omega_a$ and $\kappa_L$. In a first step, we ask that the zero of the transmission at the mirror resonance vanishes $t(\omega_d)=0$ corresponding to the point at which the mirror reflectivity is unity. This leads immediately to the identification $\mathcal{G}=\sqrt{\kappa_L\gamma_d}$. The derived coupling between the surface resonance and the cavity-confined mode is purely dissipative, in stark contrast with the situation treated in Ref.~\cite{cernotik2019cavity}, where an additional real photon exchange process between modes $a$ and $d$ occurs. In the next step, we ask that unit transmission $|t(\omega)|^2=1$ is reached at the resonance $\omega_m=\omega_d+\gamma_d/\zeta_0$. From Eq.~\eqref{tcm}, after a few calculations, we obtain a solution for the left mirror loss as
$\kappa_L=\kappa_R+\zeta_0 (\omega_a-\omega_m)+i|(\omega_a-\omega_m)-\zeta_0 \kappa_R|$. As the loss rate is defined in the real domain, we ask for the imaginary part to vanish which gives $\omega_a=\omega_m+\zeta_0 \kappa_R$. This also leads to $\kappa_L=\kappa_R (1+\zeta_0^2)$. Further fitting of the transmission profiles also identifies $\kappa_R=(\kappa_\text{flat}/2) \gamma_d \zeta_0/(\omega_\text{FSR}+\gamma_d \zeta_0)$.\\
\indent With the proper definition of the parameters appearing in Eqs.~\eqref{eq:EOM}, one can proceed in solving various problems involving hybrid cavities with quantum-emitter ensembles (eventually in the direction of Fano cavity lasing as in Ref.~\cite{mork2014photonic}) or with movable mirrors. In the direction of quantum optomechanics, Ref.~\cite{cernotik2019cavity} has shown that the Fano profile of photonic crystal mirror cavities can be constructively utilized to lead to cooling in the absence of heating. The mechanism is based on sideband resolution by fitting a Stokes, heating sideband inside the Fano resonance. For subwavelength emitter arrays the same can be realized by tailoring the relationship between the cavity length and the array resonance. In addition, the double-sided hybrid cavity provides a scenario in which both sidebands can be inhibited as it presents symmetrically placed Fano dips.\\

\subsection{Superradiant lasers}
\label{Sec5C}

\begin{figure}[t]
\center
\includegraphics[width=0.8\columnwidth]{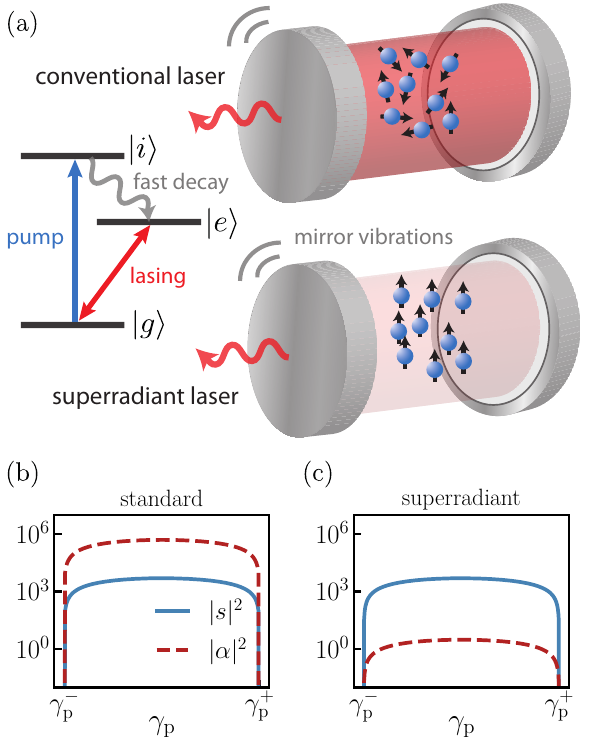}
\caption{(a) A gain medium is represented by an incoherently pumped (via intermediate state $\ket{i}$) three-level system with lasing occuring on the $\ket{e}$-$\ket{g}$-transition.   In a standard laser, stimulated emission produces a large intracavity field amplitude exhibiting coherence past threshold. In a superradiant laser, the intracavity field amplitude is instead kept small and coherence is stored in the collective atomic dipole thus protecting the laser linewidth from thermal effects. (b) Behavior of intracavity photon number $|\alpha|^2$ and coherence $|s|^2$ in the standard, good cavity regime ($\kappa=0.1 g$) and (c) in the superradiant, bad cavity regime ($\kappa=40g$). Other parameters are $\gamma=10^{-2}g$ and $\mathcal{N}=200$. Notice that the upper threshold $\gamma\ts{p}^+\approx \mathcal{N}g^2/\kappa$ is very different in both regimes.}
\label{fig53}
\end{figure}

The Tavis-Cummings Hamiltonian introduced in Sec.~\ref{Sec4A} together with the collective spin algebra from Sec.~\ref{Sec2D} are sufficient for a basic understanding of the main advantages presented by superradiant lasers (as treated in Refs.~\cite{bohnet2012asteadystate,bohnet2014linear,norcia2016superradiance}). First, we will assume a minimal model for a laser comprised of a cavity containing an incoherently pumped gain medium. Under certain conditions (implying the existence of a pumping threshold) a non-zero intracavity field amplitude can be obtained, which signifies a coherent-light output, i.e., lasing. The distinction between the good and bad cavity regimes (behavior quantified by the cavity losses) will then be seen to characterize the difference between standard and superradiant lasers. Finally, we will make the connection between superradiant lasing and superradiance as described by Eq.~\eqref{Dicke} in Sec.~\ref{Sec2D} by deriving an effective master equation for the gain medium after eliminating the cavity field.\\
\indent The minimal model assumes $\mathcal{N}$ identical emitters undergoing (independent) decay with the standard Lindblad term (as defined in Eq.~\eqref{Lstandard}) $\mathcal{L}_\gamma$ (each emitter has a decay rate $\gamma$ with corresponding collapse operator $\sigma_i$). Additionally, the incoherent pump assumes a Lindblad term $\mathcal{L}_{\gamma_\text{p}}$ (each emitter is pumped at a rate $\gamma_\text{p}$ with corresponding collapse operator $\sigma_i^\dagger$ which brings population to the excited state). The pump can for example be realized as illustrated in Fig.~\ref{fig53}a by excitation into an intermediate level $\ket{i}$ followed by fast decay to the lasing level $\ket{e}$. For $\gamma\ts{p}>\gamma$ population inversion can be produced, however without exciting the emitter dipoles $\braket{\sigma_j}=0$ as all processes are incoherent. To produce a non-zero average dipole moment, which would correspond to the generation of nonthermal light, the coupling to the optical resonator is crucial. From the Hamiltonian in Eq.~\eqref{HTC} (with $g_j=g$ and $\omega_0=\omega_c$) supplemented with the two Lindblad terms (and in terms of collective operators defined in Sec.~\ref{Sec2D}), we can derive the equations of motion:
\begin{subequations}
\label{lasing}
\begin{align}
\braket{\dot{a}}&=-\kappa\braket{a}-ig\braket{S},\\
\braket{\dot{S}}&= -(\gamma+\gamma_\text{p})\braket{S}+2ig\braket{a S_z},\\
\braket{\dot{S_z}}&= -2(\gamma+\gamma_\text{p})\braket{S_z}+ig\braket{a^\dagger S-a S^\dagger}+\mathcal{N}(\gamma_\text{p}-\gamma).
\end{align}
\end{subequations}
The above set of equations is generally not solvable. However, an expansion around averages $a=\alpha+\delta a$, $S=s+\delta S$ and $S_z=s_z+\delta S_z$ where all $\delta a, \delta S$ and $\delta S_z$ are zero-averaged quantum noise terms, can lead to a simplification. The observation (which one can eventually infer from numerical simulations) is that $\braket{\delta a \delta S}\ll \alpha s$ (and similar for all other two-operator correlations) in the limit that $\mathcal{N}$ is large. In such a case, a much simpler set of equations can be obtained for averages
\begin{subequations}
\label{lasing}
\begin{align}
\dot{\alpha}&=-\kappa\alpha-igs,\\
\dot{s}&= -(\gamma+\gamma_\text{p})s+2ig\alpha s_z,\\
\dot{s}_z&= -2(\gamma+\gamma_\text{p})s_z+ig(\alpha^* s-\alpha s^*)+\mathcal{N}(\gamma_\text{p}-\gamma).
\end{align}
\end{subequations}
Let us now assume that a steady state can be reached (by setting all derivatives to zero) in which case the solution of the above equations yields a population difference $s_z=1/(2\mathcal{C}_\text{p})$, with $\mathcal{C}_\text{p}=g^2/[\kappa(\gamma+\gamma_\text{p})]$, the cavity-field intensity (photon number)
\begin{align}
|\alpha|^2&=\frac{1}{\kappa}\left[\frac{\mathcal{N}(\gamma_\text{p}-\gamma)}{2}-\frac{\gamma+\gamma_\text{p}}{2\mathcal{C}_\text{p}}\right],
\end{align}
and the dipole coherence $|s|/|\alpha|=\kappa/g$. The expression above is positive only above a lasing threshold (signifying the existence of a lasing steady state),  which can be derived by finding the required values for the pumping strength $\gamma_\text{p}$. Requiring that in the lasing regime one has $|\alpha|^2>0$, yields a quadratic equation with upper and lower thresholds given by
\begin{align}
\gamma_\text{p}^{\pm}=\frac{g^2\mathcal{N}}{2\kappa}-\gamma\pm\sqrt{\frac{g^2\mathcal{N}}{\kappa}\left(\frac{g^2\mathcal{N}}{4\kappa}-2\gamma\right)}.
\end{align}
If  $\mathcal{N}g^2/\kappa$ is large compared to $\gamma$, the thresholds can be approximated as $\gamma_\text{p}^-\approx\gamma$, $\gamma_\text{p}^+\approx\mathcal{N}g^2/\kappa$. Note that the lower threshold corresponds to achieving population inversion while the upper threshold implies that a too large incoherent pumping rate $\gamma_\text{p}$ prevents the formation of coherence in the system.\\
\indent While the analysis above is valid in any regime, the great advantage brought on by superradiant lasers is that the coherence is stored in the gain medium which means that the laser linewidth is unaffected by common problems in optical resonators operating at large intracavity powers (such as thermal vibrations of the end mirrors). This can be seen in the ratio $|s|/|\alpha|=\kappa/g$ which indicates that for $\kappa\gg g$ the intracavity field can be negligible while the gain medium coherence is very large [see \fsref{fig53}b,c].\\
\indent Let us now provide an alternative understanding of the superradiant lasing problem, by connecting to Sec.~\ref{Sec2D} which we do by focusing on sketching the derivation of an effective master equation for the $\mathcal{N}$ emitters in the case of a lossy cavity. We assume at time $t=0$ the emitters to be fully inverted, while the cavity mode is in the $\ket{\text{vac}}$ vacuum state. We neglect spontaneous emission into free space as the cavity will provide a dominant channel of energy loss by imposing that its decay rate $\kappa$ is much larger than $g\sqrt{\mathcal{N}}$. The evolution of the system is then described by
\begin{equation}
\dot{ \rho}_\text{tot}(t) =i\left[ \rho_{\text{tot}}(t), \mathcal{H}\right] + \mathcal{L}_{\text{cav}}[ \rho_{\text{tot}}].
\end{equation}
We notice first that the collapse operator $a$ only couples to the total spin ladder operators $S$ and $S^\dagger$ which insures that the dynamics only takes place within the symmetric manifold with maximal spin length $\mathcal{N}/2$. The second observation is that, owing to the quick loss rate $\kappa$ which is much larger than the coherent light-matter exchange rate $g\sqrt{\mathcal{N}}$, the cavity field is permanently in a very low occupancy state. This allows us to apply the standard master equation procedure (see Appendix~\ref{A}) where we assume the crude factorization $ \rho_\text{tot}(t) =  \rho(t) \otimes \ket{\text{vac}}\bra{\text{vac}}$ at all times. The next step is in evaluating two-time correlation terms with the new time-dependent field operator $\mathcal{F}(t)=g a(t)$ (equivalent to the expression in Eq.~\eqref{Fspon}). The situation is now much simpler than in the free-space spontaneous emission case, as one can use the two-time correlations $\braket{a(t')a^\dagger(t'')}=e^{-\kappa(t'-t'')}$ to find an effective damping rate $g^2/\kappa$.
The dynamics can then be reduced to the atomic subspace and incorporated in a master equation
\begin{align}
\dot{\rho}(t)=i[\rho(t),\mathcal{H}_0]+\mathcal{L}_e^{_{\text{cav}}}[ \rho],
\end{align}
with $\mathcal{H}_0=\omega_0 \left[S_z+(\mathcal{N}/2)\mathbbm{1}\right]$ and the cavity-induced collective spontaneous emission
\begin{align}
\mathcal{L}_e^{_{\text{cav}}}[\rho]=\frac{g^2}{\kappa}\left(2S \rho S^\dagger-S^\dagger S \rho- \rho S^\dagger S\right).
\end{align}
The resulting dynamics is of course identical to the one described in Sec.~\ref{Sec2D} showing the emergence of quick bursts of light as superradiant pulses but with a rate established by the cavity $g^2/\kappa$. The effect is the spontaneous synchronization of dipoles by the cavity to give rise to the superradiant lasing regime. Here, the lossy cavity acts mainly as a communication bus to drive the synchronization between the atoms and the average photon number inside the cavity  is typically well below one. For example, Ref.~\cite{bohnet2012asteadystate} reports the realization of a superradiant laser with an average photon number of less than $0.2$ and a single-atom cooperativity $\mathcal{C}=7.7\cdot 10^{-3}$, while the lasing regime achieved in  Ref~\cite{norcia2016superradiance} uses a MHz linewidth optical clock transition as its gain medium and is envisioned to act as an active atomic clock insensitive to fluctuations in reference cavity length (a fundamental limitation in conventional lasers).

\subsection{Further remarks}
Composite systems made up of flat, standard mirrors and two-dimensional single emitter-thick regular arrays can find a multitude of applications in the direction of quantum technologies. When used in a standard cavity QED scenario, they can act as quick phase switchers with applications in precision spectroscopy of quantum network characterization~\cite{plankensteiner2017cavity} or in hybrid quantum optomechanical setups with enhanced photon-phonon couplings~\cite{dantan2014hybrid}. In the strong reflectivity limit, subradiant arrays have been interfaced with two-dimensional semiconductor monolayers to show enhanced quantum nonlinear optical properties~\cite{wild2018quantum}. The identification of dark collective states between the two mirrors has been proposed to allow for the preparation of Bell superpositions states between the two subradiant layers~\cite{guimond2019subradiant}, in a double-mirror hybrid setup as discussed above. As such subradiant mirrors are of extremely small mass, their zero-point motion is much larger than that of traditional dielectric mirrors used in standard optomechanics: this opens new opportunities for quantum optomechanics at the single photon-phonon level~\cite{shahmoon2019collective,shahmoon2020quantum}.\\

\section{Quantum optics with molecules}
\label{Sec6}
A relatively recent novel direction of research with great promise in the direction of quantum technologies is the engineering of \textit{cavity-dressed materials}. In the field of molecular polaritonics, embedding organic semiconductors in optical cavities allows for the design of novel materials with enhanced properties such as exciton and charge transport~\cite{orgiu2015conductivity,schachenmayer2015cavity,feist2015extraordinary,hagenmuller2017cavity,hagenmuller2018cavity}, superconductive behavior~\cite{sentef2018cavity, thomas2019exploring} or modified chemical reactivity \cite{hutchinson2012modifying,galego2016suppressing,herrera2016cavity,martinezmartinez2018can,kampschulte2018cavity} due to light-modified energy potential surfaces. This approach is also extended to study, control and design phase transitions in quantum materials by quantum light~\cite{martin2019manipulating,xiao2019cavity,rohn2020Ising,sentef2020quantum,yuto2020quantum} etc.\\
\indent Quantum emitters widely utilized in such experiments are of a much more complex nature than the two-level system approximation. We exemplify here how the previously introduced methods (master equation, quantum Langevin equations) and concepts can be extended to additionally include couplings between electronic transitions and vibrations or phonon modes. While this approach is in principle amenable to a wide range of solid-state emitters where phononic couplings play a crucial role, such as quantum dots or vacancy centers in diamond, we focus here on the specific case of molecules. In particular, organic molecules have emerged as tunable and efficient light-matter interfaces due to their relatively large dipole moments, wide range of transition frequencies and narrow linewidths at cryogenic temperatures~\cite{toninelli2021single}. As such, single-molecule impurities embedded in solid-state host matrices hold promise as single-photon sources~\cite{basche1992photon, demartini1996single, lounis2000single, pazzagli2018selfassembled} or nonlinear optical elements~\cite{maser2016few}. There also is interest in exploiting the strong inherent coupling between electronic transitions and the molecular vibrational degrees for the realization of quantum optomechanical effects at the single molecule level~\cite{roelli2016molecular, neuman2019quantum, benz2016single}.\\
\indent In this section, we will mainly follow the approach introduced in Ref.~\cite{reitz2019langevin} by considering the time evolution of a \textit{polaron} operator (i.e.~an ``effective'', vibrationally-dressed electronic dipole operator). We proceed with a first-principle derivation of the Holstein Hamiltonian for electron-vibron interactions and derive the standard Franck-Condon physics and absorption/emission properties of molecules. We then illustrate how one can circumvent a major downfall of molecular systems, i.e., they do not have closed transitions, by employing the Purcell effect: a cavity-dressed molecule can then behave as an ideal quantum emitter. Next we analyze molecular dimers as good candidates for the observation and exploration of controllable near-field couplings. Finally we introduce and analyze cooperative processes occurring between near-field coupled disordered molecules leading to what is known as the F\"{o}rster resonance energy transfer.\\

\subsection{Optical response of a single molecule}
\label{Sec6A}

\begin{figure}[t]
\center
\includegraphics[width=0.60\columnwidth]{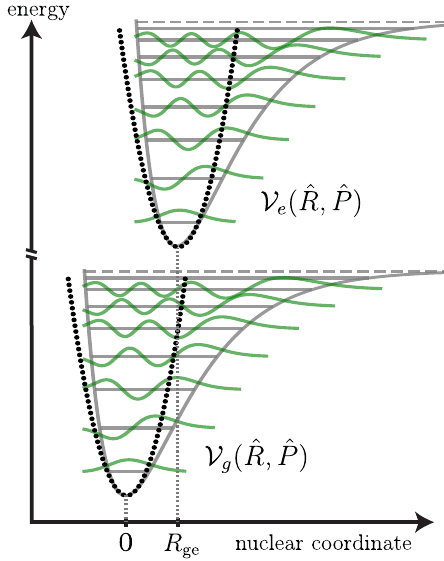}
\caption{Displaced oscillator model used to derive the Holstein Hamiltonian. The Morse potential energy landscapes (solid gray) of electronic ground and excited state are approximated as harmonic oscillators (dashed black) and their eigenstates are illustrated in green. The position mismatch between ground and excited state potential energy surfaces $R_{\text{ge}}$ gives rise to vibronic coupling.}
\label{fig61}
\end{figure}
\begin{figure*}[t]
\center
\includegraphics[width=1.8\columnwidth]{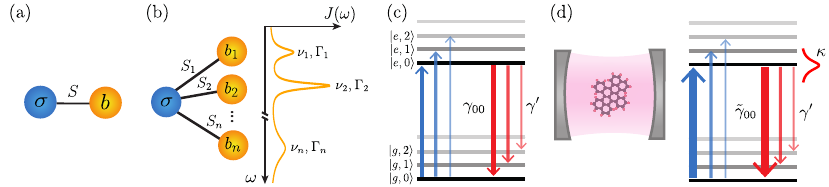}
\caption{(a) The dynamics of a molecule with a single nuclear coordinate can be seen as the coupling of an electronic transition operator $\sigma$ to a single vibrational mode $b$ with a strength given by the Huang-Rhys factor $S$. (b) For many vibrational modes, a molecule can be constructed by coupling an electronic transition operator $\sigma$ to $n$ vibrational modes with strengths $S_k$, frequencies $\nu_k$ and linewidths $\Gamma_k$ described by the (Lorentzian) spectral density function $J(\omega)$. (c) Energy eigenstates of the electron-vibron system denoted by $\ket{g,n}$ and $\ket{e,n}$. The zero phonon line is the $\ket{g,0}$ to $\ket{e,0}$ transition. For small $S<1$ the width of the arrows indicate the strength of the optical transitions. For a molecule with many vibrational levels, spontaneous emission outside the zero phonon line at summed rate $\gamma'$ can be larger than into it at rate $\gamma_{00}$. (d) The enhancement of spontaneous emission via resonant coupling to an optical cavity can lead to a great modification of the strength of the zero phonon line thus turning a molecule into a closed quantum system.}
\label{fig62}
\end{figure*}

To include the electron-vibration (\textit{vibronic}) coupling into our formalism, we start with a first-principle derivation for a single electron coupled to a single nuclear coordinate of reduced mass $\mu$. Notice that, for the simplest case of a homonuclear diatomic molecule made of nuclei each with mass $m$, a single vibrational mode exists corresponding to relative motion with effective mass $\mu = m/2$; the equilibria coordinates then correspond to the bond length in the ground and excited states. Under the Born-Oppenheimer approximation, the ground and excited state potentials are obtained by solving the Schr\"{o}dinger equation for each given fixed internuclear distance. We assume that, along the nuclear coordinate, the equilibrium positions for ground (coordinate $R_{\text{g}}$, state vector  $\ket{g}$) and excited (coordinate $R_{\text{e}}$, state vector $\ket{e}$) electronic orbitals are slightly shifted with respect to each other (as illustrated in Fig.~\ref{fig61}). We can then write the total Hamiltonian $\mathcal{H}_{\text{mol}}=\mathcal{H}_{\text{el}}\otimes\mathcal{H}_{\text{vib}}$ as
\begin{equation}
\label{molham}
\mathcal{H}_{\text{mol}}=\mathcal{V}_e(\hat{R},\hat{P})\otimes\sigma^\dagger\sigma + \mathcal{V}_g(\hat{R},\hat{P})\otimes\sigma\sigma^\dagger. \nonumber
\end{equation}
where the potential surfaces are described by the quantized position and momentum operator satisfying $[\hat{R},\hat{P}]=i$. The dynamics takes place now in an infinite-dimensional Hilbert space: in the electronic subspace the algebra is that of a spin $1/2$ and is described by projectors $\sigma^\dagger \sigma$ (into the excited state) and $\sigma \sigma^\dagger$ (into the ground state) while in the motional subspace one can for example describe the dynamics in terms of plane waves. However, a great simplification can be obtained by performing a harmonic expansion of the potential surfaces around the minima
\begin{subequations}
\begin{align}
\mathcal{V}_e(\hat{R},\hat{P})&=\omega_0+\frac{\hat{P}^2}{2\mu}+\frac{1}{2}\mu\nu^2\left(\hat{R}-R_e\right)^2,\\
\mathcal{V}_g(\hat{R},\hat{P})&=\frac{\hat{P}^2}{2\mu}+\frac{1}{2}\mu\nu^2\left(\hat{R}-R_g\right)^2.
\end{align}
\end{subequations}
Finally, we can introduce small oscillations around the equilibria $\hat{Q}=\hat{R}-R_{\text{g}}$ and subsequently $\hat{R}-R_{\text{e}}=\hat{Q}+R_{\text{g}}-R_{\text{e}}=:\hat{Q}-R_{\text{ge}}$ leading to
\begin{align}
\mathcal{H}_{\text{mol}}&=\left(\omega_0+\frac{1}{2}\mu\nu^2 R_{\text{ge}}^2 \right)\mathds{1}_{\text{vib}}\otimes \sigma^\dagger\sigma\\
&+\left(\frac{\hat{P}^2}{2\mu}+\frac{1}{2}\mu\nu^2 \hat{Q}^2\right)\otimes \mathds{1}_{\text{el}} -\mu\nu^2 R_{\text{ge}} \hat{Q}\otimes\sigma^\dagger\sigma.
\end{align}
The last term is a renormalization of the bare electronic transition frequency energy which will naturally go away when diagonalizing the Hamiltonian via the polaron transformation resulting in $\omega_0$ as the natural electronic transition frequency. We can now rewrite the momentum and position operators in terms of bosonic operators $\hat{Q}=r_{\text{zpm}}(b^\dagger+b)$, $\hat{P}=ip_{\text{zpm}}(b^\dagger-b)$ by introducing the zero-point motion $r_{\text{zpm}}=1/\sqrt{2\mu\nu}$ and $p_{\text{zpm}}=\sqrt{\mu\nu/2}$. Reexpressing the terms above via the factor $\sqrt{S}=\mu\nu R_{\text{ge}}r_{\text{zpm}}$ ($S$ is the Huang-Rhys factor) yields the Holstein-Hamiltonian \cite{holstein1959study}
\begin{equation}
\mathcal{H}_{\text{mol}}=(\omega_0+S\nu)\sd\s+\nu \bd b-\sqrt{S}\nu(\bd+b)\sd \s,
\label{holstein}
\end{equation}
A molecular box illustration is shown in Fig.~\ref{fig62}a as a minimal model for a molecule.\\
\indent One can bring this Hamiltonian into diagonal form  via the polaron transformation $U_{\text{pol}}^{\dagger}=(\Dd)^{\sd\s}=\ket{g}\bra{g}+\Dd\ket{e}\bra{e}$ where the displacement is defined as $\D=\exp(-i\sqrt{2S} p)=\exp[\sqrt{S} (\bd-b)]$ with the dimensionless momentum quadrature $p=i(\bd-b)/\sqrt{2}$ as generator. The position quadrature is defined analogously as $q=(\bd+b)/\sqrt{2}$. The displacement operator creates a coherent state with amplitude $\sqrt{S}$ when applied to the vibrational ground state $\D\ket{0}_{\text{vib}}=\ket{\sqrt{S}}_\text{vib}$. Note that the vibrational creation and annihilation operators transform in the polaron picture as
\begin{subequations}
\begin{align}
U_{\text{pol}}^\dagger b U_{\text{pol}} &= b+\sqrt{S}\sigma^\dagger\sigma,\\
U_{\text{pol}}^\dagger b^\dagger U_{\text{pol}} &= b^\dagger +\sqrt{S}\sigma^\dagger\sigma,
\end{align}
\end{subequations}
and furthermore the following commutation relations are fulfilled $[p,\Dd]=0$, $[q,\Dd]=-\sqrt{2S}\Dd$. The Hamiltonian in the diagonal basis then becomes $\tilde{\mathcal{H}}_{\text{mol}}=\omega_0\sigma^\dagger\sigma+\nu b^\dagger b$ and has eigenvectors $\{\ket{g,n},\ket{e,n}\}$ while the eigenvectors in the original basis can be obtained by an inverse  polaron transform and are given by $\{\ket{g, n }$, $\D\ket{e,n}\}$.\\

\noindent \textbf{Franck-Condon physics} - Let us now ask what is the optical response of the molecule to a weak laser excitation modelled via $\mathcal{H}_\ell=i\eta(\sigma^\dagger e^{-i\omega_\ell t}-\sigma e^{i\omega_\ell t})$. In an interaction picture, reached after performing a transformation of the Hamiltonian with $e^{i\omega_\ell\sigma^\dagger\sigma t}$, one has
\begin{align}
\label{holsteinpolaron}
\tilde{\mathcal{H}} = (\omega_0-\omega_\ell)\sigma^\dagger\sigma+\nu b^\dagger b+ i\eta\left(\sigma^\dagger\Dd-\s\D\right).
\end{align}
Starting with the molecule in the absolute ground state $\ket{g,0}$, the result of the drive is to excite the electron to state $\ket{e}$ while exciting motion on the nuclear coordinate to a coherent motional state. One can immediately compute the probability of absorption
\begin{align}
P_{\text{abs}}=|\braket{e,n|\tilde{\mathcal{H}}|g,0}|^2=\braket{e,n|\sigma^\dagger\Dd|g,0}|^2= f^n_S,
\end{align}
showing a Poissonian distribution $f^n_S= e^{-S}S^{n}/n!$ of the occupancies of number states $\ket{n}$ with coefficients known as Franck-Condon factors. In particular, the so-called zero-phonon line (transition corresponding to $n=0$) is reduced by a factor $e^{-S}$ at zero temperature. The Huang-Rhys factor $S$ should be interpreted as the average number of vibrational quanta created upon excitation or deexciation of the molecule. The probability of the emission process is similarly computed between $\ket{e,0}$ and final states $\ket{g,n}$ and yields the same Poissonian distribution in $n$. This is a known result in molecular spectroscopy where the absorption and emission spectra are mirror images of each other. For small $S<1$ the physics of absorption and emission in a molecule is illustrated in Fig.~\ref{fig62}c where the width of the arrows indicate the strength of the optical transitions between eigenstates of the free Hamiltonian (without the Holstein interaction).  \\
\indent The Holstein Hamiltonian in Eq.~\eqref{holstein} can be generalized to a more complex molecular box (is illustrated in Fig.~\ref{fig62}b) involving $n_v$ independent nuclear coordinates with frequencies $\nu_k$ and coupling strengths $\sqrt{S_k}$. The total Hamiltonian is then expressed as
\begin{align}
\mathcal{H}_{\text{mol}}=\tilde{\omega}_0 \sigma^\dagger\sigma+\sum_{k=1}^{n_v} \nu_k b_k^\dagger b_k -\sum_{k=1}^{n_v}\sqrt{S}_k\nu_k(b_k^\dagger+b_k)\sigma^\dagger\sigma,
\end{align}
where the accumulated frequency shift is $\tilde{\omega}_0=\omega_0+\sum_{k=1}^{n_v} S_k\nu_k$.
Diagonalization of this Hamiltonian is straightforward as it involves a collective displacement for all nuclear coordinates $U_{\text{pol}}^\dagger = (\prod_k \mathcal{D}_k^\dagger )^{\sd\s}$ with $\mathcal{D}_k=\exp(-i\sqrt{2S_k} p_k)$. The expectation value of the displacement operator $\braket{\D_k}=\braket{\D_k^\dagger}=e^{-S_k\braket{p_k^2}}$ (assuming the vibrational mode to be in a thermal state) can be expressed as $e^{-S_k(\bar{n}_k+1/2)}$ (expanding the displacement operator and making use of the fact that Gaussian states are characterized by their second-order moments and odd moments are vanishing)
where the average thermal occupancy is given by $2\bar{n}_k=\coth(\beta\nu_k/2)-1$ with $\beta=1/(k_B T)$. \\

\noindent \textbf{Branching ratio manipulation} - The branching ratio quantifies the rate of emission into the zero-phonon line versus all other lines. For ideal quantum emitters this is automatically unity in the absence of coupling to any other channels of de-excitation. For an isolated molecule this is instead given by $\alpha=\braket{\D}^2=\prod_k \braket{\D_k}^2=e^{-\sum_k S_k}$ (at zero temperature) stemming from loss of excitation into other vibrational levels. Complex molecules have many vibrations: even for small Huang-Rhys factors the sum in the exponential gives a considerable reduction of the oscillator strength of the zero-phonon line. The Purcell effect (introduced in Sec.~\ref{Sec4A}), stemming from the resonant interaction of the zero-phonon line transition with a lossy confined optical resonance has been experimentally proven to turn molecular emitters into almost perfect two-level quantum emitters~\cite{wang2019turning}. Assuming that the cavity linewidth is narrow compared to the vibrational frequency $\kappa\ll\nu$ and only couples to the zero-phonon line transition, we can define by $g_{00}=g\braket{\D}$ the effective reduced coupling between the cavity and the zero-phonon line with a corresponding cooperativity $\mathcal{C}_{00}=g_{00}^2/(\kappa\gamma)$. This will lead to an enlarged zero-phonon transition linewidth $\tilde{\gamma}_{00}=\gamma_{00}(1+\mathcal{C}_{00})$ as illustrated in Fig.~\ref{fig62} while other transitions are unperturbed and sum up to $\gamma'=\gamma-\gamma_{00}$. The cavity-modified branching ratio can then be expressed as
\begin{align}
\alpha_\text{cav}=\frac{\tilde{\gamma}_{00}}{\tilde{\gamma}_{00}+\gamma'}=\frac{(1+\mathcal{C}_{00})e^{-\sum_k S_k}}{1+\mathcal{C}_{00}e^{-\sum_k S_k}},
\end{align}
which reproduces the bare molecule result for $\mathcal{C}_{00}=0$ and goes to unity in the limit that $\mathcal{C}_{00}\to\infty$. Experimentally, values of $\mathcal{C}_{00}$ close to $100$ have been reached~\cite{wang2019turning}, showing that almost unity branching ratios can be achieved.\\

\noindent \textbf{Quantum Langevin equations for polarons} - We will now proceed by deriving a quantum Langevin equation for a polaron operator $\tilde{\sigma}=\sigma\Dd$ (i.e., a vibrationally dressed electronic dipole operator). This will allow us to include all coherent and incoherent effects at the level of the equations of motion. The collective displacement operator $\D^\dagger$ can refer to a large set of molecular vibrational modes $\Dd=\prod_{k=1}^{n_v}\D_k^\dagger$. The method can also be extended to include coupling to an (eventually infinite) set of phonon modes of the host environment~\cite{reitz2020molecule,clear2020phonon}. We will assume a Brownian noise dissipation model (see Appendix~\ref{E}) for the molecular vibrational modes. Starting with the Holstein Hamiltonian one can derive the following equations of motion for vibrations
\begin{subequations}
\begin{align}
\dot{q}_k &= \nu_k p_k,\\
\dot{p}_k &= -2\Gamma_k p_k - \nu_k q_k - \sqrt{2S_k}\nu_k\sigma^\dagger\sigma + \xi_k.
\end{align}
\end{subequations}
In addition to the environment induced damping at rates $2\Gamma_k$ and associated thermal input noise $\xi_k$ (see Appendix~\ref{E} for noise properties), vibrations are driven by their coupling to the electronic degree of freedom. For the electronic transition, the polaron operator can be computed from  $\dot{\tilde{\sigma}}=\dot{\s}\mathcal{D}^\dagger+\s\dot{\mathcal{D}}^\dagger$ where the full Langevin equation for $\sigma$ is derived from the Holstein Hamiltonian and the Linbdlad form radiative emission with the transformation indicated by the relation in Eq.~\eqref{diagonal_HLE}. We obtain then
\begin{align}
\dot{\tilde{\s}}=&-[\gamma-i(\omega_\ell-\omega_0)]\tilde{\sigma}-2i\sum_k \sqrt{2 S_k}\Gamma_k p_k\tilde{\s}\\\nonumber
&+\eta\Dd+\sqrt{2\gamma}\tilde{\sigma}_{\text{in}}+i\sum_k\sqrt{2S_k}\xi_k\tilde{\sigma}.
\end{align}
The equation is derived under the assumption of weak driving $\eta\ll\gamma$ and therefore small occupancy of the electronic excited state $\braket{\sigma^\dagger\sigma}\ll1$ and with special attention devoted to the rules for taking the time derivative of any exponential operator $e^{A(t)}$ with $[\dot{A}(t), A(t)]\neq 0$.
Moreover, some of the terms cancel (see Ref.~\cite{reitz2019langevin} for full details) and an extremely simple equation of motion for the polaron operator is obtained
\begin{align}
\dot{\tilde{\sigma}}=-[\gamma-i(\omega_\ell-\omega_0)]\tilde{\sigma}+\sqrt{2\gamma}\tilde{\Sigma}_{\text{in}}.
\end{align}
Here we have combined the input fields affecting the electronic transition into the expression ${\Sigma}_{\text{in}}=\eta/\sqrt{2\gamma}+\sigma_{\text{in}}$ and the displaced input field (drive plus noise) is $\tilde{\Sigma}_{\text{in}}={\Sigma}_{\text{in}}\Dd$.
A solution of the bare electronic dipole operators can be obtained by a formal integration as a sum $\s(t)=\s_\text{tr}(t)+\s_\text{ss}(t)$ between transient and steady state solutions
\begin{subequations}
\begin{align}
\s_\text{tr}(t) &=\s(0)\D(t)\Dd(0)e^{-\left[\gamma-i(\omega_\ell-\omega_0)\right]t},\\
\s_\text{ss}(t)&= \sqrt{2\gamma}\int_{0}^t dt' e^{-\left[\gamma-i(\omega_\ell-\omega_0)\right](t-t')}\Sigma_\text{{in}}(t')\D(t)\Dd(t').
\end{align}
\label{pauli}
\end{subequations}
To derive the optical response of the electronic degree of freedom in the presence of its coupling to vibrations one then has to evaluate two-time correlation functions of the displacement operator. This is factorizable as we consider that all vibrational modes are independent from each other, i.e., $\braket{\mathcal{D}(t)\Dd(t')}=\prod_k\braket{\mathcal{D}_k(t)\Dd_k(t')}$. By expanding the displacement operators and applying the Isserlis' theorem (or Wick's probability theorem for multivariate normal distributions) one obtains
\begin{align}
\label{displacementcorr}
\braket{\mathcal{D}_k(t)\D_k^\dagger(t')}=e^{-2S_k(\braket{p_k^2}-\braket{p_k(t)p_k(t')})},
\end{align}
where we made use of the fact that the momentum variance is stationary, i.e., $\braket{p_k^2}=\braket{p_k(t)^2}=1/2+\bar{n}_k$. This can be easily computed by using the properties for the Brownian noise model (see Appendix~\ref{E})
\begin{align}
\braket{p_k (t) p_k (t')}=\left[\left(\bar{n}_k+\frac{1}{2}\right)\cos(\nu_k \tau)-\frac{i}{2}\sin(\nu_k \tau)\right]e^{-\Gamma_k\tau},
\end{align}
with $\tau=t-t'$.
While expressions can be derived for any temperature of the bath, in the following we will, for the sake of simplicity, focus on the zero temperature case for all vibrational modes. This is typically a very good approximation as molecular vibrations typically lie in the range of THz frequencies and are therefore barely occupied as the condition $\hbar\nu_k\ll k_B T$ is typically fulfilled. The displacement correlation function in this case simplifies to
\begin{align}
\label{zerotemp}
\braket{\mathcal{D}_k(t)\D_k^\dagger(t')}=e^{-S_k}e^{S_k\exp\left[-(\Gamma_k+i\nu_k)\tau\right]},
\end{align}
which we will make use of in the next subsection to derive analytical expressions for absorption and emission lines and linewidths.\\

\noindent \textbf{Absorption and emission} - The absorption and emission spectra of a driven molecule can be readily obtained from Eq.~\eqref{pauli}. Considering large times $t\gg 1/\gamma$, the first term in Eq.~\eqref{pauli} goes towards zero and the average dipole moment can be expressed as (taking the average over both electronic and vibrational degrees of freedom)
\begin{align}
\braket{\sigma (t)}=\eta\int_0^t dt' e^{-(\gamma-i\Delta)(t-t')}\braket{\D (t)\Dd (t')},
\end{align}
with the laser detuning $\Delta=\omega_\ell-\omega_0$. The population of the electronic excited state evolves according to $\partial_t\braket{{\sigma^\dagger\sigma}}=-2\gamma\braket{\sigma^\dagger\sigma}+2\eta\Re\braket{\sigma (t)}$ meaning that the steady state absorption profile can be directly computed from the expectation value of the coherence $\braket{\sigma(t)}$. We define $S_\text{abs}(\Delta)=\braket{\sigma^\dagger\sigma}$ as the absorption profile for a varying laser frequency with expression
\begin{align}
S_\text{abs}(\Delta_\ell)=2\eta\int_{-\infty}^t dt' e^{-2\gamma (t-t')}\Re\braket{\sigma (t')}.
\end{align}
To evaluate this integral, we expand the displacement correlation function from Eq.~(\ref{zerotemp}) in a power series
\begin{align}
\braket{\D_k (t) \D_k^\dagger (t')}= e^{-S_k}\sum_{n_k=0}^\infty\frac{S_k^{n_k}}{n_k !}e^{-n_k(\Gamma_k+i\nu_k)(t-t')}.
\end{align}
The absorption spectrum can then be readily expressed as a sum over all indices $\{n_k\}=n_1,\hdots,n_{n_v}$ weighted by the Franck-Condon factors $f_{n_k}^{S_k}$
\begin{align}
S_\text{abs}(\Delta)=\frac{\eta^2}{\gamma}\sum_{\{n_k\}=0}^\infty\frac{f^{n_1}_{S_1}\hdots f^{n_{n_v}}_{S_{n_v}}\left[\gamma+\Gamma_{\{n_k\}}\right]}{\left[\gamma+\Gamma_{\{n_k\}}\right]^2+\left[\Delta-\nu_{\{n_k\}}\right]^2},
\end{align}
where the denominator indicates a series of blue-shifted absorption sidebands with increased linewidths $\Gamma_{\{n_k\}}=\sum_{\{n_k\}}n_k\Gamma_k$ at positions $\omega_\ell =\omega+\sum_k\nu_{\{n_k\}}$ with $\nu_{\{n_k\}}=\sum_{\{n_k\}}n_k\nu_k$.\\
\indent For the definition of the emission spectrum one can consider the transient dynamics of a molecule which is partially excited at time $t=0$: $\braket{\sigma^\dagger\sigma (0)}=p_0$. The emission spectrum is then defined as the Fourier transform $S_\text{em}(\omega)=2\Re\int_0^\infty d\tau \braket{\mathcal{T}\left\{\sigma^\dagger (\tau)\sigma(0)\right\}}e^{-i\omega\tau}$ (where $\mathcal{T}$ denotes time ordering) and can be computed as (normalized by $p_0$)
\begin{align}
S_\text{em}(\omega)=\sum_{\{n_k\}=0}^\infty\frac{2f^{n_1}_{S_1}\hdots f^{n_{n_v}}_{S_{n_v}}\left[\gamma+\Gamma_{\{n_k\}}\right]}{\left[\gamma+\Gamma_{\{n_k\}}\right]^2+\left[\omega-\omega_0+\nu_{\{n_k\}}\right]^2}.
\end{align}
The denominator indicates a series of red-shifted Stokes lines at frequencies $\omega=\omega_0-\nu_{\{n_k\}}$.
\subsection{Near-field coupled molecules}
\label{Sec6B}

Let us now consider two electronically coupled molecules. We start by discussing the vibronic dimer, a well-known model to describe the interplay between electronic and vibrational interactions in molecular aggregates which has been studied both theoretically~\cite{wittkowski1960electronic, riley1990thesymmetric, eisfeld2005vibronic} and experimentally~\cite{lippitz2004coherent, diehl2014emergence}. As opposed to the pure electronic dimer discussed in Sec.~\ref{Sec2B}, this model gives rise to a more complex energy landscape with states possessing both electronic and vibrational character.  Effects showing quantum coherence in electronically excited dimers have been experimentally investigated in a combination of single molecule spectroscopy and quantum chemistry techniques~\cite{diehl2014emergence}. We then move on to describe the dipole-dipole mediated unidirectional energy transfer process (FRET) between two molecules.\\

\noindent \textbf{Vibronic dimer model} - The vibronic dimer model consists of two molecules (indexed here by $1$ and $2$), each with a single electronic ground and excited state and each coupled to a single harmonic nuclear coordinate $q_{1/2}$ with Huang-Rhys factor $S$. The molecules are coupled to each other electronically (e.g.~via dipole-dipole interaction) with strength $\Omega$. Assuming that the excited state energies of the two molecules are identical and using the single excitation manifold with states $\ket{e}_1\otimes \ket{g}_2$ and $\ket{g}_1\otimes \ket{e}_2$, the Hamiltonian of the dimer can be represented by
\begin{figure}[t]
\includegraphics[width=0.62\columnwidth]{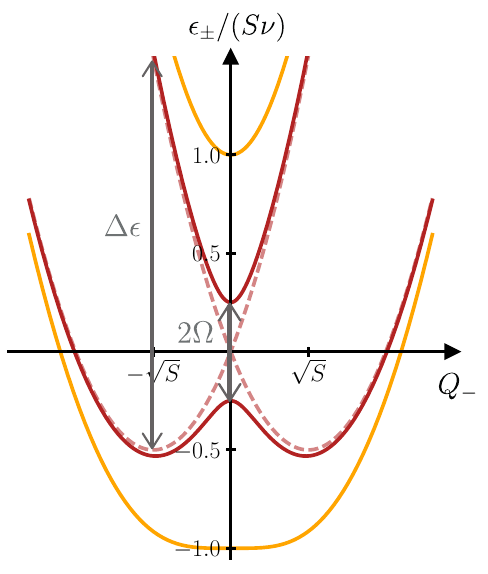}
\caption{Potential energy surfaces $\epsilon_\pm$ of the excited state manifold of the vibronic dimer model as a function of the (dimensionless) relative motion coordinate $Q_-$ for $\Omega=0$ (red dashed), $2\Omega=\Delta\epsilon/4$ (red solid) and $2\Omega=\Delta\epsilon$ (orange).}
\label{fig63}
\end{figure}
\begin{align}
\label{hamdimer}
\mathcal{H}_\text{dim}\!=\!\begin{pmatrix}
- \sqrt{2S}\nu q_1 & \Omega\\
\Omega & -\sqrt{2S}\nu q_2
\end{pmatrix}+\frac{\nu}{2}\sum_{i=1,2}\left(p_i^2+q_i^2\right)\mathbb{1},
\end{align}
where $\mathbb{1}$ is the $2\times 2$ unity matrix. Introducing symmetrized coordinates $Q_{\pm}=(q_1\pm q_2)/\sqrt{2}$ and electronic operators $\sigma_{\pm}=(\sigma_1\pm \sigma_2)/\sqrt{2}$, the Hamiltonian transforms to
\begin{align}
\label{dimertrans}
\tilde{\mathcal{H}}_\text{dim}&=\begin{pmatrix}
\Omega & -\sqrt{S}\nu Q_-\\
-\sqrt{S}\nu Q_- & -\Omega
\end{pmatrix}-\sqrt{S}\nu Q_+\mathbb{1}\\\nonumber
&+\frac{\nu}{2}\left(P_+^2+P_-^2+Q_+^2+Q_-^2\right)\mathbb{1},
\end{align}
from which one can see that only the coordinate $Q_-$ couples between electronic states in the collective basis. The coordinate $Q_+$ instead leads to a constant shift and can eventually be eliminated by a proper choice of the origin. The Hamiltonian in Eq.~(\ref{dimertrans}) can be easily diagonalized leading to the eigenvalues for the excited state potential energy surfaces
\begin{align}
\epsilon_{\pm}=\frac{\nu}{2}\left(Q_+^2+Q_-^2\right)-\sqrt{S}\nu Q_+\pm\sqrt{S\nu^2 Q_-^2+\Omega^2}.
\end{align}
The potential energy surfaces are plotted in Fig.~\ref{fig63} for various coupling strengths $\Omega$. One can see that the coupling lifts the degeneracy of the potential surfaces at $Q_-=0$ and leads to a splitting of $2\Omega$. Depending on $\Omega$, the lower energy surface either has two minima or just a single minimum.  An important quantity is the reorganization energy $\Delta\epsilon=2S\nu$ defined as the vertical energy separation between the potential curves at each minimum for zero coupling $\Omega=0$ [as illustrated in Fig.~\ref{fig63}]. One finds that a double minimum of the lower energy surface located at $Q_-^{\pm}=\pm\sqrt{S[1-\Omega^2/(S^2\nu^2)]}$ exists when the level splitting $2\Omega$ is smaller than the reorganisation energy $|2\Omega|<\Delta\epsilon$.\\

\begin{figure}[t]
\includegraphics[width=0.99\columnwidth]{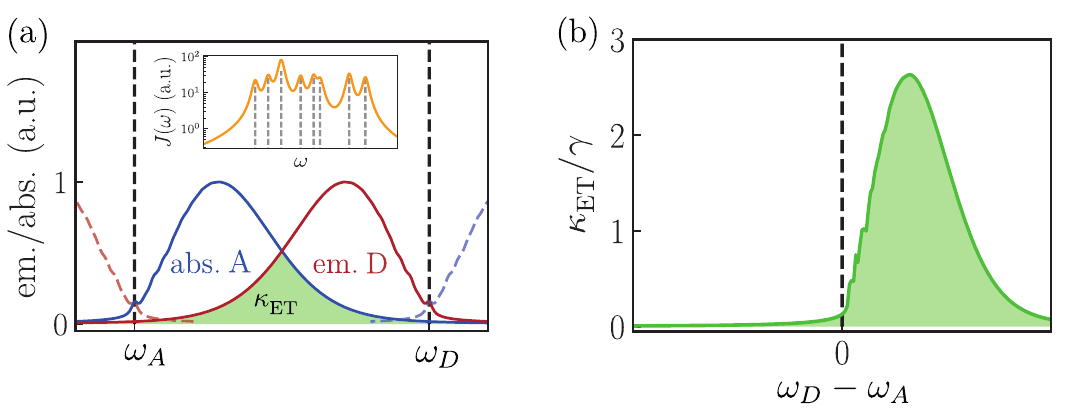}
\caption{Energy transfer in donor-acceptor configurations. (a) Overlap between acceptor emission spectrum and donor absorption spectrum leads to the FRET process. The results assume a given vibrational spectral energy $J(\omega)$ (assumed here to be identical for both molecules) shown in the inset, for $8$ vibrational modes considered. (b) Numerical analysis of the energy transfer rate (for $\Omega=30\gamma$) shows a unidirectional flow from the higher energy molecule (donor) to the lower energy one (acceptor) optimized around a detuning corresponding to the overlap of the donor electronic state to a high density of vibrational states in the acceptor excited state manifold. }
\label{fig64}
\end{figure}

\noindent \textbf{F\"{o}rster resonance energy transfer} - F{\"o}rster resonance energy transfer (FRET) is the main excitation transfer mechanism in photosynthetic light harvesting complexes, where energy is transfered with high efficiency from a light-harvesting antenna through a network of molecules (chromophores) to a reaction centre~\cite{foerster1948zwischen, blankenship2014molecular}. Although  a lot of progress has been made in understanding and unravelling the reasons for the high efficiency of the photosynthetic energy transfer occuring in nature over the last years, there is still ongoing debate e.g.~about the role of quantum coherence or entanglement in the energy transfer process~\cite{engel2007evidence, streltsov2017colloquium}. In the sense of this tutorial, FRET can be seen as a truly \textit{cooperative} process since it involves a complex interplay between electronic and nuclear degrees of freedom as well as both coherent and incoherent couplings. Generally speaking, the requirement for FRET is a spectral overlap between the emission profile of the donor molecule and the absorption profile of the acceptor, giving rise to a resonant transfer process typically accompanied by quick vibrational relaxation \cite{may2011charge}.\\
\indent Here we exemplify the application of the QLEs formalism to derive a perturbative expression for the FRET rate between two near-field coupled molecules $D$ and $A$ with energy mismatch $\Delta=\omega_D-\omega_A$, first by considering a simple configuration where each molecule has a single vibrational mode $b_D$ and $b_{A}$ (with corresponding vibrational frequencies $\nu_D$ and $\nu_A$ and Huang-Rhys factors $S_D$ and $S_A$). We will assume an initially excited donor molecule which can undergo a  resonant exchange with the acceptor's excited state vibrational manifold. In a standard scenario, a unidirectional process emerges as the acceptor quickly relaxes to its vibrational ground state $\Gamma_D\gg \Omega$. We start with the  equations of motion for the polaron-transformed dipole operators $\tilde{\sigma}_D=\mathcal{D}_D^\dagger\sigma_D$ and $\tilde{\sigma}_{A}=\mathcal{D}_{A}^\dagger\sigma_{A}$ with the displacement operators $\mathcal{D}_D=e^{\sqrt{S_D} (b_D^\dagger-b_D)}$ and $\mathcal{D}_A=e^{\sqrt{S_A} (b_A^\dagger-b_A)}$ which shows coherent coupling induced by the dipole-dipole near-field exchange of virtual photons
\begin{subequations}
\label{eom}
\begin{align}
\dot{\tilde{\sigma}}_D=-\left(\gamma_D+i\omega_D\right)\tilde{\sigma}_D-i\Omega\tilde{\sigma}_{A}\mathcal{D}_{A}\mathcal{D}_D^\dagger+\sqrt{2\gamma}\tilde{\sigma}_D^{\text{in}},\\
\dot{\tilde{\sigma}}_{A}=-\left(\gamma_A+i\omega_A\right) \tilde{\sigma}_{A}-i\Omega\tilde{\sigma}_D\mathcal{D}_D\mathcal{D}_{A}^\dagger+\sqrt{2\gamma}\tilde{\sigma}_{A}^{\text{in}}.
\end{align}
\end{subequations}
To derive a perturbative expression for the energy transfer rate, we consider a scenario with an initial full excitation of $D$ and no excitation of $A$. Then from above one can derive an equation of motion for the acceptor's population
\begin{align}
\dot{P}_{A}=-2\gamma_A P_{A}+2\Omega\Im\braket{\sigma_{A}^\dagger\sigma_D}+\sqrt{2\gamma_A}\braket{\sigma_{A}^\dagger\sigma_{A}^{\text{in}}+\sigma_{A}^{\dagger,\text{in}}\sigma_{A}},
\end{align}
which shows how the two-particle correlations $\braket{\sigma_{A}^\dagger\sigma_D}$ appear as source terms for the acceptor population. The procedure to compute $2\Omega\Im\braket{\sigma_{A}^\dagger\sigma_D}$ involves then formal integration of the equation of motion for the acceptor polaron operator (as detailed in the Appendix~\ref{F}) under the assumption that $\Gamma_D,\Gamma_A\gg \gamma_D,\gamma_A$. Finally, one arrives at an expression for the energy transfer rate showing proportionality to the donor population $2\Omega\braket{\sigma_{A}^\dagger\sigma_D}=\kappa_\text{ET}P_D (t)$
\begin{align}
\kappa_\text{ET}=\sum_{n_D,n_{A}}\frac{2\Omega^2 f_{S_D}^{n_D} f_{S_{A}}^{n_A}(n_D\Gamma_D+n_{A}\Gamma_A)}{(n_D\Gamma_D+n_{A}\Gamma_A)^2+[\Delta-n_{D}\nu_D-n_{A}\nu_A]^2},
\end{align}
where, for simplicity, we have set $\gamma_D=\gamma_A$. The denominator of the above expression asks that the resonance condition $\omega_D-n_D\nu_D=\omega_A+n_A\nu_A$ is fulfilled, i.e., fluorescence lines of the donor $\ket{e_D,0}\to \ket{g_D,n_D}$ have to overlap with absorption lines of the acceptor $\ket{g_A,0}\to \ket{e_A,n_A}$, both of which are weighted by the respective Franck-Condon factors.\\
\indent In the case of many vibrational modes $n$ for donor and acceptor, we can generalize the result by writing general displacements $\mathcal{D}_{A}=\prod_{k=1}^{n_v} \mathcal{D}_{A}^k$ and $\mathcal{D}_D=\prod_{k=1}^{n_v} \mathcal{D}_D^k$ for all vibrational modes. The rate is computed by assuming multiple paths of energy transfer between the two molecules involving all vibrational modes. We assume an initially electronically excited state with no vibrations present $\ket{e_D;0_1,0_2...0_{n_v}}$ of molecule $D$ and ground state without vibrations $\ket{g_{A};0_1,0_2...0_{n_v}}$ for molecule $A$. The emission of molecule $D$ leads it into state $\ket{g_D;m_1,...m_k,...m_{n_v}}$ and resonant interactions can occur with state $\ket{e_{A};l_1,...l_{k'},...l_{n_v}}$ of molecule $A$. Summing over all these processes leads to an analytical cumbersome expression for $\kappa_\text{ET}$ (listed in Appendix~\ref{F}) seen simply as a generalization of the energy transfer rate from above.
Such an analytical result is the discrete version of the well established integral formulation~\cite{may2011charge} describing the overlap between the emission spectrum of molecule $D$ and absorption spectrum of molecule $A$ as illustrated in Fig.~\ref{fig64}a for a donor-acceptor pair with $8$ vibrational modes and spectral density $J(\omega)=\sum_k 2S_k\nu_k^2\Gamma_k/[\Gamma_k^2+(\omega-\nu_k)^2]$ (assumed to be identical for both molecules) stemming from the coupling of the molecular vibrations to some external phonon bath allowing for vibrational relaxation at rates $\Gamma_k$. The process is unidirectional as shown in Fig.~\ref{fig64}b as $\omega_D-\omega_A$ dictates the direction of the energy flow.\\

\subsection{Further remarks}
We have covered here the electron-vibron interaction and described the optical response to a weak classical optical drive. We have however hinted that coupling to confined optical modes can lead to strong modifications on the matter side, such as the described model of a molecule turned into an efficient quantum emitter~\cite{wang2019turning}. Indeed, many experimental and theoretical efforts have been recently targeted into the direction of cavity charge and energy transport enhancement~\cite{orgiu2015conductivity,schachenmayer2015cavity,feist2015extraordinary,hagenmuller2017cavity,hagenmuller2018cavity}, F\"{o}rster resonance energy transfer (FRET) enhancement~\cite{zhong2016non,zhong2017energy,feist2017long,du2018theory,reitz2018energy}, modified chemical reactivity~\cite{hutchinson2012modifying,galego2016suppressing,herrera2016cavity,martinezmartinez2018can,kampschulte2018cavity}, polariton dynamics~\cite{kansanen2019theory, schwartz2013polariton,durand2015non}, etc. On the theory side a large effort is aimed at translating standard cavity QED concepts (as introduced in Sec.~\ref{Sec4A}) with two-level systems into the realm of molecules. To this end, theoretical efforts consider a generalized light-electronic-vibrations problem modeled as a Holstein-Tavis-Cummings Hamiltonian. Investigations aim at providing an understanding of the vibrationally induced cavity polariton asymmetry~\cite{durand2015non,neuman2018origin}, vibrationally dressed polaritons~\cite{zeb2018exact}, dark vibronic polaritons ~\cite{herrera2017dark,herrera2018theory}, developing a cavity Born-Oppenheimer theory~\cite{flick2017atoms,flick2017cavityBorn} or deriving relevant simplified models for large scale numerics in the mesoscopic limit~\cite{pino2018tensor}.\\
\indent We also remark that molecules are natural quantum mechanical platforms interfacing electronic and vibrational degrees of freedom via an interaction that resembles the radiation pressure Hamiltonian (via a boson-spin replacement) utilized in quantum optomechanics; as vibration at terahertz frequencies are practically in the ground state even at room temperature and the strength of the coherent coupling can be comparable to the vibrational frequency, molecules could be a good platform for studying quantum state transfer between light and motion \cite{benz2016single}. Moreover, molecules are good single photon emitters as they are characterized by large emission rates and they allow for highly efficient collection schemes, thus utilizable in applications for photon-photon entanglement generation and optical quantum computing \cite{lettow2010quantum}.

\section{Conclusions}
Quantum systems, comprised of many particles, inherently coupled via common reservoirs (the electromagnetic vacuum, optical resonators, optical waveguides, etc.) can exhibit \textit{cooperative} behavior with some specific properties scaling more favorably than what would be expected from a collection of uncoupled particles. In order to analytically and numerically understand the cooperativity of light-matter platforms, we have detailed two alternative approaches based on either the time evolution of the density operator (master equation approach) or of system operators (stochastic quantum Langevin equations). Based on this toolbox, this tutorial introduced a set of equations successfully applied to chains, rings and arrays of quantum emitters or ensembles of molecules within the confined mode of optical resonators.\\
\indent A first set of applications that we have tackled are based on effects stemming from the dipole-dipole interaction occurring in dense quantum emitter ensembles. This interaction allows for the hopping of excitations in one or two-dimensional regular arrays and for the engineering of energy bands with special properties such as band gaps or Dirac points. Thus, such systems are ideal platforms for the investigation of topological quantum photonics with built-in nonlinearities. Their subradiant properties allow for applications in quantum metrology, in the design of high-fidelity photon-storage platforms and in the generation of entangled Bell states of two or more quantum emitter arrays (hinting towards quantum networks of spatially distant subwavelength arrays). Their enhanced reflectivity properties render them useful as extremely light mechanical resonators for quantum nano-optomechanical applications or as metasurfaces with enhanced optical nonlinearities at the level of a single photon. Also, subradiant rings illuminated by incoherent light can provide a natural gain medium via supported waveguide resonances leading to thresholdless nanolasers.\\
\indent A second set of applications that we have discussed are based on the enhancement of the photon-emitter coupling in cavity QED. We have demonstrated that the light-matter cooperativity can be greatly enhanced by addressing collective subradiant states of one or two-dimensional arrays which are characterized by narrow antiresonances in transmission. A hybrid cavity approach, where flat mirrors are replaced with frequency-dependent mirrors (such as subwavelength arrays or photonic crystal mirrors) can lead to narrow cavity resonances useful for example in resolved sideband optomechanics. We provided an input-output theory, valid outside Markovian regimes with wide applicability. In the same context of cavity QED, we provided a simple theory of superradiant lasers with potential applications in enhancing the stability of atomic clocks.\\
\indent As molecular systems, quantum dots, vacancy centers in solid-state hosts are increasingly important for quantum technology applications, it is crucial to understand and to tailor the electron-vibration interaction. We provided here a first-principle derivation of the Holstein Hamiltonian for electron-vibron interactions in simple molecular systems and extended the QLEs approach to polaron physics. This theory leads to very simple analytical insight in the understanding of molecular spectral lines, their absorption and emission profiles and processes such as FRET migration of energy in donor-acceptor configurations. We also made the analytical connection between the molecular dimer model, widely used in quantum chemistry for studying vibronic quantum coherence and the dipole-dipole interaction Hamiltonian. \\

\section*{Acknowledgments}
We acknowledge useful discussion with A.~Dantan. We acknowledge financial support from the Max Planck Society and from the German Federal Ministry of Education and Research, co-funded by the European Commission (project RouTe), project number 13N14839 within the research program ``Photonik Forschung Deutschland". This work was funded by the Deutsche Forschungsgemeinschaft (DFG, German Research Foundation) -- Project-ID 429529648 -- TRR 306 QuCoLiMa
(``Quantum Cooperativity of Light and Matter''). M.~R. acknowledges financial support from the International Max Planck Research School - Physics of Light (IMPRS-PL).

\bibliography{TutorialPRXQ}
\onecolumngrid

\appendix

\section*{Appendix}
\label{Appendix}
\setcounter{equation}{0}
\renewcommand{\theequation}{A\arabic{equation}}
\renewcommand{\thesubsection}{A\arabic{subsection}}

\tocless\subsection{Master equations for open-system dynamics}
\label{A}
We provide here a short review of the master equation approach in open-system dynamics closely following Ref.~\cite{breuer2002theory} but also extensively covered in, among others, Refs.~\cite{loudon1973the,weiss1999quantum,scully1997quantum,gardiner2004quantum,vogel2006quantum,walls2012quantum}. In particular we illustrate how this approach deals with the process of spontaneous emission of a single and many quantum emitters, with photon loss of an optical cavity mode and how it can be used to show the occurrence of cavity Dicke superradiant decay.\\

\noindent \textbf{General formalism}  \\

\noindent Let us consider a system $S$ which is weakly coupled to a bath (or reservoir) $B$. The Hamiltonian of the total system is given by
\begin{align}
\Hm=\Hm_0+\Hm_\text{int}=\Hm_S\otimes\mathbb{1}_B+\mathbb{1}_S\otimes \Hm_B+\Hm_\text{int},
\end{align}
where $\Hm_S$ and $\Hm_B$ denote the free Hamiltonians of the system and bath, respectively, and $\Hm_\text{int}$ accounts for the interaction between the two. The symbols $\mathbb{1}_B$ and $\mathbb{1}_S$ represent the unity in the $B$ and $S$ Hilbert spaces. In the Schr\"{o}dinger picture, the density operator satisfies $\dot{\rho}=-i[\Hm,\rho]$. We remove the free Hamiltonian by going into an interaction picture with a unitary transformation $\rho_I (t)=\mathcal{U}_0^\dagger(t)\rho(t)\mathcal{U}_0$ where $\mathcal{U}_0(t)=e^{-i\Hm_0 t}$. The von Neumann equation in the interaction picture then takes the form
\begin{align}
\label{vonneumann}
\frac{d}{dt}\rho(t)=-i[\Hm_\text{int}(t),\rho(t)],
\end{align}
where the interaction Hamiltonian is now time-dependent $\Hm\ts{int}(t)=\mathcal{U}^\dagger_0 (t)\Hm\ts{int}\mathcal{U}_0(t)$ and we removed the index $I$ which served to indicate the interaction picture. The formal solution of this is given by
\begin{align}
\rho(t)=\rho(0)-i\int_0^t ds\,[\Hm_\text{int}(s),\rho(s)].
\end{align}
Inserting this back into Eq.~\eqref{vonneumann} and taking the trace over the bath $\rho_S=\mathrm{Tr}_B[\rho]$, one finds
\begin{align}
\frac{d}{dt}\rho_S (t)=-\int_0^t ds \,\mathrm{Tr}_B[\Hm_\text{int}(t), [\Hm\ts{int}(s),\rho(s)]],
\end{align}
with $\mathrm{Tr}_B[\Hm\ts{int},\rho(0)]=0$. The above expression still contains the density operator of the total system on its right-hand side. The density matrix of the total system should show deviations on the order of $\Hm\ts{int}$ from an uncorrelated state $\rho(t)=\rho_S (t)\otimes \rho_B+\mathcal{O}(\Hm\ts{int})$.
Assuming the coupling between system and bath to be weak (Born approximation), one can neglect terms higher than second order in $\Hm\ts{int}$ and obtain an integro-differential equation for the reduced system operator
\begin{align}
\frac{d}{dt}\rho_S (t)=-\int_0^t ds \,\mathrm{Tr}_B[\Hm_\text{int}(t), [\Hm\ts{int}(s),\rho_S(s)\otimes\rho_B]].
\end{align}
A more detailed discussion of the Born approximation can also be found in Ref.~\cite{haake2006statistical}. This equation of motion can be brought in time-local form by replacing $\rho_S(s)$ with $\rho_S (t)$ which is then called the Redfield equation. Finally, to obtain a Markovian master equation, one first performs a change of variables $s\to t-s$. Under the assumption that the bath correlations decay quickly as compared to the time scale over which the state of the system varies (as induced by the bath), one may replace the upper bound of the integral by infinity and obtain
\begin{align}
\label{markov}
\frac{d}{dt}\rho_S (t)=-\int_0^\infty ds \,\mathrm{Tr}_B[\Hm_\text{int}(t), [\Hm\ts{int}(t-s),\rho_S(t)\otimes\rho_B]].
\end{align}
\\

\noindent \textbf{Spontaneous emission of a single atom}  \\

\noindent Let us consider an interaction of the form $\Hm\ts{int}=\sigma^\dagger\mathcal{F}(t)+\mathcal{F}^\dagger(t)\sigma$. For spontaneous emission of a single atom, the term driving the emitter expresses as $\mathcal{F}(t)=\sum_\mathbf{k}g_\mathbf{k}a_\mathbf{k}e^{i(\omega_0-\omega_k)t}$, where the coupling is given by
\begin{align}
g_{\mathbf{k}}=\sqrt{\frac{\omega_k}{2\epsilon_0\mathcal{V}}}\boldsymbol{\epsilon}_\mathbf{k}^{(\lambda)}\cdot\mathbf{d}\ts{eg}.
\end{align}
Note that in our notation, the sum over $k$ vectors implicitly contains a sum over the two orthogonal polarizations $\boldsymbol{\epsilon}_\mathbf{k}^{(1)}$ and $\boldsymbol{\epsilon}_\mathbf{k}^{(2)}$. Inserting the interaction Hamiltonian into Eq.~\eqref{markov}, using the cyclic property of the trace (e.g., $\mathrm{Tr}_B[\mathcal{F}(t)\rho_B(0)\mathcal{F}^\dagger(t-s)]=\mathrm{Tr}_B[\mathcal{F}^\dagger(t-s)\mathcal{F}(t)\rho_B(0)]=\braket{\mathcal{F}^\dagger(t-s)\mathcal{F}(t)}
$) and assuming a bath $\rho_B=\ket{\mathrm{vac}}\bra{\mathrm{vac}}$, the only remaining terms are
\begin{align}
\dot{\rho}(t)=\int_0^\infty ds \left\{\left[\sigma\rho(t)\sigma^\dagger-\sigma^\dagger\sigma\rho(t)\right]\braket{\mathcal{F}(t)\mathcal{F}^\dagger(t-s)}+\mathrm{h.c.}\right\}.
\end{align}
In consistence with the main text of the tutorial, we now denote the system density matrix simply by $\rho$. One is left with evaluating the term $\braket{\mathcal{F}(t)\mathcal{F}^\dagger(t-s)}=\sum_\mathbf{k}|g_\mathbf{k}|^2e^{i(\omega_0-\omega_k)s}$ since $\braket{a_{\mathbf{k}}a_{\mathbf{k}'}^\dagger}=\delta_{\mathbf{k}{\mathbf{k}'}}$. Identifying the minimal volume in $k$-space which is occupied by two modes of orthogonal polarization and same $k$ vector as $(2\pi/\ell)^3$, we can replace the sum by an integral and go into spherical coordinates
\begin{align}
\frac{1}{\mathcal{V}}\sum_\mathbf{k}\to \int\frac{d^3 k}{(2\pi)^3}=\frac{1}{(2\pi c)^3}\int_0^\infty d\omega_k\omega_k^2\int d\Omega_k.
\end{align}
Additionally, one still has to account for the sum over the two orthogonal polarizations which can be performed as $\sum_\lambda(\boldsymbol{\epsilon}_\mathbf{k}^{(\lambda)}\cdot\mathbf{d}\ts{eg})^2=d\ts{eg}^2-{(\mathbf{k}\cdot\mathbf{d}\ts{eg})^2}/k^2$. For the integration over $s$, one can make use of the Sokhotski-Plemelj theorem
\begin{align}
\label{intexp}
\int d\omega_k\int_0^\infty ds e^{\pm i(\omega_k-\omega_0)s}=\int d\omega_k\left[\pi\delta(\omega_k-\omega_0)\pm i\mathcal{P}\left(\frac{1}{\omega_k-\omega_0}\right)\right],
\end{align}
where $\mathcal{P}$ denotes the Cauchy principal value. While the real part of the above expression gives rise to decay, the imaginary part corresponds to the Lamb shift which will eventually give rise to a small frequency renormalization of the atom. Neglecting this small correction and only considering the decay, one can finally arrive at a master equation in Lindblad form
\begin{align}
\dot{\rho}(t)=\gamma\left(2\sigma\rho\sigma^\dagger-\{\sigma^\dagger\sigma, \rho\}\right),
\end{align}
where $\gamma=(\omega_0^3 d\ts{eg}^2)/(6\pi c^3\epsilon_0)$ is the spontaneous emission rate.
\\

\noindent \textbf{The master equation for collective dynamics of closely spaced quantum-emitter ensembles} \\

\noindent Let us now consider the case of $\mathcal{N}$ identical emitters located at positions $\mathbf{R}_j$ mutually coupled to the electromagnetic vacuum described by the interaction $\Hm\ts{int}=\sum_j( \sigma_j^\dagger\mathcal{F}_j(t)+\mathcal{F}_j^\dagger(t)\sigma_j)$ with $\mathcal{F}_j(t)=\sum_\mathbf{k}g_\mathbf{k} a_\mathbf{k}e^{i\mathbf{k}\mathbf{R}_j}e^{-i(\omega_k-\omega_0)t}$, where we assume equal orientation of all transition dipoles $\mathbf{d}\ts{eg}^j=\mathbf{d}\ts{eg}$. In this case, the master equation takes the form
\begin{align}
\dot{\rho}(t)=\sum_{j,j'=1}^\mathcal{N}\int_0^\infty  ds \left\{\sigma_j\rho(t)\sigma_{j'}^\dagger\left(\mathcal{C}_{j' j}(s)+\mathcal{C}_{jj'}^*(s)\right)-\sigma_{j}^\dagger\sigma_{j'}\rho (t)\mathcal{C}_{jj'}(s)-\rho(t)\sigma_j^\dagger\sigma_{j'}\mathcal{C}_{j'j}^*(s)\right\},
\end{align}
where we denote the correlations as
\begin{align}
\label{corrf}
\mathcal{C}_{jj'}(s)=\braket{\mathcal{F}_j(t)\mathcal{F}_{j'}^\dagger(t-s)}=\sum_\mathbf{k}|g_\mathbf{k}|^2e^{i\mathbf{k}(\mathbf{R}_j-\mathbf{R}_{j'})}e^{i(\omega_0-\omega_k)s},
\end{align}
and used furthermore that $\braket{\mathcal{F}_j (t)\mathcal{F}_{j'}^\dagger(t-s)}=\braket{\mathcal{F}_{j'}(t-s)\mathcal{F}_j^\dagger(t)}^*$. After summing over the two orthogonal polarizations we now again perform the continuum limit as in the previous paragraph and obtain for the correlation function Eq.~\eqref{corrf}
\begin{align}
\braket{\mathcal{F}_j(t)\mathcal{F}_{j'}^\dagger(t-s)}=\frac{d\ts{eg}^2}{4\pi^2c^3\epsilon_0}\int d\omega_k\omega_k^3 F(k R_{jj'})e^{i(\omega_0-\omega_k)s},
\end{align}
where the function $F(kR)$ is defined as the solid angle integral $F(kR)=1/(4\pi)\int d\Omega_k (1-(\mathbf{d}\ts{eg}\cdot\mathbf{k})^2)/(d\ts{eg}^2 k^2)\cdot e^{i\mathbf{k}\mathbf{R}}$. This integral can be calculated as
\begin{align}
\label{functionf}
F(kR)=\frac{1}{4\pi}\left(1+\frac{\left(\mathbf{e}_\mathbf{d}\cdot\nabla_\mathbf{R}\right)^2}{k^2}\right)\int_0^\pi d\theta_k e^{ikR\cos\theta_k}\int_0^{2\pi}d\phi_k=\left[1+\frac{(\mathbf{e}_\mathbf{d}\cdot\nabla_\mathbf{R})^2}{k^2}\right]\frac{\sin(kR)}{kR}.
 \end{align}
 Finally, resolving the integral over $s$ by means of Eq.~\eqref{intexp} results in
 \begin{align}
 \int_0^\infty ds \braket{\mathcal{F}_j (t)\mathcal{F}_{j'}^\dagger (t-s)}=\frac{3\gamma}{2}F(k_0 R_{jj'})-i\frac{3\gamma}{2} G(k_0 R_{jj'})=:\gamma_{jj'}+i\Omega_{jj'},
 \end{align}
 where
 \begin{align}
\label{cauchyprincip}
 G(k_0 R)=\frac{1}{\pi\omega_0^3}\mathcal{P}\int d\omega_k \frac{\omega_k^3}{\omega_k-\omega_0}F(kR).
 \end{align}
All together, this yields the master equation Eq.~\eqref{rhored} of the main text.
 \\

 \noindent \textbf{The master equation for cavity-induced Dicke superradiance} \\

\noindent For $\mathcal{N}$ identical emitters placed within the volume of an optical resonator, the interaction Hamiltonian is given by $\Hm\ts{int}=S^\dagger\mathcal{F}(t)+\mathcal{F}^\dagger(t)S$ with $\mathcal{F}(t)=g a(t)$. Using that the correlation of the field operator is given by (assuming the cavity field to be in the vacuum field due to its fast decay)
\begin{align}
\braket{\mathcal{F}(t)\mathcal{F}^\dagger (t-s)}=g^2 e^{-\kappa s},
\end{align}
the term appearing in the master equation is
\begin{align}
\dot{\rho}(t)=\frac{g^2}{\kappa}\left(2 S\rho S^\dagger-\{S^\dagger S, \rho\}\right).
\end{align}

\noindent \textbf{Damped cavity mode}   \\

\noindent The tunneling of photons between a confined cavity mode $a$ to the outside continuum (modes on the left $b(\omega)$ and modes on the right $c(\omega)$) can be phenomenologically described via the interaction Hamiltonian $\Hm\ts{int}=a^\dagger\mathcal{F}(t)+\mathcal{F}^\dagger (t) a$ where $\mathcal{F}(t)=\int d\omega[V(\omega)b(\omega)+W(\omega)c(\omega)]e^{-i(\omega-\omega_c)t}$ with frequency-dependent rates $V(\omega)$ and $W(\omega)$. Evaluating the correlation (approximating the outside field as $\rho_B=\rho_B^{(L)}\otimes\rho_B^{(R)}$ with $\rho_B^{(L,R)}=\ket{\mathrm{vac}}_{L,R}\bra{\mathrm{vac}}_{L,R}$)
\begin{align}
 \braket{\mathcal{F}(t)\mathcal{F}^\dagger(t-s)}=\int d\omega [V(\omega)^2+W(\omega)^2]e^{-i(\omega-\omega_c)s},
\end{align}
and subsequently the integral over $s$, gives rise to loss rates via left and right mirrors $\kappa_L=\pi V(\omega_c)^2$, $\kappa_R=\pi W(\omega_c)^2$ and a master equation in Lindblad form
\begin{align}
\dot{\rho}(t)=\kappa\left(2 a\rho a^\dagger-\{a^\dagger a, \rho\}\right),
\end{align}
where $\kappa=\kappa_L+\kappa_R$ sums the loss rates via the left and right mirror.

\tocless\subsection{Functional dependence of $F(kR)$, $G(kR)$}
\label{B}

The expression for $F(kR)$ in Eq.~\eqref{functionf} can be brought into a more convenient form widely used in the literature by expressing the nabla operator in spherical coordinates
\begin{align}
\nabla_\mathbf{R}=\partial_R\mathbf{e}_R+\frac{1}{R}\partial_\theta\mathbf{e}_\theta+\frac{1}{R\sin\theta}\partial_\phi\mathbf{e}_\phi.
\end{align}
Assuming the dipoles to be aligned along the $z$ direction $\mathbf{e}_\mathbf{d}=\mathbf{e}_z$ with $\mathbf{e}_z=\cos\theta\mathbf{e}_R-\sin\theta\mathbf{e}_\theta$, we can express
\begin{align}
\mathbf{e}_\mathbf{d}\cdot\nabla_\mathbf{R}=\cos\theta\,\partial_R-\frac{\sin\theta}{R}\partial_\theta,
\end{align}
Calculating the derivatives, one obtains
\begin{align}
F(kR)=(1-\cos^2\theta)\frac{\sin(kR)}{kR}+(1-3\cos^2\theta)\left[\frac{\cos(kR)}{(kR)^2}-\frac{\sin(kR)}{(kR)^3}\right].
\end{align}
The calculation of $G(kR)$ involves evaluation of the principal value integral Eq.~\eqref{cauchyprincip}. We just specify the result here for brevity
\begin{align}
G(kR)=(1-\cos^2\theta)\frac{\cos(kR)}{kR}-(1-3\cos^2\theta)\left[\frac{\sin(kR)}{(kR)^2}+\frac{\cos(kR)}{(kR)^3}\right].
\end{align}
\\

\tocless\subsection{Lattice sum of spherical waves}
\label{C}

The sum over spherical waves of a 2D array with atoms periodically arranged at positions $\mathbf{R}_j$ in the $x$-$y$ plane can be expressed along the $z$ direction by going into the continuum as
\begin{align}
\mathcal{S}=\sum_j \frac{e^{ik|z\mathbf{e}_z-\mathbf{R}_j|}}{|z\mathbf{e}_z-\mathbf{R}_j|}=\sum_j \int d\mathbf{R}_\parallel\frac{e^{ik\sqrt{\mathbf{R}_\parallel^2+z^2}}}{\sqrt{\mathbf{R}_\parallel^2+z^2}}\delta(\mathbf{R}_\parallel-\mathbf{R}_j),
\end{align}
where we denote the in-plane components by $\mathbf{R}_\parallel=(x,y)$. The spherical wave can be expanded by means of a Weyl expansion as \cite{novotny2006principles}
\begin{align}
\mathcal{S}=\frac{i}{2\pi}\sum_j \int d\mathbf{R}_\parallel\int d\mathbf{k}_\parallel e^{-i\mathbf{k}_\parallel\cdot \mathbf{R}_\parallel}\frac{e^{ik_z|z|}}{k_z}\delta(\mathbf{R}_\parallel-\mathbf{R}_j),
\end{align}
with in-plane momentum $\mathbf{k}_\parallel=(k_x,k_y)$ and $k_z=\sqrt{k^2-|\mathbf{k}_\parallel|^2}$. Poisson's summation formula allows one to turn the sum over the real-space lattice into a sum over the reciprocal lattice
\begin{align}
\sum_j \int d\mathbf{R}_\parallel e^{-i\mathbf{k}_\parallel\cdot\mathbf{R}_\parallel}\delta(\mathbf{R}_\parallel-\mathbf{R}_j)=\frac{(2\pi)^2}{a^2}\sum_{m,n} \delta (\mathbf{k}_\parallel-\mathbf{q}_{m,n}),
\end{align}
with reciprocal lattice vectors given by $\mathbf{q}_{m,n}=(q_m,q_n)=\frac{2\pi}{a}(m,n)$ for a square lattice with lattice constant $a$. This yields for the sum
\begin{align}
\mathcal{S}=\frac{2\pi i}{a^2}\sum_{m,n}\frac{e^{i\sqrt{k^2-|\mathbf{q}_{m,n}|^2}|z|}}{\sqrt{k^2-|\mathbf{q}_{m,n}|^2}}.
\end{align}

\tocless\subsection{One-dimensional transfer-matrix theory}
\label{D}

\begin{figure}[b]
\includegraphics[width=0.25\columnwidth]{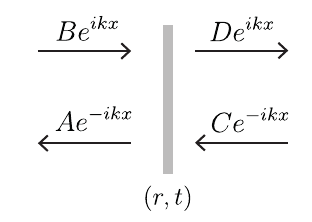}
\caption{In- and outgoing waves for an arbitrary scatterer with reflectivity and transmission coefficients $(r,t)$.}
\label{figapp1}
\end{figure}

For an arbitrary scatterer (mirror, atom, etc.) at a fixed position, we assume the modes on the left as  $A e^{-ikx}$ (left propagating) and $Be^{ikx}$ (right propagating) and on the right as  $C e^{-ikx}$ (left propagating) and $D e^{ikx}$ (right propagating), as illustrated in \fref{figapp1}. We assume the scatterer to have a complex reflectivity $r$ and transmittivity $t$. The two are connected as $t=1+r$ and in the absence of absorption one has $|t|^2+|r|^2=1$. The outgoing fields can be related to the incoming fields as
\begin{subequations}
\begin{align}
A&=rB+tC,\\
D&=tB+rC.
\end{align}
\end{subequations}
Note that  in the case where one has no incoming wave from the right ($C=0$), this simply gives $D=tB$ and $A=rB$. The fields on the left can be connected to the fields on the right via
\begin{align}
\begin{pmatrix} A \\ B \end{pmatrix}=\frac{1}{t}\begin{pmatrix}
t^2-r^2&r  \\
-r & 1
\end{pmatrix}\begin{pmatrix} C \\ D \end{pmatrix}.
\end{align}
\\
For the two-mirror arrangement considered in Sec.~\ref{Sec5B}, each mirror can be characterized by its polarizability $\zeta_j$, $j=R,L$, where the polarizability generally depends on frequency $\zeta_j=\zeta_j(\omega)$. The transmission and reflection coefficients for each mirror are given by $t_j=1/(1-i\zeta_j)$ and $r_j=i\zeta_j/(1-i\zeta_j)$, respectively and the transfer matrix of each mirror reads
\begin{align}
\mathbf{T}_j &= \begin{pmatrix}
1+i\zeta_j&i\zeta_j  \\
-i\zeta_j & 1-i\zeta_j
\end{pmatrix},
\end{align}
while the matrix for free-space propagation is given by $\mathbf{T}_f=\mathrm{diag}(e^{i\theta}, e^{-i\theta})$ where $\theta=\omega\ell/c$. The transfer matrix of the whole system is then obtained by multiplying the three individual matrices
\begin{align}
\mathbf{T}&=\mathbf{T}_R\mathbf{T}_f\mathbf{T}_L\\\nonumber
&=\begin{pmatrix}
(1+i\zeta_R)(1+i\zeta_L)e^{i\theta}+\zeta_R\zeta_L e^{-i\theta}& i(1-i\zeta_L)\zeta_R e^{-i\theta}+i\zeta_L(1+i\zeta_R)e^{i\theta} \\
- i(1+i\zeta_L)\zeta_R e^{i\theta}-i\zeta_L(1-i\zeta_R)e^{-i\theta}& (1-i\zeta_R)(1-i\zeta_L)e^{-i\theta}+\zeta_R\zeta_L e^{i\theta}
\end{pmatrix},
\end{align}
from which the cavity transmission coefficient can be obtained as $t=1/T_{22}$.\\

\tocless\subsection{Brownian noise model for vibrational relaxation}
\label{E}
A particular case of dissipation arises in the case of a quantum harmonic oscillator with free energy $\mathcal{H}=\nu(p^2+q^2)/2$ interacting with a heat bath of (mutually independent) harmonic oscillators. In the Brownian motion model
\begin{subequations}
\begin{align}
\dot{q}&=\nu p,\\
\dot{p}&=-2\Gamma p -\nu q +\xi,
\end{align}
\end{subequations}
the decay $2\Gamma$ stems from the correlations of the stochastic input noise $\xi$ which is only affecting the momentum quadrature. This model is a consequence of displacement-displacement interactions between the oscillator and an infinite surrounding bath $\mathcal{H}_{\text{int}}=-q\sum_k\alpha_k  {Q}_k$ with coupling coefficients $\alpha_k$ \cite{weiss1999quantum}.  The correlations of the input noise $\xi$ at a given bath temperature in the time domain read~\cite{genes2008ground}
\begin{equation}
\label{corrbrown}
\braket{\xi(t)\xi(t')}=\int_{-\infty}^{\infty}\frac{d\omega}{2\pi} e^{-i\omega(t-t')} S_\text{th}(\omega).
\end{equation}
where we defined the thermal noise spectrum $S_{\text{th}}(\omega)$
\begin{equation}
S_\text{th}(\omega)=\frac{2\Gamma\omega}{\nu}\left[\coth{\frac{\hbar \omega}{2k_\text{B} T}}+1\right].
\end{equation}
While generally, the noise is not delta-correlated, one can approximate Eq.~(\ref{corrbrown}) in the limit of large temperatures as (using that in this limit $\bar{n}+1/2\approx k_B T/(\hbar \nu)$ with the average thermal occupancy $\bar{n}=\left[\mathrm{exp}(\hbar\nu/k_B T)-1\right]^{-1}$ and $\coth(x)\approx 1/x$ for $x\ll 1$)
\begin{equation}
\braket{\xi(t)\xi(t')}\approx 2\Gamma (2\bar{n}+1)\delta(t-t')+i\frac{2\Gamma}{\nu}\delta'(t-t'),
\end{equation}
where $\delta'(t-t')$ is the time derivative of the delta function.
From the spectrum evaluated at $\pm \nu$ one can obtain the cooling and heating contributions: $S_\text{th}(\nu)=4\Gamma(\bar{n}+1)$ and $S_\text{th}(-\nu)=4\Gamma\bar{n}$. Generally, from these properties one can always estimate the damping rate as $2\Gamma=(S_\text{th}(\nu)-S_\text{th}(-\nu))/2$ and the equilibrium thermal occupancy as $\bar{n}=S_\text{th}(-\nu)/[S_\text{th}(\nu)+S_\text{th}(-\nu)]$. In the case of small temperatures $k_B T/(\hbar\nu)\ll 1$, the thermal spectrum becomes very asymmetric $S_\text{th}(\omega)=\left[4\omega \Gamma/\nu\right] \theta(\omega)$ as it vanishes for negative frequencies (we introduced the Heaviside function $\theta(\omega)$). Generally, one can proceed with a Fourier domain analysis of the steady state of the equations above which leads to $p(\omega)=\epsilon(\omega)\xi(\omega)$ and $q(\omega)=i\nu p(\omega)/\omega$ with the mechanical susceptibility describing the response of the phonon bath
\begin{align}
\epsilon(\omega)=\frac{i\omega}{\omega^2-\nu^2+2i\Gamma \omega}.
\end{align}
One can also make use of the fact that the noise is always delta-correlated in Fourier domain $\braket{\xi(\omega)\xi\omega')}=S_{\text{th}}(\omega)\delta(\omega+\omega')$. One can then proceed by computing the time-domain correlations (using that $\epsilon(-\omega)=\epsilon^*(\omega)$)
\begin{subequations}
\label{corrfunctions}
\begin{align}
\braket{p(t)p(t')}&=\frac{1}{2\pi}\int_{-\infty}^\infty d\omega e^{-i\omega (t-t')}|\epsilon(\omega)|^2S_{\text{th}}(\omega),\\
\braket{q(t)q(t')}&=\frac{1}{2\pi}\int_{-\infty}^\infty d\omega  e^{-i\omega (t-t')}\frac{\nu}{\omega}|\epsilon(\omega)|^2S_{\text{th}}(\omega).
\end{align}
\end{subequations}
Generally, these expressions can be solved by a contour integral in the complex plane. We note that the integral of the momentum correlation function diverges and generally requires the introduction of a cutoff frequency $\Lambda>\nu$ to keep it finite. However, for $\Gamma\ll\nu$, the expression for $|\epsilon(\omega)|^2S_{\text{th}}(\omega)$ is sharply peaked around $\pm\nu$. In this case, one can expand around the poles $\omega=\nu+\delta$ and $\omega=\nu-\delta$ with $|\delta|\ll|\nu|$ and only keep leading order terms in $\delta$. The integrals in Eq.~(\ref{corrfunctions}) can then be approximated as the Fourier transforms of Lorentzian lineshapes which is much easier to solve and results in (for $t>t'$)
\begin{align}
\braket{p(t)p(t')}=\left[\left(\bar{n}+\frac{1}{2}\right)\cos(\nu\tau)-\frac{i}{2}\sin(\nu\tau)\right]e^{-\Gamma|\tau|},
\end{align}
with $\tau=t-t'$.
\tocless\subsection{Derivation of the FRET rate}
\label{F}

Let us evaluate the term $2\Omega\Im\braket{\sigma_{A}^\dagger\sigma_D}$ which is the one that give rise to energy transfer between the two molecules. Formal integration of the equation of motion for the acceptor gives
\begin{align}
\sigma_{A}^\dagger (t)=\sigma_{A}(0)e^{-(\gamma_A-i\omega_A) t}\mathcal{D}_{A}(0)\mathcal{D}_{A}^\dagger (t)+\int_{0}^t dt' e^{-(\gamma_A-i\omega_A)(t-t')}\left[i\Omega\sigma_D^\dagger(t')+\sqrt{2\gamma}\sigma_{A}^{\dagger,\text{in}}(t')\right]\mathcal{D}_{A} (t')\mathcal{D}_{A}^\dagger (t).
\end{align}
Due to the quick vibrational relaxation, we can make a great simplification by neglecting the backaction of the acceptor's dipole moment onto the donor and assume free evolution for the donor
\begin{align}
\dot{\tilde{\sigma}}_D\approx-\left(\gamma_D+i\omega_D\right)\tilde{\sigma}_D+\sqrt{2\gamma}\tilde{\sigma}_D^{\text{in}}.
\end{align}
Integrating this, one can calculate the term $2\Omega\Im\braket{\sigma_{A}^\dagger\sigma_D}$ which requires the correlation function (assuming $t\geq t' \geq 0$ and $t\gg1/\Gamma_D$)
\begin{align}
\braket{\D_D (0) \D_D^\dagger (t') \D_D (t) \D_D^\dagger (0)}=e^{-S_D}e^{S_De^{-(\Gamma_D-i\nu_D)(t-t')}} e^{-S_De^{-(\Gamma_D+i\nu_D) t'}}e^{S_De^{-(\Gamma_D-i\nu_D) t'}},
\end{align}
where the four-operator correlation function can be reduced to a two-operator correlation function by commuting $\D_D^\dagger (0)$ with $\D_D (t)$ and $\D_D^\dagger (t')$  and using that $\D_D (0)\D^\dagger_D(0)=\mathds{1}$.
 Under the assumption of $\Gamma_D,\Gamma_A\gg \gamma_D,\gamma_A$, one can finally arrive at an expression for the energy transfer rate showing proportionality to the donor population $2\Omega\braket{\sigma_{A}^\dagger\sigma_D}=\kappa_\text{ET}P_D (t)$ (assuming $\gamma_D=\gamma_A$):
\begin{align}
\kappa_\text{ET}=\sum_{n_D,n_{A}}\frac{2\Omega^2 f_{S_D}^{n_D} f_{S_{A}}^{n_A}(n_D\Gamma_D+n_{A}\Gamma_A)}{(n_D\Gamma_D+n_{A}\Gamma_A)^2+[\Delta-n_{D}\nu_D-n_{A}\nu_A]^2}.
\end{align}
The denominator of the above expression asks that the resonance condition $\omega_D-n_D\nu_D=\omega_A+n_A\nu_A$ is fulfilled, i.e., fluorescence lines of the donor $\ket{e_D,0}\to \ket{g_D,n_D}$ have to overlap with absorption lines of the acceptor $\ket{g_A,0}\to \ket{e_A,n_A}$, both of which are weighted by the respective Franck-Condon factors.
In the case of many vibrational modes $n$ for donor and acceptor, we can generalize the result by writing general displacements $\mathcal{D}_{A}=\prod_{k=1}^{n} \mathcal{D}_{A}^k$ and $\mathcal{D}_D=\prod_{k=1}^{n} \mathcal{D}_D^k$ for all vibrational modes. The equations of motion can then be expressed in the same form as in Eqs.~(\ref{eom})
\begin{subequations}
\begin{align}
\dot{\tilde{\sigma}}_D=-\left(\gamma+i\omega_D\right)\tilde{\sigma}_D-i\Omega\tilde{\sigma}_{A}\mathcal{D}_{A}\mathcal{D}_D^\dagger+\sqrt{2\gamma}\tilde{\sigma}_D^{\text{in}},\\
\dot{\tilde{\sigma}}_{A}=-\left(\gamma+i\omega_A\right) \tilde{\sigma}_{A}-i\Omega\tilde{\sigma}_D\mathcal{D}_D\mathcal{D}_{A}^\dagger+\sqrt{2\gamma}\tilde{\sigma}_{A}^{\text{in}},
\end{align}
\end{subequations}
except that now the displacement is produced by all modes. We assume that the two molecules have the same vibrational properties $\braket{\mathcal{D}_{A}^k(t')\mathcal{D}_{A}^{k,\dagger} (t)}=\braket{\mathcal{D}_{D}^k(t')\mathcal{D}_{D}^{k,\dagger} (t)}$. This leads to a generalized energy transfer rate
\begin{align}
\label{FRETgen}
\kappa_{\text{ET}}=\sum_{\{m_k=0\}}^\infty\sum_{\{l_k=0\}}^\infty 2\prod_{k=1}^n e^{-2S_k}\frac{S_k^{(m_k+l_k)}}{m_k!l_k!}\frac{\sum_{k=1}^n(m_k+l_k)\Gamma_k\Omega^2}{[\sum_{k=1}^n (m_k+l_k)\Gamma_k]^2+[\omega_D-\omega_A-\sum_{k=1}^n(m_k+l_k)\nu_k]^2}.
\end{align}
Here the sums go over all indices $\{m_k\}=m_1,\hdots,m_n$ and $\{l_k\}=l_1,\hdots,l_n$ referring to the possible occupancies of all $n$ vibrational modes of the two involved molecules.

\end{document}